\documentclass[prd,nofootinbib,reprint,preprintnumbers,showkeys]{revtex4-1}
\usepackage{xcolor}
\usepackage{amsmath,amsfonts,bm}
\usepackage{graphicx, epstopdf,scrextend}
\usepackage{mathrsfs,mathtools}
\usepackage{enumitem}
\usepackage{tikz}
\usepackage{tikz-feynman}

\definecolor{mygreen}{rgb}{0,0.6,0}
\definecolor{myorange}{rgb}{0.8,0.5,0}
\definecolor{darkgreen}{rgb}{0,0.4,0} %% This is for hyperref.
\definecolor{darkblue}{rgb}{0,0,0.6} %% This is for hyperref.
\usepackage[colorlinks=true,linkcolor=darkblue,citecolor=darkgreen]{hyperref}

\usepackage{pgfplots,pgfplotstable}
\usepgfplotslibrary{fillbetween}
\usetikzlibrary{intersections}
\pgfplotsset{compat=newest}

\newcommand{\as}{\alpha_{\mathrm{s}}}

\newcommand{\LA}{\mathrm{A}}

\newcommand{\LC}{\mathrm{C}}
\newcommand{\scC}{\textsc{c}}
\newcommand{\scD}{\textsc{d}}
\newcommand{\LE}{\mathrm{E}}
\newcommand{\scE}{\textsc{e}}
\newcommand{\LF}{\mathrm{F}}

\newcommand{\scH}{\textsc{h}}

\newcommand{\LL}{\mathrm{L}}

\newcommand{\LR}{\mathrm{R}}
\newcommand{\scR}{\textsc{r}}

\newcommand{\scS}{\textsc{s}}
\newcommand{\LT}{\mathrm{T}}
\newcommand{\La}{\mathrm{a}}
\newcommand{\Lb}{\mathrm{b}}
\newcommand{\Lc}{\mathrm{c}}

\newcommand{\Lf}{\mathrm{f}}
\newcommand{\Lg}{\mathrm{g}}

\newcommand{\Ls}{\mathrm{s}}

\newcommand{\scV}{\textsc{v}}

\newcommand{\mpone}{{m\!+\!1}}

\newcommand{\GeV}{\ \mathrm{GeV}}
\newcommand{\TeV}{\ \mathrm{TeV}}

\newcommand{\mur}{\mu_\scR}
\newcommand{\mus}{\mu_{\textsc s}}
\newcommand{\muf}{\mu_{\rm f}}

\newcommand{\cD}{\mathcal{D}}

\newcommand{\cO}{\mathcal{O}}
\newcommand{\cP}{\mathcal{P}}

\newcommand{\cS}{\mathcal{S}}

\newcommand{\cU}{\mathcal{U}}
\newcommand{\cV}{\mathcal{V}}

\definecolor{red}{rgb}{1,0,0}

\def\ket#1{\big|{#1}\big\rangle}
\def\bra#1{\big\langle{#1}\big|}
\def\brax#1{\big\langle{#1}}   % No "|"
\def\<>#1{\big\langle{#1}\big\rangle}
\def\[]#1{\big[{#1}\big]}

\def\sket#1{\big|{#1}\big)}
\def\sbra#1{\big({#1}\big|}
\def\sbrax#1{\big({#1}}        % No "|"

\def\isbra#1{({#1}|}

\usepackage{pgfplots,pgfplotstable}

\newcommand{\errorband}[6][]{
  \pgfplotstableread{#3}\datatable
  \addplot[name path=pluserror,draw=none,no markers,forget plot]  
  table [x={#4},y expr=\thisrow{#5}+\thisrow{#6}]{\datatable};
  \addplot[name path=minuserror,draw=none,no markers,forget plot] 
  table [x={#4},y expr=\thisrow{#5}-\thisrow{#6}]{\datatable};
  \addplot[forget plot,#2] fill between[on layer={},of=pluserror and minuserror];
  \addplot [#1,no markers] table [x={#4},y={#5}]{\datatable};
}

\pgfplotsset{every axis/.style={
    width=8.2cm,
    height=8.2cm,
    grid=both,
    scaled ticks=false,
    yticklabel style={/pgf/number format/.cd, fixed,precision=5}
  }
}

\newif\ifusefigs
\usefigstrue
\begin{document}

\title{Multivariable evolution in final state parton shower algorithms}

\author{Zolt\'an Nagy}

\affiliation{
 Deutsches Elektronen-Synchrotron DESY, 
 Notkestr.\ 85, 22607 Hamburg, Germany
}

\email{Zoltan.Nagy@desy.de}

\author{Davison E.\ Soper}

\affiliation{
Institute for Fundamental Science,
University of Oregon,
Eugene, OR  97403-5203, USA
}

\email{soper@uoregon.edu}

\begin{abstract}

One can use more than one scale variable to specify the family of surfaces in the space of parton splitting parameters that define the evolution of a parton shower. Considering $e^+e^-$ annihilation, we use two variables, with shower evolution following a special path in this two dimensional space. In addition, we treat in a special way the part of the splitting function that has a soft emission singularity but no collinear singularity. This leads to certain advantages compared to the usual shower formulation with only one scale variable.

\end{abstract}

\date{19 January 2022}

\preprint{DESY-22-009}

\maketitle

%-------------------------------------------------

\section{Introduction}
\label{sec:introduction}

In a parton shower event generator, one can view the parton state as evolving according to an operator based renormalization group equation. Starting with a state with just a few partons, the shower evolves as a scale $\mus$ changes from a large value $\mu_\scH$ characteristic of the hard scattering state at the start of the shower to a low value $\muf$ on the order of $1 \GeV$. As the shower evolves, more and more partons are emitted. The function of the shower scale $\mus$ is to divide possible parton splittings into {\em resolvable} splittings, with scales $\mu > \mus$, and {\em unresolvable} splittings, with scales $\mu < \mus$. There is substantial freedom to choose exactly what this means. The space of possible splittings is divided into the resolvable and unresolvable regions by a surface labelled by $\mus$. Many different choices are possible for defining this surface. For instance, one can use a measure of the transverse momentum in the splitting to define the surface or one can use a measure of the virtuality in the splitting.

In this paper, we explore the possibility of using more than one variable to define a family of surfaces. Instead of one $\mu_\scS$, we use $\vec\mu = (\mu_{1}, \mu_{2}, \dots)$. Then evolution means moving from large values of the component scales $\mu_{n}$ to small values along a path $\vec\mu(t)$ with $0 < t < \infty$. Defining this path is then part of defining the shower algorithm. 

There is an additional freedom available when multiple scales are involved. It may be possible to divide the shower splitting functions into separate terms such that one of the terms is not sensitive to one of the scales in the sense that no singularity is encountered when this scale approaches zero. When this happens, we can modify the definition of the unresolved region for this term in a way that makes this term exactly independent of this scale. This redefinition can simplify the shower evolution.

In this paper, we explore the additional freedom obtained by using two scales instead of one.

This general concept works for proton-proton, $e^\pm$-proton, and $e^+e^-$ collisions. The simplest case is $e^+e^-$ collisions, so we consider $e^+e^-$ collisions in this paper, reserving cases involving incoming hadrons for future work.

We begin in Secs.~\ref{sec:UnresolvedAllOrder} and \ref{MultipleScales} with a general description of multivariable evolution in the framework of shower splittings defined at any order of perturbation theory and matched to perturbative QCD at any order of perturbation theory \cite{NSAllOrder}. This will help us to understand the path dependence in general. In existing parton shower programs \cite{Herwig, Pythia, Sherpa, Dire}, including ours \cite{NSI, NSII, NSspin, NScolor, Deductor, ShowerTime, NSThreshold, NSThresholdII, NSNewColor, GapColor, NSColoriPi, ISevolve, NSThrustSum}, the shower splitting operators are only defined at order $\as^1$, although some higher order contributions may be included by adjusting the scale argument of $\as$.

In Sec.~\ref{sec:SplittingFirstOrder}, we turn to parton splitting operators truncated to order $\as^1$, possibly with more than one scale. Then in Sec.~\ref{sec:UnresolvedOneScale}, we define the unresolved region for first order splittings with one scale, based either on transverse momentum or on virtuality or on angle. In Sec.~\ref{sec:UnresolvedTwoScales} we generalize this to two scales. One of these scales is one of the previously considered scales based on transverse momentum, virtuality, or angle. We make a specific useful choice for the second scale, a special choice for the unresolved region for part of the splitting operator,  and a corresponding specific choice for the path. In Sec.~\ref{sec:EvolutionTwoScales}, we examine the form of shower evolution in this scheme. In particular, we find that this gives us a different way of understanding an angular ordered shower within the context of the general formalism of Ref.~\cite{NSAllOrder}. In Sec.~\ref{sec:BetterColor}, we find that the choices made in the previous sections give us a substantially improved treatment of SU(3) color within the context of a first order parton shower. In Sec.~\ref{sec:morecomplex}, we discuss the possibility of a more complex path $\vec\mu(t)$ within the two scale space previously defined. In Sec.~\ref{sec:10TeV}, we provide a numerical example for $e^+e^-$ annihilation at 10 TeV. Finally, we provide a short summary in Sec.~\ref{sec:conclusions}. There are two appendices, \ref{sec:deductor} with details about kinematics and splitting functions and \ref{sec:I2angle} with some results about the summation of large logarithms.

%-------------------------------------------------
\section{General structure of unresolved regions}
\label{sec:UnresolvedAllOrder}

This paper generally concerns the definition of the unresolved region in the space of parton momenta in a parton shower. We will concentrate in the following sections on a single emission in a first order shower, but we begin with a discussion of the general case of a shower algorithm at an arbitrary order of perturbation theory. We use the general framework presented in Ref.~\cite{NSAllOrder}. This general framework allows for substantial freedom in choosing the functions that define a particular parton shower algorithm. We have developed particular realizations of these choices for a first order shower \cite{NSI}. It remains an open problem to realize these choices for a parton shower with splitting functions beyond order $\as$. The general theory applies to hadron-hadron collisions, lepton-hadron collisions, and $e^+e^-$ collisions, but in this paper we restrict our analysis to $e^+ e^-$ annihilation so as to present the methods that we have in mind in the simplest possible context.

Denote by $Q$ the total momentum of the electron and positron. At some stage in the shower, there are $m$ partons with momenta and flavors $\{p,f\}_m = \{p_1,f_1; p_2,f_2,\dots,p_m,f_m\}$, with
\begin{equation}
\label{eq:ISmomentumsum}
\sum_{i=1}^m p_i = Q
\;.
\end{equation}
We consider operators that create parton splittings and the exchange of virtual partons. After the action of one of these operators, we have partons with momenta and flavors $\{\hat p,\hat f\}_{\hat m}$ with $\hat m \ge m$. Momentum is conserved, so that 
\begin{equation}
\label{eq:FSmomentumsum}
\sum_{i=1}^{\hat m} \hat p_i = Q
\;.
\end{equation}

The general theory is expressed using linear operators that act on a vector space that we call the statistical space. Basis vectors for this space have the form $\sket{\{p,f,c, c', s, s'\}_m}$. Here $(c,c')$ and $(s,s')$ represent the quantum colors and spins of the $m$ partons. We use the apparatus of quantum statistical mechanics, with the color and spin part of $\sket{\{p, f, c, c', s, s'\}_m}$ representing the density matrix $\ket{\{c,s\}_m} \bra{\{c',s'\}_m}$.

\begin{figure}[t]
\ifusefigs % then we include the figure
  \centering
  \def\finalstatecut {
    % integral sign
    \draw (0.3,3.5) .. controls (0.2, 3.4) and (0,3) .. (0,2.5);
    \draw (0.0,2.5) -- (0,-2.5);
    \draw (-0.3,-3.5) .. controls (-0.2, -3.4) and (0,-3) .. (0,-2.5);
  }  

  \def\aampl#1 {
    \begin{scope}[#1]
      \path [name path=blob](0,0) ellipse (0.5 and 2); 
      
      % guidelines
      \coordinate(o) at(-10,0);
      \coordinate(o1) at(5,0);
      \coordinate(o2) at(0,0);
      
      % projection guidelines
      \coordinate(p0) at($(3.5,3)-(2.2,0)$);
      \coordinate(p1) at($(3.5,-3)-(2.2,0)$);
      \coordinate(p2) at($(3.5,3)-(2.45,0)$);
      \coordinate(p3) at($(3.5,-3)-(2.45,0)$);
      
      \path [red!30!white, name path=pr,draw] (p0)--(p1);
      \path [red!30!white, name path=pr1] (p2)--(p3);
      \path [red!30!white, name path=p] (3.25,3)--(3.25,-3);

      % guidelines of the outgoing particles
      \coordinate[rotate around={9:(o)}](a0) at(0,0);
      \coordinate[rotate around={3:(o)}](b0) at(0,0);
      \coordinate[rotate around={0:(o)}](c0) at(0,0);
      \coordinate[rotate around={-3:(o)}](d0) at(0,0);
      \coordinate[rotate around={-9:(o)}](e0) at(0,0);
      
      \coordinate[rotate around={9:(o)}](a00) at(o1);
      \coordinate[rotate around={3:(o)}](b00) at(o1);
      \coordinate[rotate around={0:(o)}](c00) at(o1);
      \coordinate[rotate around={-3:(o)}](d00) at(o1);
      \coordinate[rotate around={-9:(o)}](e00) at(o1);

      \path [gray!30!white, name path=g1] (a0)--(a00);
      \path [gray!30!white, name path=g2] (b0)--(b00);
      \path [gray!30!white, name path=g3] (c0)--(c00);
      \path [gray!30!white, name path=g4] (d0)--(d00);
      \path [gray!30!white, name path=g5] (e0)--(e00);

      % quark line
      \path[name intersections={of=g1 and blob, by={bi}}];
      \path[name intersections={of=g1 and p, by={b1i}}];
      \path[name intersections={of=g1 and pr, by={b2i}}];
      \path[name intersections={of=g1 and pr1, by={b3i}}];

      \coordinate(b') at(bi);
      \coordinate(b1') at(b1i);
      \coordinate(b2') at(b2i);
      \coordinate(b3') at(b3i);
      
      \vertex [] at(b') (a);
      \vertex [] at(b1')(a1);
      \vertex [crossed dot] at(b2')(a2){};
      \vertex [] at(b3')(a3);

      % anti-quark line
      \path[name intersections={of=g2 and blob, by={bi}}];
      \path[name intersections={of=g2 and p, by={b1i}}];
      \path[name intersections={of=g2 and pr, by={b2i}}];
      \path[name intersections={of=g2 and pr1, by={b3i}}];

      \coordinate(b') at(bi);
      \coordinate(b1') at(b1i);
      \coordinate(b2') at(b2i);
      \coordinate(b3') at(b3i);
      
      \vertex [] at(b')(b);
      \vertex [] at(b1')(b1);
      \vertex [crossed dot] at(b2')(b2){};
      \vertex [] at(b3')(b3);

      % real gluon emmision
      \vertex [dot] at ($(a2)!0.2!(a1)$)(r){};
      \vertex [] at ($(a1)!0.5!(b1)$)(r1);

      % three other hard gluons   
      \path[name intersections={of=g3 and blob, by={bi}}];
      \path[name intersections={of=g3 and p, by={b1i}}];
      \path[name intersections={of=g3 and pr, by={b2i}}];
      \path[name intersections={of=g3 and pr1, by={b3i}}];

      \coordinate(b') at(bi);
      \coordinate(b1') at(b1i);
      \coordinate(b2') at(b2i);
      \coordinate(b3') at(b3i);
      
      \vertex [] at(b')(c);
      \vertex [] at(b1')(c1);
      \vertex [crossed dot] at(b2')(c2){};
      \vertex [] at(b3')(c3);

      \path[name intersections={of=g4 and blob, by={bi}}];
      \path[name intersections={of=g4 and p, by={b1i}}];
      \path[name intersections={of=g4 and pr, by={b2i}}];
      \path[name intersections={of=g4 and pr1, by={b3i}}];

      \coordinate(b') at(bi);
      \coordinate(b1') at(b1i);
      \coordinate(b2') at(b2i);
      \coordinate(b3') at(b3i);
      
      \vertex [] at(b')(d);
      \vertex [] at(b1')(d1);
      \vertex [crossed dot] at(b2')(d2){};
      \vertex [] at(b3')(d3);

      \path[name intersections={of=g5 and blob, by={bi}}];
      \path[name intersections={of=g5 and p, by={b1i}}];
      \path[name intersections={of=g5 and pr, by={b2i}}];
      \path[name intersections={of=g5 and pr1, by={b3i}}];

      \coordinate(b') at(bi);
      \coordinate(b1') at(b1i);
      \coordinate(b2') at(b2i);
      \coordinate(b3') at(b3i);
      
      \vertex [] at(b')(e);
      \vertex [] at(b1')(e1);
      \vertex [crossed dot] at(b2')(e2){};
      \vertex [] at(b3')(e3);

      \diagram* {
        (r) -- [gluon](r1),
       % (c) -- [gluon](c3),
        (c2) -- [gluon](c1),
       % (d) -- [gluon](d3),
        (d2) -- [gluon](d1),
       % (e) -- [gluon](e3),
        (e2) -- [gluon](e1),
      };	
    \end{scope}
  }

  \begin{tikzpicture}[scale=1, transform shape]
    \begin{feynman}[small]
      \aampl{}
      % loop gluon	
      \vertex [dot] at ($(a)!0.8!(a1)$)(l){};	
      \coordinate(ll) at($(l) - (0,5)$);
      \vertex[dot] at(intersection of l--ll and b--b1)(l1){};
      
      \diagram*{
        %(a) -- [fermion](a3),
        (a2) -- [fermion](a1),
        %(b) -- [anti fermion](b3),
        (b2) -- [anti fermion](b1),
        (l) -- [gluon](l1),
      };	
      
      \coordinate(w0) at(p1);
     
      \aampl{xscale=-1, xshift=-7cm}
      \diagram*{
        %(a) -- [anti fermion](a3),
        (a2) -- [anti fermion](a1),
        %(b) -- [ fermion](b3),
        (b2) -- [ fermion](b1),
      };	
      \coordinate(w1) at(p1);

      \draw [decoration={brace}, decorate] (w1) -- (w0)
      node [pos=0.5, below] {$\left[\frac{\as(\mu_\scR^2)}{2\pi}\right]^2\cD^{(1,1)}(\mu_\scR, \vec\mu)$};

    \end{feynman}

    % integral sign
    \begin{scope}[yscale = 0.8, xshift=3.5cm]
      \finalstatecut
    \end{scope}

  \end{tikzpicture}
  
\else % We use the pdf figure.
\begin{center}
\includegraphics[width = 8.2 cm]{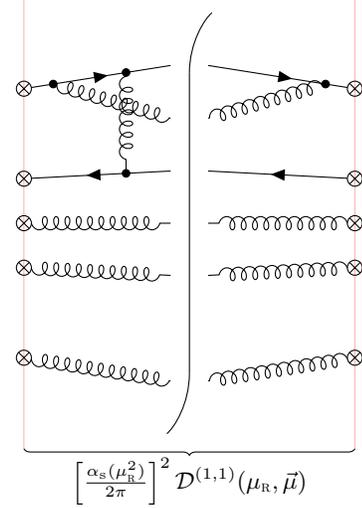}
\end{center}
\fi  
\caption{
A contribution to $\cD^{(1,1)}$. The partons next to the final state cut are on shell with momenta $\{\hat p\}_\mpone$ and spins $\{\hat s\}_\mpone$ on the left and $\{\hat s'\}_\mpone$ on the right. At the $\otimes$ symbols, some of the parton lines are off shell propagators. The $\otimes$ vertices connect these propagators to partons with momenta $\{p\}_m$ and spins $\{s\}_m$ on the left and $\{s'\}_m$ on the right.
}
\label{fig:D11}
\end{figure}

The general theory of Ref.~\cite{NSAllOrder} is based on what is called the infrared sensitive operator $\cD(\mu_\scR^2,\mu_\scS^2)$. Here $\mu_\scR$ is the standard renormalization scale and $\mu_\scS$ is called the shower scale. In this paper, we contemplate the possibility of having more than one independent shower scale, $\vec\mu = (\mu_{1}, \mu_{2}, \dots)$. The infrared sensitive operator is expanded in operators $\cD^{(n_\scR,n_\scV)}(\mu_\scR, \vec\mu)$,
\begin{equation}
\label{eq:cDfullexpansion}
\cD(\mu_\scR, \vec\mu) = 1+
\sum_{n = 1}^k \left[\frac{\as(\mu_\scR^2)}{2\pi}\right]^n
\mathop{\sum_{n_\scR = 0}^n \sum_{n_\scV = 0}^n}_{n_\scR + n_\scV = n}\,
\cD^{(n_\scR,n_\scV)}(\mu_\scR, \vec\mu)
\;.
\end{equation}
The operator $\cD^{(n_\scR,n_\scV)}(\mu_\scR, \vec\mu)$ creates $n_\scR$ real emissions and $n_\scV$ virtual exchanges. An example graph for $\cD^{(1,1)}$ is illustrated in Fig.~\ref{fig:D11}.

%%%%%%%%%%%%%%%%%%%% FIGURE %%%%%%%%%%%%%%%%%%%%%%%%%%%
% -------------------- Figure -----------------------------
% -------------------- Figure -----------------------------

%%%%%%%%%%%%%%%%%%% END FIGURE %%%%%%%%%%%%%%%%%%%%%%%%

\begin{widetext}
We propose a straightforward structure for the operator $\cD^{(n_\scR,n_\scV)}$:
\begin{equation}
\begin{split}
\label{eq:cDstructure}
\big(\{\hat p, \hat f, \hat c, \hat c', \hat s,& \hat s'\}_{m+n_\scR}\big|
\cD^{(n_\scR, n_\scV)}(\mu_\scR, \vec\mu)\sket{\{p,f,c, c', s, s'\}_m} 
\\= {}& 
\sum_{G\in\mathrm{Graphs}}\ \sum_{I\in\mathrm{Terms(G)}}
\int d^d\{\ell\}_{n_\scV} \,
\sbra{\{\hat p, \hat f\}_{m+n_\scR}}\cP(G,I)\sket{\{p,f\}_m}
\\&\times
\!\!{\phantom\langle}_\scD \!
\bra{\{\hat c, \hat s\}_{m+n_\scR}}\bm{V}_\LL(G,I; 
\{\hat p, \hat f\}_{m+n_\scR}, \{\ell\}_{n_\scV},\mur)\ket{\{c,s\}_m}
\\&\times
\bra{\{c,s\}_m}\bm{V}_\LR^\dagger(G,I; \{\hat p, \hat f\}_{m+n_\scR}, 
\{\ell\}_{n_\scV},\mur)\ket{\{\hat c, \hat s\}_{m+n_\scR}}_\scD
\\&\times
\Theta(G,I; \{\hat p, \hat f\}_{m+n_\scR}, \{\ell\}_{n_\scV}; 
{\vec\mu})
\;.
\end{split}
\end{equation}
\end{widetext}
There are $n_\scV$ virtual exchanges, so there is an integration over the space of loop momenta $\ell$ for these exchanges. There is a sum over Feynman graphs like that in Fig.~\ref{fig:D11}. It may be desirable to break the Feynman graphs into separate terms with different sorts of singularity structures. For this reason, there is a sum over terms $I$ of each graph. The factor $\sbra{\{\hat p, \hat f\}_{m+n_\scR}}\cP(G,I)\sket{\{p,f\}_m}$ consists of delta functions that fix $\{p,f\}_m$ in terms of $\{\hat p, \hat f\}_{m+n_\scR}$ according to the momentum mapping chosen for the shower, as in Appendix \ref{sec:deductor} at first order. The effects of the graphs acting on the ket state (L) and the bra state (R) are encoded in $\bm{V}_\LL$ and $\bm{V}_\LR^\dagger$, which are operators on the quantum color and spin space.\footnote{The subscripts D denote dual basis vectors, ${\phantom\langle}_\scD\!\brax{c's'}\ket{c,s} = \delta_{c',c}\, \delta_{s',s}$ \cite{NSI}.} An example at first order is worked out in Ref.~\cite{NSI}, while providing examples beyond first order remains an open problem.

\begin{figure}
\begin{center}
\ifusefigs % then we include the figure

\begin{tikzpicture}

\begin{axis}[
ymin = 0,
ymax = 1,
xmin = 0,
xmax = 1,
xticklabels = {,,},
yticklabels = {,,},
legend cell align=left,
every axis legend/.append style = {
    at={(0.95,0.95)},
    anchor=north east}
]

\addplot [name path=plotAupper,blue,thick,samples=300,domain=0.22:0.78]
{0.4 + 0.2*x + 0.4*sqrt((x-0.22)*(0.78 - x)*(2+sin(600*x)))};
\addplot [name path=plotAlower,blue,thick,samples=300,domain=0.22:0.78]
{0.4 + 0.2*x - sqrt((x-0.22)*(0.78 - x))};

\addplot [name path=plot1L,black,domain=0:0.22]{1};
\addplot [name path=plot1C,black,domain=0.22:0.78]{1};
\addplot [name path=plot1R,black,domain=0.78:1]{1};

\addplot [name path=plot0L,black,domain=0:0.22]{0};
\addplot [name path=plot0C,black,domain=0.22:0.78]{0};
\addplot [name path=plot0R,black,domain=0.78:1]{0};

\addplot[fill opacity=0.2, blue] fill between[of = plotAupper and plotAlower];
\addplot[fill opacity=0.2, yellow] fill between[of = plot1L and plot0L];
\addplot[fill opacity=0.2, yellow] fill between[of = plot1R and plot0R];
\addplot[fill opacity=0.2, yellow] fill between[of = plot1C and plotAupper];
\addplot[fill opacity=0.2, yellow] fill between[of = plot0C and plotAlower];

\addplot [red,ultra thick] coordinates{(0.5,0.35) (0.6,0.5)};
\addplot [red,ultra thick] coordinates{(0.5,0.35) (0.4,0.55)};

\end{axis}
\end{tikzpicture}

\else % We use the pdf figure.
\begin{center}
\includegraphics[width = 8.2 cm]{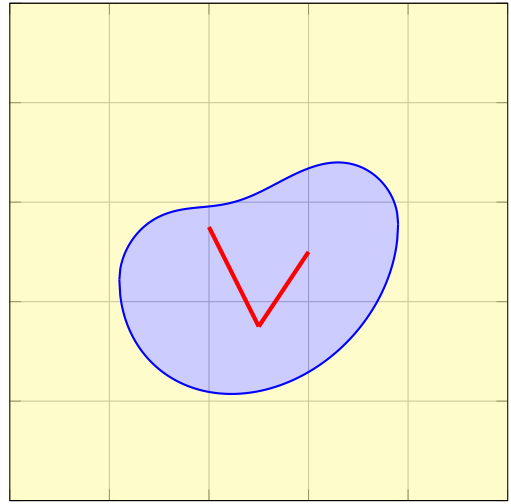}
\end{center}
\fi
\end{center}
\caption{
Resolved and unresolved regions. The red lines represent the singularities. The unresolved region for the momenta is the blue region. 
}
\label{fig:unresolvedgeneral}
\end{figure}

The final factor in Eq.~(\ref{eq:cDstructure}) is of most interest for this paper. It defines the unresolved region. The parton splitting functions $\bm{V}_\LL$ and $\bm{V}_\LR^\dagger$ are singular in a surface in the space of momenta $\{\{\hat p\}_{m+n_\scR}, \{\ell\}_{n_\scV}\}$ in which some of the momenta are exactly collinear to each other or some are zero. We illustrate this singular surface conceptually by the red lines in Fig.~\ref{fig:unresolvedgeneral}. The unresolved region is a region in the space of momenta that surrounds the singular surface. We illustrate this unresolved region by the blue area in Fig.~\ref{fig:unresolvedgeneral}. The singular surface must not extend outside of the unresolved region. That is, to borrow a phrase from general relativity, there can be {\em no naked singularity}. The idea behind this is that a measurement using an infrared safe measurement algorithm (such as a jet algorithm) cannot distinguish between a single parton and a set partons, some of which are carry very small momenta and the others of which carry momenta that are very nearly collinear. One can then say that the difference between these two states is unresolvable. The parton shower version of an unresolvable region incorporates this idea without referring to any specific infrared safe observable. In designing a shower algorithm, there is then great freedom in choosing the unresolved region. We let the boundary of the unresolved region depend on one or more scale parameters $\mu_{\scS,i}$ such that increasing any of the $\mu_{\scS,i}$ makes the unresolved region larger and decreasing $\mu_{\scS,i}$ makes the unresolved region smaller. The function $\Theta(G,I; \{\hat p, \hat f\}_{m+n_\scR}, \{\ell\}_{n_\scV};  {\vec\mu})$, equals 1 when $\{\{\hat p\}_{m+n_\scR}, \{\ell\}_{n_\scV}\}$ is in the unresolved region and equals 0 otherwise. The complement of the unresolved region is the resolved region, colored yellow in Fig.~\ref{fig:unresolvedgeneral}.

%%%%%%%%%%%%%%%%%%%% FIGURE %%%%%%%%%%%%%%%%%%%%%%%%%%%
%%%%%%%%%%%%%%%%%%% END FIGURE %%%%%%%%%%%%%%%%%%%%%%%%

%-------------------------------------------------
\section{Multiple shower scales}
\label{MultipleScales}

With multiple shower scales $\vec\mu$, the singular operator depends on these scales and on the renormalization scale $\mu_\scR$. We can let the renormalization scale $\mu_\scR$ be some function of the shower scales,\footnote{In a first order shower, one can modify the shower splitting functions by adjusting the argument of $\as$ to be not $\mu_\scR^2$ but an approximation to $k_\LT^2$. This is intended to incorporate terms from higher order splitting functions into the first order splitting function, but it is separate from the general formalism discussed here.}
\begin{equation}
\label{eq:muR}
\mu_\scR = \mu_\scR(\vec\mu)
\;.
\end{equation}
Then we can simplify the notation by writing
\begin{equation}
\cD(\vec\mu) = 
\cD(\mu_\scR(\vec\mu), \vec\mu)
\;.
\end{equation}
The shower will evolve from hard scales $\vec\mu_\scH$ that are characteristic of the hard scattering that initiates the shower to soft scales $\vec\mu_\Lf$ that are on the order of $1 \GeV$. 

For $e^+e^-$ annihilation, we define the shower evolution operator by
\begin{equation}
\label{eq:cUdef}
\cU(\vec\mu_2,\vec\mu_1) = \cD^{-1}(\vec\mu_2)\,\cD(\vec\mu_1)
\;.
\end{equation}
We should note that this is special to $e^+e^-$ annihilation. For hadron-hadron collisions, there are initial state singularities and we need parton distribution functions. Then there is a mismatch between the evolution equation for the parton distribution functions and the shower evolution. This mismatch requires the introduction of an operator $\cV$ that accounts for threshold logarithms \cite{NSAllOrder, NSThreshold, NSThresholdII}. For $e^+e^-$ annihilation, one can arrange that $\cV = 1$. We will return to the analysis of multiple scale evolution for hadron-hadron collisions in a later paper. In this paper, we restrict the analysis to $e^+e^-$ annihilation.

Using Eq.~(\ref{eq:cUdef}), to go from the hard scales to the final soft scales we have
\begin{equation}
\cU(\vec\mu_\Lf,\vec\mu_\scH) = \cD^{-1}(\vec\mu_\Lf)\,\cD(\vec\mu_\scH)
\;.
\end{equation}
We choose a path $\vec\mu(t)$ from $\vec\mu_\scH$ to $\vec\mu_\Lf$,
\begin{equation}
\begin{split}
\vec\mu(0) ={}& \vec\mu_\scH
\;,
\\
\vec\mu(t_\Lf) ={}& \vec\mu_\Lf
\;.
\end{split}
\end{equation}
We define the shower splitting operator
\begin{equation}
\label{eq:cSdef}
\cS(t) = 
-\cD^{-1}(\vec\mu(t))
\frac{d}{d t}\,
\cD(\vec\mu(t))
\;.
\end{equation}
That is
\begin{equation}
\label{eq:cSfromcSj}
\cS(t) = -\sum_j \frac{d\mu_j(t)}{dt}\,\cS_j(\vec\mu(t))
\;,
\end{equation}
where
\begin{equation}
\label{eq:cSjdef}
\cS_j(\vec\mu) = 
\cD^{-1}(\vec\mu)\,
\frac{\partial}{\partial \mu_j}\,
\cD(\vec\mu)
\;.
\end{equation}
(Following the notation in Ref.~\cite{NSAllOrder} when there is only one shower scale $\mu$, we would have defined $\mu^2(d/d\mu^2) \cD = \cD \cS$, but with more than one scale it is more convenient to use a simple derivative with respect to $\mu_j$.)

%%%%%%%%%%%%%%%%%%%% FIGURE %%%%%%%%%%%%%%%%%%%%%%%%%%%
\begin{figure}
\begin{center}
\ifusefigs % then we include the figure

\begin{tikzpicture}

\begin{axis}[
xlabel={$\mu_1$}, 
ylabel={$\mu_2$},
ymin = 0,
ymax = 1,
xmin = 0,
xmax = 1,
xticklabels = {,,},
yticklabels = {,,},
legend cell align=left,
every axis legend/.append style = {
    at={(0.95,0.95)},
    anchor=north east}
]

\addplot [mygreen,thick,samples=300,domain=0.05:0.85, postaction={decorate, decoration={markings,
    mark=at position 0.5 with {\arrow[scale=1.1,>=Latex]{<};},}}]
{0.05 + 1.1*(x-0.05) + 0.3*((x-0.05)*(0.85 - x)*(2+sin(600*x)))};

\addplot [blue,thick,samples=300,domain=0.05:0.85, postaction={decorate, decoration={markings,
    mark=at position 0.5 with {\arrow[scale=1.1,>=Latex]{<};},}}]
{0.05 + 1.1*(x-0.05) - 0.3*((x-0.05)*(0.85 - x)*(2+sin(400*x)))};

\addplot [
scatter,only marks,point meta=explicit symbolic, 
scatter/classes={a={mark=*,black,mark size=2.5pt}
},
] table [meta=label] {
x y label 
0.05 0.05 a 
0.85 0.93 a 
};

\node[]at({0.2,0.06}){$\vec\mu(t_\Lf)$};
\node[]at({0.87,0.85}){$\vec\mu(0)$};

\end{axis}
\end{tikzpicture}

\else % We use the pdf figure.
\begin{center}
\includegraphics[width = 8.2 cm]{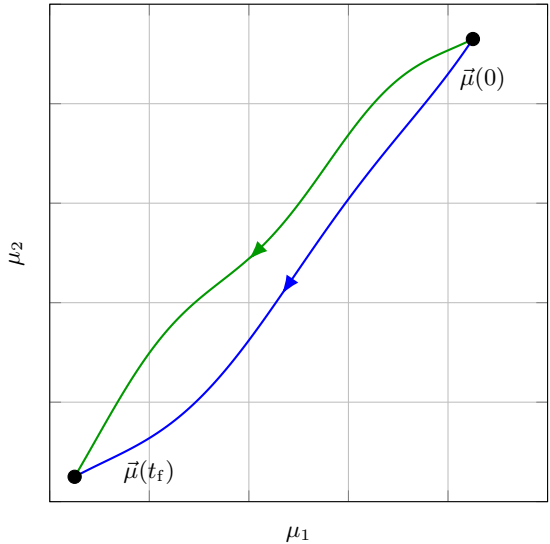}
\end{center}
\fi
\end{center}
\caption{
Two paths in the space of scales from  $\vec\mu(0)$ to $\vec\mu(t_\Lf)$.
}
\label{fig:twopaths}
\end{figure}
%%%%%%%%%%%%%%%%%%% END FIGURE %%%%%%%%%%%%%%%%%%%%%%%%

Then
\begin{equation}
\label{eq:cUdiffeqn}
\frac{d}{dt}\,\cU(\vec\mu(t),\vec\mu (t_1))
= \cS(t)\,
\cU(\vec\mu(t),\vec\mu(t_1))
\;.
\end{equation}
We can write the solution of this differential equation with boundary condition $\cU(\vec\mu(t_1),\vec\mu(t_1)) = 1$ as
\begin{equation}
\cU\big(\vec\mu(t_2),\vec\mu(t_1)\big) 
=\cU\big(t_2,t_1\big) 
\;,
\medskip
\end{equation}
where, using $\mathbb{T}$ to indicate ordering of operators along the path,
\begin{equation}
\label{eq:cUexponential}
\cU\big(t_2,t_1\big)
= \mathbb{T}\exp\!\left(\int_{t_1}^{t_2}\! dt\
\cS(t)\right)
.
\end{equation}
The operator $\cU(t_2,t_1)$ depends on the chosen path. We illustrate schematically the possibility of two paths in Fig.~\ref{fig:twopaths} between $\vec\mu(0) = \vec\mu_\scH$ and $\vec\mu(t_\Lf) = \vec\mu_\Lf$. If $\cS(t)$ is defined exactly according to Eq.~(\ref{eq:cSdef}), then $\cU\big(t_\Lf,0\big)$ does not depend on $\vec\mu(t)$ for intermediate values of $t$, $0 < t < t_\Lf$. This follows simply because of the definition Eq.~(\ref{eq:cUdef}) of $\cU$. However, it is often useful to use an approximation for $\cS(t)$, for instance by using only a finite number of terms in its perturbative expansion. If $\cS(t)$ is approximated in any way and $\cU\big(t_\Lf,0\big)$ is obtained by solving the differential equation (\ref{eq:cUdiffeqn}), then $\cU\big(t_\Lf,0\big)$ can depend on the whole path.

To understand the dependence on the path, we can consider an altered path between the same endpoints as illustrated in Fig~\ref{fig:twopaths}:
\begin{equation}
\mu_j(t;\varepsilon) = \mu_j(t) + \varepsilon \eta_j(t)
\;,
\end{equation}
where
\begin{equation}
\eta_j(0) = \eta_j(t_\Lf) = 0
\;.
\end{equation}
We can let $\cU_{\varepsilon}\big(t_\Lf,0\big)$ denote the shower evolution operator over the path that has been deformed by an amount $\varepsilon$. We evaluate this operator between the two fixed points at which the deformation vanishes. Then after a little analysis we find
\begin{widetext}
\begin{equation}
\left[\frac{d}{d\varepsilon}\,\cU_{\varepsilon}
\big(t_\Lf,0\big)\right]_{\varepsilon = 0} =  
\int_0^{t_\Lf}\!dt\ 
\cU\big(t_\Lf,t\big)
\sum_{i,j} \frac{d\mu_i(t)}{dt}\,\eta_j(t)
\left[
\frac{\partial S_j(\vec\mu(t))}{\partial \mu_i}
- \frac{\partial S_i(\vec\mu(t))}{\partial \mu_j}
+ [S_i(\vec\mu(t)),S_j(\vec\mu(t))]
\right]
\cU\big(t,0\big)
\;.
\end{equation}
\end{widetext}
The expression in square brackets vanishes if we use Eq.~(\ref{eq:cSjdef}) exactly to define $S_j(\vec\mu(t))$, but not otherwise. If the perturbative expansion of $S_j(\vec\mu(t))$ is truncated at order $\as^N$, then the expression in square brackets will be of order $\as^{N+1}$.

In this paper, we work with a first order shower, in which the perturbative expansion of $S_j(\vec\mu(t))$ is truncated at order $\as^1$. The most straightforward choice of path in the first order shower is computationally difficult because of noncommuting color matrices. We use the freedom to specify a path $\vec\mu(t)$ to create a first order shower algorithm that is computationally simpler than with the more straightforward choice of path. The computationally difficult parts of the more straightforward approach are eliminated because, with the chosen path, they would appear only at order $\as^2$.

%-------------------------------------------------
\section{Splitting at first order}
\label{sec:SplittingFirstOrder}

We now turn to the description of the unresolved region for $e^+ e^-$ annihilation with total momentum $Q$ in a first order parton shower, such as \textsc{Deductor}. In the description that we use in this paper, the partons carry color\footnote{We describe the color treatment in somewhat more detail in Sec.~\ref{sec:BetterColor}.}, but, as in the current version of \textsc{Deductor}, we average over spins so that there are no spin states represented in the parton states. We begin with $m$ partons, in a state $\sket{\{p, f, c, c'\}_m}$. The singular operator $\cD(\vec\mu)$ has a perturbative expansion
\begin{equation}
\label{cDexpansion}
\cD(\vec\mu) = 1 + \cD^{[1]}(\vec\mu) + \cO(\as^2)
\;,
\end{equation}
where $\cD^{[1]}(\vec\mu)$ contains a factor of $\as$,
\begin{equation}
\cD^{[1]}(\vec\mu) = \frac{\as(\mur^2(\vec\mu))}{2\pi}\,\cD^{(1)}(\vec\mu)
\;.
\end{equation}
The operator $\cD^{[1]}(\vec\mu)$ consists of two terms,
\begin{equation}
\label{eq:cD10cD01}
\cD^{[1]}(\vec\mu) = \cD^{[1,0]}(\vec\mu) + \cD^{[0,1]}(\vec\mu)
\;.
\end{equation}
In $\cD^{[1,0]}(\vec\mu)$ one of the partons splits into two. In $\cD^{[0,1]}(\vec\mu)$, a virtual parton is exchanged, leaving the number of partons unchanged. 

In the real emission operator $\cD^{[1,0]}(\vec\mu)$, let $l$ be the label of the parton that splits, so that $p_l$ is its momentum. This splitting produces two partons, which we label $l$ and $\mpone$. These partons carry momenta $\hat p_l$ and $\hat p_\mpone$. In \textsc{Deductor}, the momenta of the other partons after the splitting, $\hat p_i$, are adjusted by means of a small Lorentz transformation so that momentum is conserved, as in Eq.~(\ref{eq:FSmomentumsum}). We can describe the splitting by splitting variables $(y,z,\phi)$. Here $\phi$ is the azimuthal angle of $\hat p_\mpone$ about the $p_l$ axis in the rest frame of $Q$. The momentum fraction $z$ is defined by
\begin{equation}
\label{eq:zdef}
\frac{1-z}{z} = \frac{\hat p_\mpone \cdot n_l}{\hat p_l \cdot n_l}
\;,
\end{equation}
where the lightlike vector $n_l$ is
\begin{equation}
\label{eq:nldef}
n_l = \frac{2 p_l\cdot Q}{Q^2}\,Q - p_l
\;.
\end{equation}
Finally, $y$ is the dimensionless virtuality variable
\begin{equation}
\label{eq:ydef}
y =  \frac{2\hat p_l \cdot \hat p_\mpone}{2 p_l\cdot Q}
\;.
\end{equation}

The default ordering variable in \textsc{Deductor} is $\Lambda^2$, defined by \cite{ShowerTime}
\begin{equation}
\label{eq:Lambda}
\Lambda^2 = y Q^2 = 
a_l\, 2\hat p_l \cdot \hat p_\mpone
\;,
\end{equation}
where $a_l$ is a dimensionless measure of the inverse of the energy of the mother parton,
\begin{equation}
\label{eq:aldef}
a_l = \frac{Q^2}{2 p_l \cdot Q}
\;.
\end{equation}
Momentum conservation implies that $a_l \ge 1$.

Parton splittings are often described by the squared transverse momentum $k_\LT^2$ in the splitting.  Then with the kinematic definitions used in \textsc{Deductor}, as outlined in Appendix \ref{sec:deductor},
\begin{equation}
\label{eq:kTsq}
k_\LT^2 = \frac{z (1-z)}{a_l} \, \Lambda^2
\;.
\end{equation}

We can also describe the parton splitting using the angle variable
\begin{equation}
\vartheta = \frac{1}{2}\,[1 - \cos(\theta)]
\;,
\end{equation}
where $\theta$ is the angle between the daughter parton momenta in the rest frame of $Q$. That is
\begin{equation}
\label{eq:vartheta}
\vartheta = \frac{\hat p_l \cdot \hat p_{\mpone}\ Q^2}
{2 \hat p_l\cdot Q \ \hat p_\mpone\cdot Q}
\;.
\end{equation}

The variables $k_\LT^2$, $\Lambda^2$, and $\vartheta Q^2$ are measures of the hardness of a splitting. We can relate these variables. We relate $k_\LT^2$ to $\Lambda^2$ using Eq.~(\ref{eq:kTsq}). To relate $\vartheta$ to $\Lambda^2$ we can use the definition (\ref{eq:vartheta}),
\begin{equation}
\vartheta\, Q^2 = a_l\, \frac{p_l\cdot Q}{\hat p_l\cdot Q} 
\frac{p_l\cdot Q}{\hat p_\mpone\cdot Q}\ \Lambda^2 
\;.
\end{equation}
For small angle splittings, in which $\hat p_l \approx z p_l$ and $\hat p_\mpone \approx (1-z) p_l$, this is
\begin{equation}
\label{eq:AfromLambda}
\vartheta\, Q^2 \approx  \frac{a_l}{z(1-z)}\, \Lambda^2 
\;.
\end{equation}
(The exact relationship is in Eq.~(\ref{eq:aperpdef}) or Eq.~(\ref{eq:xlthetal}).) Thus $\Lambda^2$ lies between $k_\LT^2$ and $\vartheta Q^2$: $k_\LT^2$ is smaller by a factor $z(1-z)/a_l$ and $\vartheta Q^2$ is larger by (approximately) the inverse of this factor.

Let $\cD^{[1,0]}(\vec\mu)$ act on a state with parton momenta and flavors $\{p,f\}_m$. Consider the contribution in which parton $l$ splits with splitting variables $(y,z,\phi)$ and flavor $\hat f_\mpone$ of the emitted parton. This contribution is proportional to an operator that we can call $\bm D_l(\{\hat p,\hat f\}_\mpone;\{p,f\}_m)$. Here $\bm D_l$ is a function of the momenta and flavors before and after the splitting but is still an operator on the color space of the partons. The relation of $\cD^{[1,0]}(\vec\mu)$ to $\bm D_l(\{\hat p,\hat f\}_\mpone;\{p,f\}_m)$ is outlined in Appendix \ref{sec:deductor}.

For $e^+ e^-$ annihilation (but not for hadron-hadron collisions), $\cD^{[0,1]}(\vec\mu)$ is determined from $\cD^{[1,0]}(\vec\mu)$ in a simple way \cite{NSThreshold}. See Appendix \ref{sec:deductor}.

We specify $\bm D_l(\{\hat p,\hat f\}_\mpone;\{p,f\}_m)$ in detail in Appendix \ref{sec:deductor}, but for now these details do not matter. What is important is that $\bm D_l$ exhibits collinear and soft singularities. To describe these singularities, it is useful to consider $\bm D_l$ at fixed $\{p\}_m$ to be a function of  the angle variable $\vartheta$, Eq.~(\ref{eq:vartheta}), and the momentum fraction $z$, Eq.~(\ref{eq:zdef}). Then $\bm D_l$ is singular in the collinear limit, $\vartheta \to 0$ with fixed $z$, in the soft limit $(1-z) \to 0$ with fixed $\vartheta$, and in the soft$\times$collinear limit, $(1-z) \to 0$ and $\vartheta\to 0$. It is of some significance that $\bm D_l$ can be decomposed into two terms,
\begin{equation}
\begin{split}
\label{eq:scsoft}
\bm D_l(\{\hat p,\hat f\}_\mpone;&\{p,f\}_m) 
\\={}& 
\bm D_l^\mathrm{sc}(\{\hat p,\hat f\}_\mpone;\{p,f\}_m) 
\\&
+ \bm D_l^\mathrm{soft}(\{\hat p,\hat f\}_\mpone;\{p,f\}_m)
\;,
\end{split}
\end{equation}
where $\bm D_l^\mathrm{sc}$ has both soft and collinear singularities, while $\bm D_l^\mathrm{soft}$ has a soft singularity but no collinear singularity (and no soft$\times$collinear singularity). An example of such a decomposition will be given in Eq.~(\ref{eq:cSjLCplus}). The decomposition of $\bm D_l$ leads to a corresponding decomposition of $\cD^{[1,0]}(\vec\mu)$, 
\begin{equation}
\label{eq:cDscandsoft}
\cD^{[1,0]}(\vec\mu)=
\cD^{[1,0]}_\mathrm{sc}(\vec\mu)
+\cD^{[1,0]}_\mathrm{soft}(\vec\mu)
\;.
\end{equation}
We will make use of this decomposition in this paper to treat the two terms differently.

%-------------------------------------------------
\section{Unresolved region with one scale}
\label{sec:UnresolvedOneScale}

We now consider the unresolved region for a splitting  in a first order shower in the standard case that there is a single shower scale $\mu_\scS$. 

The operator $\cD^{[1,0]}(\mu_\scS)$ contains an integration over splitting variables $(y,z,\phi)$ with $(y,z)$ integrated over the unresolved region defined by the scale $\mu_\scS$. The shower splitting operator is given by the first order version of Eq.~(\ref{eq:cSjdef}),
\begin{equation}
\label{eq:cSjdef1storder}
\cS^{[1,0]}(\mu_\scS) = 
\frac{d}{d\mu_\scS}\,
\cD^{[1,0]}(\mu_\scS)
\;.
\end{equation}
Integrating between a scale $\mu_{\scS,1}$ and a slightly smaller scale $\mu_{\scS,2}$ gives the exponent in Eq.~(\ref{eq:cUexponential}) for shower evolution between these two scales,
\begin{equation}
\label{eq:cSintegral}
\int_{\mu_{\scS,2}}^{\mu_{\scS,1}}\!d\mu_\scS\
\cS^{[1,0]}(\mu_\scS)
= \cD^{[1,0]}(\mu_{\scS,1})
-\cD^{[1,0]}(\mu_{\scS,2})
\;.
\end{equation}
Thus we integrate over the unresolved region for the larger scale omitting the unresolved region for the smaller scale.

There are several possibilities for how the unresolved region depends on the scale. In each of three cases that we consider, we adopt a different name for the shower scale, $\mu_\Ls^2 = \mu_\perp^2$, $\mu_\Ls^2 = \mu_\Lambda^2$, and $\mu_\Ls^2 = \mu_\angle^2$.

One possibility is to define the unresolved region for a splitting by $k_\LT^2 < \mu_\perp^2$, where $k_\LT^2$ was defined in Eq.~(\ref{eq:kTsq}). The angle variable $\vartheta$ is related to $k_\LT^2$ and $z$ in any kinematically allowed splitting by
\begin{equation}
\label{eq:Afromkappa}
\vartheta = a_\perp(z,k_\LT^2)
\;,
\end{equation}
where
\begin{widetext}%--------------------
\begin{equation}
\label{eq:aperpdef}
a_\perp(z,k_\LT^2)
=
\begin{cases}
\displaystyle{\frac{a_l^2 k_\LT^2/Q^2}
{(z(1-z) + a_l k_\LT^2/Q^2)^2 +  a_l^2  (1 - 4 z(1-z))k_\LT^2/Q^2} }
& \mathrm{for}\ z(1-z) > c_z k_\LT^2/Q^2
\\
1 & \mathrm{otherwise}
\end{cases}
\;.
\end{equation}
\end{widetext}%--------------------
Here $a_l$ was defined in Eq.~(\ref{eq:aldef}) and
\begin{equation}
c_z = a_l\,
(\sqrt{a_l} + \sqrt{a_l-1})^2
\;.
\end{equation}
Only splittings with $z(1-z) > c_zk_\LT^2/Q^2$ allow the variable $\lambda(y)$ in Eq.~(\ref{eq:lambdahpmdefs}) to be defined, so only these splittings are kinematically possible. We have set $a_\perp(z,k_\LT^2) = 1$ in the case that $k_\LT^2$ is too large to allow a splitting with momentum fraction $z$.

%%%%%%%%%%%%%%%%%%%% FIGURE %%%%%%%%%%%%%%%%%%%%%%%%%%%
\begin{figure}
\begin{center}
\ifusefigs % then we include the figure

\begin{tikzpicture}

\newcommand\al{2.0}
\newcommand\kappaA{0.002}
\newcommand\ymaxinverse{(sqrt(\al)+sqrt(\al-1))^2}
\newcommand\xxmin{0.5*(1 - sqrt(1 - 4*\al*\kappaA*\ymaxinverse))}
\newcommand\xxmax{1 - \xxmin}

\newcommand\anglekT{\al^2*\kappaA/((x*(1-x) + \al*\kappaA)^2 
+ \al^2*\kappaA*(1 - 4*x*(1-x)))};

\newcommand\kappaepsilon{0.0005}
\newcommand\xepsmin{0.5*(1 - sqrt(1 - 4*\al*\kappaepsilon*\ymaxinverse))}
\newcommand\xepsmax{1 - \xepsmin}
\newcommand\angleepsilonT{\al^2*\kappaepsilon/((x*(1-x) + \al*\kappaepsilon)^2 
+ \al^2*\kappaepsilon*(1 - 4*x*(1-x)))};

\begin{axis}[
xlabel={$1-z$}, 
ylabel={$\vartheta$},
ymin = 0,
ymax = 1,
xmin = 0,
xmax = 1,
legend cell align=left,
every axis legend/.append style = {
    at={(0.95,0.95)},
    anchor=north east}
]

\addplot [name path=plot1,blue,thick,domain=\xxmin:\xxmax]
{\anglekT};
\addplot [name path=plot2,black,domain=\xxmin:\xxmax]{0};
\addplot [name path=plot3,thin, black,domain=\xxmin:\xxmax]{1};

\addplot [name path=plot1L,blue,thick,domain=0:\xxmin]{1};
\addplot [name path=plot2L,black,domain=0:\xxmin]{0};

\addplot [name path=plot1R,blue,thick,domain=\xxmax:1]{1};
\addplot [name path=plot2R,black,domain=\xxmax:1]{0};

\addplot [name path=ploteps,black,dashed,domain=\xepsmin:\xepsmax]
{\angleepsilonT};

\addplot[fill opacity=0.2, blue] fill between[of = plot1 and plot2];
\addplot[fill opacity=0.2, blue] fill between[of = plot1L and plot2L];
\addplot[fill opacity=0.2, blue] fill between[of = plot1R and plot2R];
\addplot[fill opacity=0.2, yellow] fill between[of = plot1 and plot3];

\addplot [red,ultra  thick] coordinates{(0.003,0.0) (0.003,1)};
\addplot [red,ultra  thick] coordinates{(0.0,0.003) (1.0,0.003)};

\end{axis}
\end{tikzpicture}

\else % We use the pdf figure.
\begin{center}
\includegraphics[width = 8.2 cm]{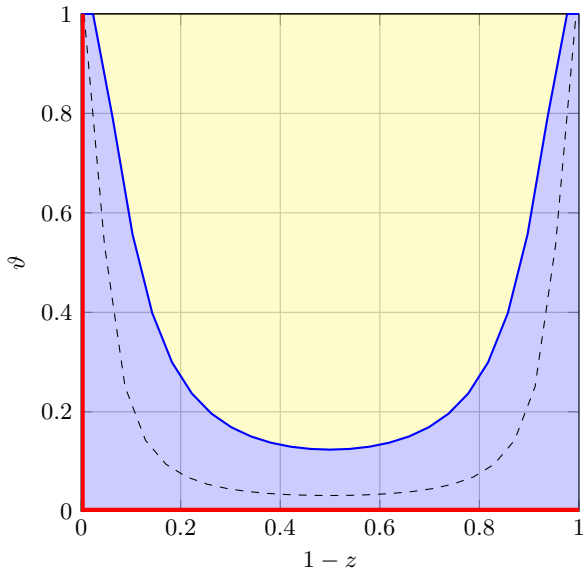}
\end{center}
\fi
\end{center}
\caption{
Resolved and unresolved regions for fixed $\mu_\perp^2$. Here $a_l = 2$, $m_\perp^2 = 0.0005\,Q^2$, and $\mu_\perp^2 = 0.002\,Q^2$. (Thus $m_\perp^2$ is too small to play a role in this figure. The curve for $m_\perp^2$ is shown as a dashed line.)
}
\label{fig:unresolvedkappa}
\end{figure}
%%%%%%%%%%%%%%%%%%% END FIGURE %%%%%%%%%%%%%%%%%%%%%%%%

We can use the function $a_\perp(z,k_\LT^2)$ to define the unresolved region specified by the singular operator $\cD^{[1,0]}$ for $k_\LT$ ordering.
We first address an issue concerning the range of $k_\LT^2$. The argument of $\as$ used in \textsc{Deductor} and other shower generators is an approximation to $k_\LT^2$.\footnote{Precisely, \textsc{Deductor} uses $k_\LT^2/z = (1-z) 2 \hat p_l \cdot \hat p_\mpone$ in $\as$. This is the same as $k_\LT^2$ for $(1-z) \to 0$. The splitting functions have no $z \to 0$ singularity.} We cannot trust perturbation theory if $\as$ is not small. For this reason, a splitting with squared transverse momentum $k_\LT^2$ must be considered unresolved if $k_\LT^2$ is smaller than a value $m_\perp^2$ of order $1 \GeV^2$. The parameter $m_\perp^2$ is not an adjustable scale parameter but rather serves as a fixed cutoff parameter. We therefore define the unresolved region corresponding to a shower scale $\mu_\scS^2 \equiv \mu_\perp^2$ by
\begin{equation}
\label{eq:unresolvedperp}
\vartheta < 
\max [a_\perp(z,\mu_\perp^2),a_\perp(z,m_\perp^2)]
\;.
\end{equation}
This is illustrated in Fig.~\ref{fig:unresolvedkappa} for the choice of shower scale parameter $\mu_\perp^2 = 0.002\,Q^2$ with $a_l = 2$ and $m_\perp^2 = 0.0005\, Q^2$. The singular surface, consisting of the lines $\vartheta = 0$ and $(1-z) = 0$ is indicated in red. The curve $\vartheta = a_\perp(z,\mu_\perp^2)$ with $\mu_\perp^2 = 0.002\,Q^2$ is indicated in blue. The unresolved region is the blue region below this curve. Note that the unresolved region includes the entire singular surface. The resolved region is the yellow region above this curve.

%%%%%%%%%%%%%%%%%%%% FIGURE %%%%%%%%%%%%%%%%%%%%%%%%%%%
\begin{figure}
\begin{center}
\ifusefigs % then we include the figure

\begin{tikzpicture}

\newcommand\al{2.0}
\newcommand\ymaxinverse{(sqrt(\al)+sqrt(\al-1))^2}
\newcommand\kappaA{0.002}
\newcommand\kappaB{0.004}
\newcommand\xminA{0.5*(1 - sqrt(1 - 4*\al*\kappaA*\ymaxinverse))}
\newcommand\xmaxA{1 - \xminA}
\newcommand\xminB{0.5*(1 - sqrt(1 - 4*\al*\kappaB*\ymaxinverse))}
\newcommand\xmaxB{1 - \xminB}

\newcommand\anglekTA{\al^2*\kappaA/((x*(1-x) + \al*\kappaA)^2 
+ \al^2*\kappaA*(1 - 4*x*(1-x)))};

\newcommand\anglekTB{\al^2*\kappaB/((x*(1-x) + \al*\kappaB)^2 
+ \al^2*\kappaB*(1 - 4*x*(1-x)))};

\begin{axis}[
xlabel={$1-z$}, 
ylabel={$\vartheta$},
ymin = 0,
ymax = 1,
xmin = 0,
xmax = 1,
legend cell align=left,
every axis legend/.append style = {
    at={(0.95,0.95)},
    anchor=north east}
]

\addplot [name path=plotAleft,blue,thick,domain=\xminA:\xminB]{\anglekTA};
\addplot [name path=plotAcenter,blue,thick,domain=\xminB:\xmaxB]{\anglekTA};
\addplot [name path=plotAright,blue,thick,domain=\xmaxB:\xmaxA]{\anglekTA};

\addplot [name path=plotBleft,blue,thick,domain=\xminA:\xminB]{1};
\addplot [name path=plotBcenter,blue,thick,domain=\xminB:\xmaxB]{\anglekTB};
\addplot [name path=plotBright,blue,thick,domain=\xmaxB:\xmaxA]{1};

\addplot[fill opacity=0.3, mygreen] fill between[of = plotAleft and plotBleft];
\addplot[fill opacity=0.3, mygreen] fill between[of = plotAcenter and plotBcenter];
\addplot[fill opacity=0.3, mygreen] fill between[of = plotAright and plotBright];

\addplot [red,ultra  thick] coordinates{(0.003,0.0) (0.003,1)};
\addplot [red,ultra  thick] coordinates{(0.0,0.003) (1.0,0.003)};

\end{axis}
\end{tikzpicture}

\else % We use the pdf figure.
\begin{center}
\includegraphics[width = 8.2 cm]{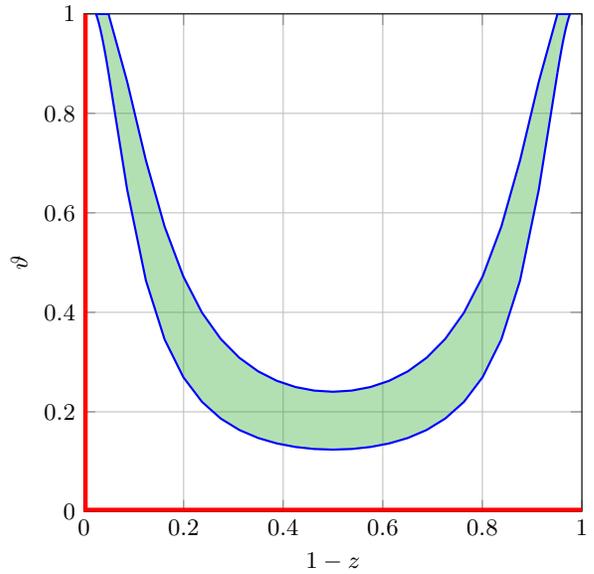}
\end{center}
\fi
\end{center}
\caption{
Evolution in $\mu_\perp^2$ for $0.002 < \mu_\perp^2/Q^2 < 0.004$. Here $a_l = 2$, $m_\perp^2 = 0.0005\,Q^2$.}
\label{fig:evolutionkappa}
\end{figure}
%%%%%%%%%%%%%%%%%%% END FIGURE %%%%%%%%%%%%%%%%%%%%%%%%

If we use a $k_\LT$-ordered shower, then shower evolution from scale $\mu_{\perp,1}^2$ to a smaller scale $\mu_{\perp,2}^2$ includes splittings in the unresolved region for the larger scale but not splittings that are unresolved at the smaller scale, as in Eq.~(\ref{eq:cSintegral}). This is illustrated in Fig.~\ref{fig:evolutionkappa} for the case $\mu_{\perp,1}^2 = 0.004\,Q^2$, $\mu_{\perp,2}^2 = 0.002\,Q^2$. The region covered is displayed in green in the figure.

The unresolved region can also be defined by $\Lambda^2 < \mu_\Lambda^2$, supplemented by a fixed cut $k_\LT^2 < m_\perp^2$. Here $\Lambda^2$ is the virtuality variable defined in Eq.~(\ref{eq:Lambda}). The angle variable $\vartheta$ is related to $y = \Lambda^2/Q^2$ and $z$ by
\begin{equation}
\label{eq:AusingaLambda}
\vartheta = a_\Lambda(z,\Lambda^2)
\;,
\end{equation}
where 
\begin{equation}
\label{eq:Afromy}
a_\Lambda(z,yQ^2)
= \frac{a_l y}{(1 + y)^2  z(1-z) +  a_l y  (1 - 4 z(1-z))}
\;.
\end{equation}
%

%%%%%%%%%%%%%%%%%%%% FIGURE %%%%%%%%%%%%%%%%%%%%%%%%%%%
\begin{figure}
\begin{center}
\ifusefigs % then we include the figure

\begin{tikzpicture}

\newcommand\al{2.0}
\newcommand\yA{0.02}
\newcommand\kappaA{0.0005}
\newcommand\ymaxinverse{(sqrt(\al)+sqrt(\al-1))^2}
\newcommand\xkappamin{0.5*(1 - sqrt(1 - 4*\al*\kappaA*\ymaxinverse))}
\newcommand\xkappamax{1 - \xkappamin}

\newcommand\zomz{\al*\kappaA/\yA}
\newcommand\xcrossL{2*\zomz/(1 + sqrt(1 - 4*\zomz))}
\newcommand\xcrossR{1-\xcrossL}

\newcommand\angley{\al*\yA/((1 + \yA)^2*x*(1-x)
+ \al*\yA*(1 - 4*x*(1-x)))};
\newcommand\anglekT{\al^2*\kappaA/((x*(1-x) + \al*\kappaA)^2 
+ \al^2*\kappaA*(1 - 4*x*(1-x)))};

\begin{axis}[
xlabel={$1-z$}, 
ylabel={$\vartheta$},
ymin = 0,
ymax = 1,
xmin = 0,
xmax = 1,
legend cell align=left,
every axis legend/.append style = {
    at={(0.95,0.95)},
    anchor=north east}
]

\addplot [name path=plotyleft,black,dashed,domain=0:\xcrossL,samples=30]
{\angley};
\addplot [name path=plotycenter,blue,thick,domain=\xcrossL:\xcrossR,samples=30]
{\angley};
\addplot [name path=plotyright,black,dashed,domain=\xcrossR:1,samples=30]
{\angley};

\addplot [name path=plotkTleft,mygreen,thick,samples=100,domain=\xkappamin:\xcrossL]
{\anglekT};
\addplot [name path=plotkTcenter,black,dashed,domain=\xcrossL:\xcrossR]
{\anglekT};
\addplot [name path=plotkTright,mygreen,thick,domain=\xcrossR:\xkappamax]
{\anglekT};

\addplot [name path=plot0LL,black,domain=0:\xkappamin]{0};
\addplot [name path=plot0LC,black,domain=\xkappamin:\xcrossL]{0};
\addplot [name path=plot0center,black,domain=\xcrossL:\xcrossR]{0};
\addplot [name path=plot0CR,black,domain=\xcrossR:\xkappamax]{0};
\addplot [name path=plot0RR,black,domain=\xkappamax:1]{0};

\addplot [name path=plot1LL,blue,domain=0:\xkappamin]{1};
\addplot [name path=plot1LC,black,domain=\xkappamin:\xcrossL]{1};
\addplot [name path=plot1center,black,domain=\xcrossL:\xcrossR]{1};
\addplot [name path=plot1CR,black,domain=\xcrossR:\xkappamax]{1};
\addplot [name path=plot1RR,blue,domain=\xkappamax:1]{1};

\addplot[fill opacity=0.2, blue] fill between[of = plot0LL and plot1LL];
\addplot[fill opacity=0.2, blue] fill between[of = plot0LC and plotkTleft];
\addplot[fill opacity=0.2, blue] fill between[of = plot0center and plotycenter];
\addplot[fill opacity=0.2, blue] fill between[of = plot0CR and plotkTright];
\addplot[fill opacity=0.2, blue] fill between[of = plot0RR and plot1RR];

\addplot[fill opacity=0.2, yellow] fill between[of = plotkTleft and plot1LC];
\addplot[fill opacity=0.2, yellow] fill between[of = plotycenter and plot1center];
\addplot[fill opacity=0.2, yellow] fill between[of = plotkTright and plot1CR];

\addplot [red,ultra  thick] coordinates{(0.003,0.0) (0.003,1)};
\addplot [red,ultra  thick] coordinates{(0.0,0.003) (1.0,0.003)};

\end{axis}
\end{tikzpicture}

\else % We use the pdf figure.
\begin{center}
\includegraphics[width = 8.2 cm]{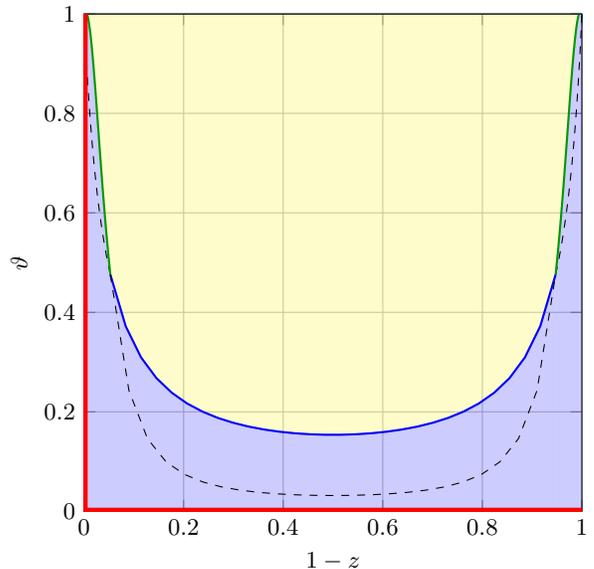}
\end{center}
\fi
\end{center}
\caption{
Unresolved regions for fixed $\mu_\Lambda^2 = 0.02\,Q^2$. Here $a_l = 2$ and $m_\perp^2 = 0.0005\,Q^2$.
}
\label{fig:unresolvedyepsilon}
\end{figure}
%%%%%%%%%%%%%%%%%%% END FIGURE %%%%%%%%%%%%%%%%%%%%%%%%

%%%%%%%%%%%%%%%%%%%% FIGURE %%%%%%%%%%%%%%%%%%%%%%%%%%%xxx
\begin{figure}
\begin{center}
\ifusefigs % then we include the figure

\begin{tikzpicture}

\newcommand\al{2.0}
\newcommand\yAlow{0.016}
\newcommand\yAhigh{0.032}
\newcommand\kappaA{0.0005}
\newcommand\ymaxinverse{(sqrt(\al)+sqrt(\al-1))^2}
\newcommand\xkappamin{0.5*(1 - sqrt(1 - 4*\al*\kappaA*\ymaxinverse))}
\newcommand\xkappamax{1 - \xkappamin}

\newcommand\zomzhigh{\al*\kappaA/\yAhigh}
\newcommand\zomzlow{\al*\kappaA/\yAlow}
\newcommand\xcrossLhigh{2*\zomzhigh/(1 + sqrt(1 - 4*\zomzhigh))}
\newcommand\xcrossRhigh{1-\xcrossLhigh}
\newcommand\xcrossLlow{2*\zomzlow/(1 + sqrt(1 - 4*\zomzlow))}
\newcommand\xcrossRlow{1-\xcrossLlow}

\newcommand\angleylow{\al*\yAlow/((1 + \yAlow)^2*x*(1-x)
+ \al*\yAlow*(1 - 4*x*(1-x)))};
\newcommand\angleyhigh{\al*\yAhigh/((1 + \yAhigh)^2*x*(1-x)
+ \al*\yAhigh*(1 - 4*x*(1-x)))};

\newcommand\anglekT{\al^2*\kappaA/((x*(1-x) + \al*\kappaA)^2 
+ \al^2*\kappaA*(1 - 4*x*(1-x)))};

\begin{axis}[
xlabel={$1-z$}, 
ylabel={$\vartheta$},
ymin = 0,
ymax = 1,
xmin = 0,
xmax = 1,
legend cell align=left,
every axis legend/.append style = {
    at={(0.95,0.95)},
    anchor=north east}
]

\addplot [name path=plotyhighleft,black,dashed,
  domain=0:\xcrossLhigh,samples=30]
  {\angleyhigh};
\addplot [name path=plotyhighLC,blue,thick,
  domain=\xcrossLhigh:\xcrossLlow,samples=30]
  {\angleyhigh};
\addplot [name path=plotyhighcenter,blue,thick,
  domain=\xcrossLlow:\xcrossRlow,samples=30]
  {\angleyhigh};
\addplot [name path=plotyhighCR,blue,thick,
  domain=\xcrossRlow:\xcrossRhigh,samples=30]
  {\angleyhigh};
\addplot [name path=plotyhighright,black,dashed,
  domain=\xcrossRhigh:1,samples=30]
  {\angleyhigh};

\addplot [name path=plotylowleft,black,dashed,
  domain=0:\xcrossLlow,samples=30]
  {\angleylow};
\addplot [name path=plotylowcenter,blue,thick,
  domain=\xcrossLlow:\xcrossRlow,samples=30]
  {\angleylow};
\addplot [name path=plotylowright,black,dashed,
  domain=\xcrossRlow:1,samples=30]
  {\angleylow};

\addplot [name path=plotkTLL,black,dashed,
  domain=\xkappamin:\xcrossLhigh]
  {\anglekT};
\addplot [name path=plotkTLC,mygreen,thick,
  domain=\xcrossLhigh:\xcrossLlow]
  {\anglekT};
\addplot [name path=plotkTcenter,black,dashed,
  domain=\xcrossLlow:\xcrossRlow]
  {\anglekT};
\addplot [name path=plotkTCR,mygreen,thick,
  domain=\xcrossRlow:\xcrossRhigh]
  {\anglekT};
\addplot [name path=plotkTRR,black,dashed,
  domain=\xcrossRhigh:\xkappamax]
  {\anglekT};

\addplot[fill opacity=0.3, mygreen] fill between[of = plotkTLC and plotyhighLC];
\addplot[fill opacity=0.3, mygreen] fill between[of = plotylowcenter and plotyhighcenter];
\addplot[fill opacity=0.3, mygreen] fill between[of = plotkTCR and plotyhighCR];

\addplot [red,ultra  thick] coordinates{(0.003,0.0) (0.003,1)};
\addplot [red,ultra  thick] coordinates{(0.0,0.003) (1.0,0.003)};

\end{axis}
\end{tikzpicture}

\else % We use the pdf figure.
\begin{center}
\includegraphics[width = 8.2 cm]{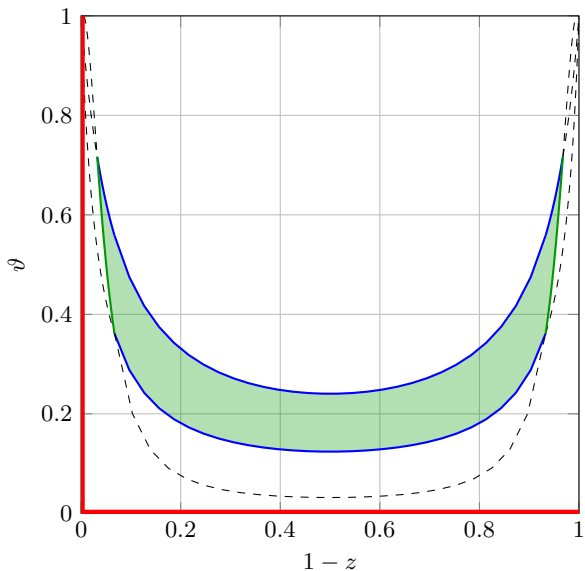}
\end{center}
\fi
\end{center}
\caption{
Evolution in $\mu_\Lambda^2$ with $m_\perp^2 = 0.0005\,Q^2$. Here $a_l = 2$ and $0.016 < \mu_\Lambda^2/Q^2 < 0.032$. This figure is analogous to Fig.~\ref{fig:evolutionkappa} for $k_\LT$ ordering.
}
\label{fig:evolutionyepsilon}
\end{figure}
%%%%%%%%%%%%%%%%%%% END FIGURE %%%%%%%%%%%%%%%%%%%%%%%%

We can use the function $a_\Lambda(z,\Lambda^2)$ to define the unresolved region specified by the singular operator $\cD^{[1,0]}$ for $\Lambda$ ordering, which is the default choice in \textsc{Deductor}. Since \textsc{Deductor} uses an approximation to $k_\LT^2$ as the argument of $\as$, we again do not allow $k_\LT^2$ to be smaller than a fixed cutoff parameter $m_\perp^2$ of order $1 \GeV^2$ in the resolved region. With this definition, the unresolved region for a given choice of the shower scale $\mu_\scS^2 \equiv \mu_\Lambda^2$ is defined by\footnote{If $\mu_\Lambda^2/Q^2 > (\sqrt{a_l} + \sqrt{a_l - 1})^{-2}$, then $a_\Lambda(z,\mu_\Lambda^2) > 1$ for $0 < z < 1$, so all splittings with $0 < \vartheta < 1$, $0 < z < 1$ are unresolved.}
\begin{equation}
\vartheta < \max [a_\Lambda(z,\mu_\Lambda^2),a_\perp(z,m_\perp^2)]
\;.
\end{equation}
This region is illustrated in Fig.~\ref{fig:unresolvedyepsilon} in the case $a_l = 2$ with $m_\perp^2 = 0.0005\,Q^2$ for the choice of shower scale parameter $\mu_\Lambda^2 = 0.02\,Q^2$. Again, the singular surface is indicated in red and the unresolved region is depicted in blue.

If we use a $\Lambda$-ordered shower with a $k_\LT^2$ cutoff at a small fixed scale $m_\perp^2 = 0.0005\,Q^2$, then shower evolution from scale $\mu_{\Lambda,1}^2$ to a smaller scale $\mu_{\Lambda,2}^2$ includes splittings in the unresolved region for the larger scale but not splittings that are unresolved at the smaller scale. This is illustrated in Fig.~\ref{fig:evolutionyepsilon} for the case $\mu_{\Lambda,1}^2 = 0.032\,Q^2$, $\mu_{\Lambda,2}^2 = 0.016\,Q^2$. The region covered is displayed in green in the figure. Note how the region of small $z(1-z)$ is removed by the $m_\perp^2$ cut.

%%%%%%%%%%%%%%%%%%%% FIGURE %%%%%%%%%%%%%%%%%%%%%%%%%%%
\begin{figure}
\begin{center}
\ifusefigs % then we include the figure

\begin{tikzpicture}

\newcommand\al{2.0}
\newcommand\Avalue{0.4}
\newcommand\kappaA{0.0005}
\newcommand\ymaxinverse{(sqrt(\al)+sqrt(\al-1))^2}
\newcommand\xkappamin{0.5*(1 - sqrt(1 - 4*\al*\kappaA*\ymaxinverse))}
\newcommand\xkappamax{1 - \xkappamin}

\newcommand\coefA{\Avalue}
\newcommand\coefB{(-2*\al*(2*\al - 1)*\Avalue*\kappaA)}
\newcommand\coefC{(-\al^2*(\kappaA*(1 - \Avalue) - \Avalue*\kappaA^2))}
\newcommand\zomz{(-\coefB + sqrt(\coefB^2 - 4*\coefA*\coefC))/2/\coefA}
\newcommand\xcrossL{2*\zomz/(1 + sqrt(1 - 4*\zomz))}
\newcommand\xcrossR{1-\xcrossL}

\newcommand\anglekT{\al^2*\kappaA/((x*(1-x) + \al*\kappaA)^2 
+ \al^2*\kappaA*(1 - 4*x*(1-x)))};

\begin{axis}[
xlabel={$1-z$}, 
ylabel={$\vartheta$},
ymin = 0,
ymax = 1,
xmin = 0,
xmax = 1,
legend cell align=left,
every axis legend/.append style = {
    at={(0.95,0.95)},
    anchor=north east}
]

\addplot [name path=plotAleft,black,dashed,domain=0:\xcrossL,samples=30]
{\Avalue};
\addplot [name path=plotAcenter,blue,thick,domain=\xcrossL:\xcrossR,samples=30]
{\Avalue};
\addplot [name path=plotAright,black,dashed,domain=\xcrossR:1,samples=30]
{\Avalue};

\addplot [name path=plotkTleft,mygreen,thick,samples=100,domain=\xkappamin:\xcrossL]
{\anglekT};
\addplot [name path=plotkTcenter,black,dashed,domain=\xcrossL:\xcrossR]
{\anglekT};
\addplot [name path=plotkTright,mygreen,thick,domain=\xcrossR:\xkappamax]
{\anglekT};

\addplot [name path=plot0LL,black,domain=0:\xkappamin]{0};
\addplot [name path=plot0LC,black,domain=\xkappamin:\xcrossL]{0};
\addplot [name path=plot0center,black,domain=\xcrossL:\xcrossR]{0};
\addplot [name path=plot0CR,black,domain=\xcrossR:\xkappamax]{0};
\addplot [name path=plot0RR,black,domain=\xkappamax:1]{0};

\addplot [name path=plot1LL,blue,domain=0:\xkappamin]{1};
\addplot [name path=plot1LC,black,domain=\xkappamin:\xcrossL]{1};
\addplot [name path=plot1center,black,domain=\xcrossL:\xcrossR]{1};
\addplot [name path=plot1CR,black,domain=\xcrossR:\xkappamax]{1};
\addplot [name path=plot1RR,blue,domain=\xkappamax:1]{1};

\addplot[fill opacity=0.2, blue] fill between[of = plot0LL and plot1LL];
\addplot[fill opacity=0.2, blue] fill between[of = plot0LC and plotkTleft];
\addplot[fill opacity=0.2, blue] fill between[of = plot0center and plotAcenter];
\addplot[fill opacity=0.2, blue] fill between[of = plot0CR and plotkTright];
\addplot[fill opacity=0.2, blue] fill between[of = plot0RR and plot1RR];

\addplot[fill opacity=0.2, yellow] fill between[of = plotkTleft and plot1LC];
\addplot[fill opacity=0.2, yellow] fill between[of = plotAcenter and plot1center];
\addplot[fill opacity=0.2, yellow] fill between[of = plotkTright and plot1CR];

\addplot [red,ultra  thick] coordinates{(0.003,0.0) (0.003,1)};
\addplot [red,ultra  thick] coordinates{(0.0,0.003) (1.0,0.003)};

\end{axis}
\end{tikzpicture}

\else % We use the pdf figure.
\begin{center}
\includegraphics[width = 8.2 cm]{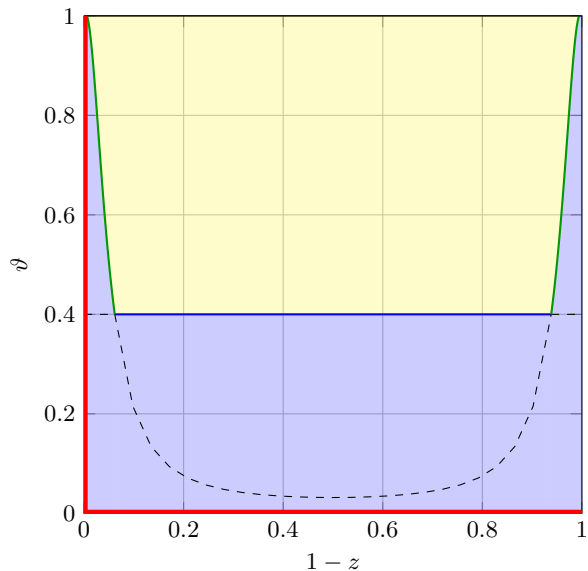}
\end{center}
\fi
\end{center}
\caption{
Unresolved regions for fixed $\mu_\angle^2$ with cutoff $m_\perp^2 = 0.0005\,Q^2$. Here $a_l = 2$ and $\mu_\angle^2 = 0.4\,Q^2$.
}
\label{fig:unresolvedAepsilon}
\end{figure}
%%%%%%%%%%%%%%%%%%% END FIGURE %%%%%%%%%%%%%%%%%%%%%%%%

Finally, we can use angular ordering and define the unresolved region by $\vartheta Q^2 < \mu_\angle^2$, supplemented by a fixed cut $k_\LT^2 < m_\perp^2$. We define
\begin{equation}
a_\angle(z, \vartheta Q^2) = \vartheta
\;.
\end{equation}
With this definition, the unresolved region for a given choice of the shower scale $\mu_\scS^2 \equiv \mu_\angle^2$ is defined by
\begin{equation}
\label{eq:unresolvedangularordered}
\vartheta < \max [a_\angle(z,\mu_\angle^2),a_\perp(z,m_\perp^2)]
\;.
\end{equation}
This region is illustrated in Fig.~\ref{fig:unresolvedAepsilon} in the case $a_l = 2$ with $m_\perp^2 = 0.0005\,Q^2$ for the choice of shower scale parameter $\mu_\angle^2 = 0.4\,Q^2$. Again, the singular surface is indicated in red and the unresolved region is depicted in blue.

There is an important difference between the unresolved regions for $\Lambda$ ordering, Fig.~\ref{fig:unresolvedyepsilon}, and angular ordering, Fig.~\ref{fig:unresolvedAepsilon}. With $\Lambda$ ordering, we could set $m_\perp^2 = 0$. There would be a problem with $\as$ with an argument proportional to $(1-z)$ when $(1-z) \ll 1$, but this problem could be eliminated by letting the argument of $\as$ be $y Q^2$.  With angular ordering, if $m_\perp^2$ were zero, there would be a naked singularity: points $((1-z), \vartheta)$ with $(1-z) = 0$ are in the resolved region when $\vartheta > \mu_\angle^2/Q^2$. Thus we need a nonzero $m_\perp^2$ with angular ordering.

We have described the unresolved region for three choices of a single ordering variable. Angular ordering is available in \textsc{Herwig} \cite{Herwig1992,Herwig}. Variants of $k_\LT$ ordering are used in \textsc{Pythia} \cite{Pythia}, \textsc{Sherpa} \cite{Sherpa}, and \textsc{Dire} \cite{Dire}. The default ordering variable in \textsc{Deductor} \cite{Deductor} is $\Lambda$. The papers \cite{PanScales, HamiltonShowerSum} have a family of ordering choices defined by a parameter $\beta$. With $\beta = 0$, the ordering variable is a transverse momentum variable, although other features of the shower are not the same as in the \textsc{Deductor} shower. With $\beta = 1/2$, the ordering variable is not among those investigated in this paper but is roughly half way between $k_\LT$ and $\Lambda$.

We note that the shower that we discuss here is a full dipole shower with interference between emitting a gluon from one parton and emitting the same gluon from a second parton. All that we do with angular ordering is to use the emission angle as the ordering variable. Thus no approximation involving averaging over the azimuthal angle of the emission is involved, as it is in \textsc{Herwig} \cite{Herwig1992,Herwig}.

%%%%%%%%%%%%%%%%%%%% FIGURE %%%%%%%%%%%%%%%%%%%%%%%%%%%
\begin{figure}
\begin{center}
\ifusefigs % then we include the figure

\begin{tikzpicture}

\newcommand\al{2.0}
\newcommand\Avalue{0.6}
\newcommand\kappaA{0.0005}
\newcommand\ymaxinverse{(sqrt(\al)+sqrt(\al-1))^2}
\newcommand\xkappamin{0.5*(1 - sqrt(1 - 4*\al*\kappaA*\ymaxinverse))}
\newcommand\xkappamax{1 - \xkappamin}

\newcommand\zomz{0.2}
\newcommand\xcrossL{2*\zomz/(1 + sqrt(1 - 4*\zomz))}
\newcommand\xcrossR{1-\xcrossL}

\newcommand\anglekT{\al^2*\kappaA/((x*(1-x) + \al*\kappaA)^2 
+ \al^2*\kappaA*(1 - 4*x*(1-x)))};

\begin{axis}[
xlabel={$1-z$}, 
ylabel={$\vartheta$},
ymin = 0,
ymax = 1,
xmin = 0,
xmax = 1,
legend cell align=left,
every axis legend/.append style = {
    at={(0.95,0.95)},
    anchor=north east}
]

\addplot [name path=plotAcenter,blue,thick,domain=\xcrossL:\xcrossR,samples=30]
{\Avalue};

\addplot [name path=plotkTleft,black,dashed,samples=100,domain=\xkappamin:\xcrossL]
{\anglekT};
\addplot [name path=plotkTcenter,black,dashed,domain=\xcrossL:\xcrossR]
{\anglekT};
\addplot [name path=plotkTright,black,dashed,domain=\xcrossR:\xkappamax]
{\anglekT};

\addplot [name path=plot0LL,black,domain=0:\xkappamin]{0};
\addplot [name path=plot0LC,black,domain=\xkappamin:\xcrossL]{0};
\addplot [name path=plot0center,black,domain=\xcrossL:\xcrossR]{0};
\addplot [name path=plot0CR,black,domain=\xcrossR:\xkappamax]{0};
\addplot [name path=plot0RR,black,domain=\xkappamax:1]{0};

\addplot [name path=plot1LL,blue,domain=0:\xkappamin]{1};
\addplot [name path=plot1LC,black,domain=\xkappamin:\xcrossL]{1};
\addplot [name path=plot1center,black,domain=\xcrossL:\xcrossR]{1};
\addplot [name path=plot1CR,black,domain=\xcrossR:\xkappamax]{1};
\addplot [name path=plot1RR,blue,domain=\xkappamax:1]{1};

\addplot [blue,thick] coordinates{(\xcrossL,\Avalue) (\xcrossL,1)};
\addplot [blue,thick] coordinates{(\xcrossR,\Avalue) (\xcrossR,1)};

\addplot[fill opacity=0.2, blue] fill between[of = plot0LL and plot1LL];
\addplot[fill opacity=0.2, blue] fill between[of = plot0LC and plot1LC];
\addplot[fill opacity=0.2, blue] fill between[of = plot0center and plotAcenter];
\addplot[fill opacity=0.2, blue] fill between[of = plot0CR and plot1CR];
\addplot[fill opacity=0.2, blue] fill between[of = plot0RR and plot1RR];

\addplot[fill opacity=0.2, yellow] fill between[of = plotAcenter and plot1center];

\addplot [red,ultra  thick] coordinates{(0.003,0.0) (0.003,1)};
\addplot [red,ultra  thick] coordinates{(0.0,0.003) (1.0,0.003)};

\end{axis}
\end{tikzpicture}

\else % We use the pdf figure.
\begin{center}
\includegraphics[width = 8.2 cm]{sample.pdf}
\end{center}
\fi
\end{center}
\caption{
Unresolved regions for fixed $\vec\mu = (\mu_\scE,\mu_\angle)$ with cutoff $m_\perp^2 = 0.0005\,Q^2$. Here $a_l = 2$. The two scales are $\mu_\scE^2 = 0.2\,Q^2$ and $\mu_\angle^2 = 0.6\,Q^2$.
}
\label{fig:unresolvedA2epsilon}
\end{figure}
%%%%%%%%%%%%%%%%%%% END FIGURE %%%%%%%%%%%%%%%%%%%%%%%%

%%%%%%%%%%%%%%%%%%%% FIGURE %%%%%%%%%%%%%%%%%%%%%%%%%%%
\begin{figure}
\begin{center}
\ifusefigs % then we include the figure

\begin{tikzpicture}

\newcommand\al{2.0}
\newcommand\yA{0.1}
\newcommand\kappaA{0.0005}
\newcommand\Avalue{0.708}
\newcommand\ymaxinverse{(sqrt(\al)+sqrt(\al-1))^2}
\newcommand\xkappamin{0.5*(1 - sqrt(1 - 4*\al*\kappaA*\ymaxinverse))}
\newcommand\xkappamax{1 - \xkappamin}

\newcommand\zomz{0.2}
\newcommand\xcrossL{2*\zomz/(1 + sqrt(1 - 4*\zomz))}
\newcommand\xcrossR{1-\xcrossL}

\newcommand\angley{\al*\yA/((1 + \yA)^2*x*(1-x)
+ \al*\yA*(1 - 4*x*(1-x)))};
\newcommand\anglekT{\al^2*\kappaA/((x*(1-x) + \al*\kappaA)^2 
+ \al^2*\kappaA*(1 - 4*x*(1-x)))};

\begin{axis}[
xlabel={$1-z$}, 
ylabel={$\vartheta$},
ymin = 0,
ymax = 1,
xmin = 0,
xmax = 1,
legend cell align=left,
every axis legend/.append style = {
    at={(0.95,0.95)},
    anchor=north east}
]

\addplot [name path=plotycenter,blue,thick,domain=\xcrossL:\xcrossR,samples=30]
{\angley};

\addplot [name path=plotkTleft,black,dashed,samples=100,domain=\xkappamin:\xcrossL]
{\anglekT};
\addplot [name path=plotkTcenter,black,dashed,domain=\xcrossL:\xcrossR]
{\anglekT};
\addplot [name path=plotkTright,black,dashed,domain=\xcrossR:\xkappamax]
{\anglekT};

\addplot [name path=plot0LL,black,domain=0:\xkappamin]{0};
\addplot [name path=plot0LC,black,domain=\xkappamin:\xcrossL]{0};
\addplot [name path=plot0center,black,domain=\xcrossL:\xcrossR]{0};
\addplot [name path=plot0CR,black,domain=\xcrossR:\xkappamax]{0};
\addplot [name path=plot0RR,black,domain=\xkappamax:1]{0};

\addplot [name path=plot1LL,blue,domain=0:\xkappamin]{1};
\addplot [name path=plot1LC,black,domain=\xkappamin:\xcrossL]{1};
\addplot [name path=plot1center,black,domain=\xcrossL:\xcrossR]{1};
\addplot [name path=plot1CR,black,domain=\xcrossR:\xkappamax]{1};
\addplot [name path=plot1RR,blue,domain=\xkappamax:1]{1};

\addplot [blue,thick] coordinates{(\xcrossL,\Avalue) (\xcrossL,1)};
\addplot [blue,thick] coordinates{(\xcrossR,\Avalue) (\xcrossR,1)};

\addplot[fill opacity=0.2, blue] fill between[of = plot0LL and plot1LL];
\addplot[fill opacity=0.2, blue] fill between[of = plot0LC and plot1LC];
\addplot[fill opacity=0.2, blue] fill between[of = plot0center and plotycenter];
\addplot[fill opacity=0.2, blue] fill between[of = plot0CR and plot1CR];
\addplot[fill opacity=0.2, blue] fill between[of = plot0RR and plot1RR];

\addplot[fill opacity=0.2, yellow] fill between[of = plotycenter and plot1center];

\addplot [red,ultra  thick] coordinates{(0.003,0.0) (0.003,1)};
\addplot [red,ultra  thick] coordinates{(0.0,0.003) (1.0,0.003)};

\end{axis}
\end{tikzpicture}

\else % We use the pdf figure.
\begin{center}
\includegraphics[width = 8.2 cm]{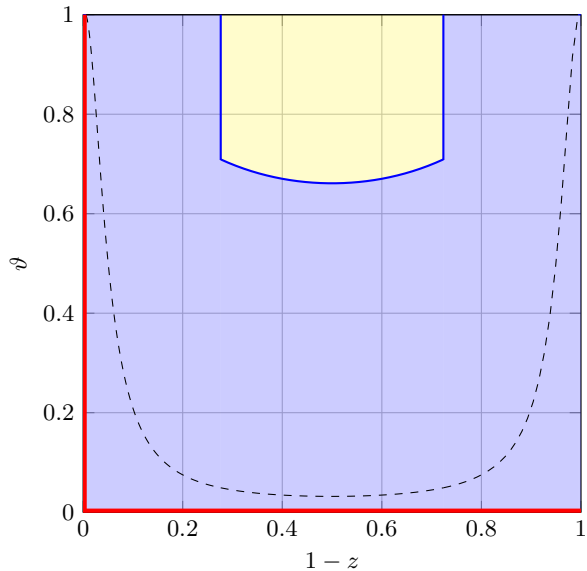}
\end{center}
\fi
\end{center}
\caption{
Unresolved regions for fixed $\vec\mu = (\mu_\scE,\mu_\Lambda)$ with cutoff $m_\perp^2 = 0.0005\,Q^2$ and $a_l = 2$. The two scales are $\mu_\scE^2 = 0.2\,Q^2$ and $\mu_\Lambda^2 = 0.1\,Q^2$.
}
\label{fig:unresolvedy2epsilon}
\end{figure}
%%%%%%%%%%%%%%%%%%% END FIGURE %%%%%%%%%%%%%%%%%%%%%%%%

%%%%%%%%%%%%%%%%%%%% FIGURE %%%%%%%%%%%%%%%%%%%%%%%%%%%
\begin{figure}
\begin{center}
\ifusefigs % then we include the figure

\begin{tikzpicture}

\newcommand\al{2.0}
\newcommand\Avalue{0.6}
\newcommand\kappaA{0.0005}
\newcommand\ymaxinverse{(sqrt(\al)+sqrt(\al-1))^2}
\newcommand\xkappamin{0.5*(1 - sqrt(1 - 4*\al*\kappaA*\ymaxinverse))}
\newcommand\xkappamax{1 - \xkappamin}

\newcommand\zomz{0.2}
\newcommand\xcrossL{2*\zomz/(1 + sqrt(1 - 4*\zomz))}
\newcommand\xcrossR{1-\xcrossL}

\newcommand\anglekT{\al^2*\kappaA/((x*(1-x) + \al*\kappaA)^2 
+ \al^2*\kappaA*(1 - 4*x*(1-x)))};

\newcommand\anglekTcut{\al^2*\kappaA/((\zomz + \al*\kappaA)^2 
+ \al^2*\kappaA*(1 - 4*\zomz))};

\begin{axis}[
xlabel={$1-z$}, 
ylabel={$\vartheta$},
ymin = 0,
ymax = 1,
xmin = 0,
xmax = 1,
legend cell align=left,
every axis legend/.append style = {
    at={(0.95,0.95)},
    anchor=north east}
]

\addplot [name path=plotycenter, black, dashed,domain=\xcrossL:\xcrossR,samples=30]
{\Avalue};

\addplot [name path=plotkTleft,black,dashed,samples=100,domain=\xkappamin:\xcrossL]
{\anglekT};
\addplot [name path=plotkTcenter,mygreen,thick,domain=\xcrossL:\xcrossR]
{\anglekT};
\addplot [name path=plotkTright,black,dashed,domain=\xcrossR:\xkappamax]
{\anglekT};

\addplot [name path=plot0LL,black,domain=0:\xkappamin]{0};
\addplot [name path=plot0LC,black,domain=\xkappamin:\xcrossL]{0};
\addplot [name path=plot0center,black,domain=\xcrossL:\xcrossR]{0};
\addplot [name path=plot0CR,black,domain=\xcrossR:\xkappamax]{0};
\addplot [name path=plot0RR,black,domain=\xkappamax:1]{0};

\addplot [name path=plot1LL,blue,domain=0:\xkappamin]{1};
\addplot [name path=plot1LC,black,domain=\xkappamin:\xcrossL]{1};
\addplot [name path=plot1center,black,domain=\xcrossL:\xcrossR]{1};
\addplot [name path=plot1CR,black,domain=\xcrossR:\xkappamax]{1};
\addplot [name path=plot1RR,blue,domain=\xkappamax:1]{1};

\addplot [blue,thick] coordinates{(\xcrossL,\anglekTcut) (\xcrossL,1)};
\addplot [blue,thick] coordinates{(\xcrossR,\anglekTcut) (\xcrossR,1)};

\addplot[fill opacity=0.2, blue] fill between[of = plot0LL and plot1LL];
\addplot[fill opacity=0.2, blue] fill between[of = plot0LC and plot1LC];
\addplot[fill opacity=0.2, blue] fill between[of = plot0center and plotkTcenter];
\addplot[fill opacity=0.2, blue] fill between[of = plot0CR and plot1CR];
\addplot[fill opacity=0.2, blue] fill between[of = plot0RR and plot1RR];

\addplot[fill opacity=0.2, yellow] fill between[of = plotkTcenter and plot1center];

\addplot [red,ultra  thick] coordinates{(0.003,0.0) (0.003,1)};
%\addplot [red,ultra  thick] coordinates{(0.0,0.003) (1.0,0.003)};

\end{axis}
\end{tikzpicture}

\else % We use the pdf figure.
\begin{center}
\includegraphics[width = 8.2 cm]{sample.pdf}
\end{center}
\fi
\end{center}
\caption{
Unresolved regions for fixed $\vec\mu = (\mu_\scE,\mu_\angle)$ with cutoff $m_\perp^2 = 0.0005\,Q^2$ for $\cD^{(1)}_\mathrm{soft}(\vec\mu)$. As in Fig.~\ref{fig:unresolvedA2epsilon}, we take $a_l = 2$, $\mu_\scE^2 = 0.2\,Q^2$ and $\mu_\angle^2 = 0.6\,Q^2$. For $\cD^{(1)}_\mathrm{soft}(\vec\mu)$, the unresolved region is independent of $\mu_\angle^2$.
}
\label{fig:unresolvedA2epsilonsoft}
\end{figure}
%%%%%%%%%%%%%%%%%%% END FIGURE %%%%%%%%%%%%%%%%%%%%%%%%

%-------------------------------------------------
\section{Unresolved region with two scales}
\label{sec:UnresolvedTwoScales}

We now consider the unresolved region for a splitting when we use two independent scale parameters. Throughout this section, we also incorporate the fixed infrared cutoff $k_\LT^2 > m_\perp^2$.

We let one scale be a collinear sensitive scale $\mu_\scC$, which could be any of $\mu_\angle$, $\mu_\Lambda$, or $\mu_\perp$. The scale $\mu_\scC$ controls at least the collinear singularity for one parton splitting into two. With $\mu_\scC = \mu_\angle$, this scale controls {\em only} the collinear singularity. The other singularity is the wide angle soft singularity, which is reached when a parton emits a gluon at a finite angle when the energy of the gluon approaches zero. We need a scale $\mu_\scE$ to control this singularity. The emitted gluon energy is proportional to $(1-z)$, so it is convenient to define an unresolved region parameterized by $\mu_\scE^2$ by using an energy variable $4z(1-z)Q^2$. The factor $z$ here is not important since there is no $z \to 0$ singularity in the splitting functions as defined in \textsc{Deductor}, but it is helpful to keep the scale definitions symmetric under $(1-z) \leftrightarrow z$. We define a function 
\begin{equation}
\label{eq:aEdef}
a_\scE(z,\mu_\scE^2)
=
\begin{cases}
1 & \mathrm{for}\ 4z(1-z)Q^2 < \mu_\scE^2
\\
0 & \mathrm{otherwise}
\end{cases}
\;.
\end{equation}
We can use this function and our previously defined function $a_\scC(z,\mu_\scC^2)$ for $\LC = \angle$, $\Lambda$, or $\perp$ to define an unresolved region for a given choice of two shower scales $\vec\mu = (\mu_\scE,\mu_\scC)$. We define the unresolved region by
\begin{equation}
\label{eq:twoscaleangle}
\vartheta < \max [
a_\scE(z,\mu_\scE^2),
a_\scC(z,\mu_\scC^2),
a_\perp(z,m_\perp^2)]
\;.
\end{equation}
This region is illustrated in Fig.~\ref{fig:unresolvedA2epsilon} in the case $\LC = \angle$, $a_l = 2$ with $m_\perp^2 = 0.0005\,Q^2$ for the choice of shower scales $\mu_\scE^2 = 0.2\,Q^2$, $\mu_\angle^2 = 0.6\,Q^2$. Again, the singular surface is indicated in red and the unresolved region is depicted in blue. A point $(1-z,\vartheta)$ is in the unresolved region if $4z(1-z) < \mu_\scE^2/Q^2$ {\em or} $\vartheta < \mu_\angle^2/Q^2$. The point is also in the unresolved region if $k_\LT^2 < m_\perp^2$, although this cutoff does not play a role in Fig.~\ref{fig:unresolvedA2epsilon}. 

The unresolved region specified by Eq.~(\ref{eq:twoscaleangle}) is illustrated in Fig.~\ref{fig:unresolvedy2epsilon} for the case $\LC = \Lambda$, with $\mu_\scE^2 = 0.2\,Q^2$, $\mu_\Lambda^2 = 0.1\,Q^2$.

As foreseen in Eq.~(\ref{eq:cDscandsoft}), we can divide $\cD^{[1,0]}(\vec\mu)$ into a part $\cD^{[1,0]}_\mathrm{sc}(\vec\mu)$ with both soft and collinear singularities and a part $\cD^{[1,0]}_\mathrm{soft}(\vec\mu)$ with only soft singularities.  Since $\cD^{[1,0]}_\mathrm{soft}(\vec\mu)$ lacks the collinear singularity, we can treat it differently.  We define the unresolved region for $\cD^{[1,0]}_\mathrm{soft}(\vec\mu)$ by
\begin{equation}
\label{eq:unresolvedsoft}
\vartheta < \max [a_\scE(z,\mu_\scE^2),
a_\perp(z,m_\perp^2)]
\;.
\end{equation}
That is, we replace $\mu_\scC^2$ by zero for $\cD^{[1,0]}_\mathrm{soft}(\vec\mu)$. This resulting unresolved region for $\cD^{[1,0]}_\mathrm{soft}(\vec\mu)$ for $\LC = \angle$ is illustrated in Fig.~\ref{fig:unresolvedA2epsilonsoft}. No lower limit for $\vartheta$ is needed for $\cD^{[1,0]}_\mathrm{soft}(\vec\mu)$ since it has no $\vartheta \to 0$ singularity. The only cutoff that applies for small $\vartheta$ is $k_\LT^2 > m_\perp^2$.

Now, to define the shower operator $\cU(t_\Lf,0)$, we need to define initial and final scales $\vec\mu(0) = \vec\mu_\scH$ and  $\vec\mu(t_\Lf) = \vec\mu_\Lf$ and a path $\vec\mu(t)$ that connects them. For the hard scales we take $\mu_{\scC,\scH}^2 = Q^2$ and $\mu_{\scE,\scH}^2 = Q^2$. For the infrared limiting values $\vec\mu_\Lf$, we could take values on the order of $\mu_{\scC,\Lf}^2 = \mu_{\scE,\Lf}^2 = 1 \GeV^2$. However, there is already a cutoff $k_\LT^2 > m_\perp^2$, so it suffices to set $\mu_{\scC,\Lf}^2 = \mu_{\scE,\Lf}^2 = 0$. 

%%%%%%%%%%%%%%%%%%%% FIGURE %%%%%%%%%%%%%%%%%%%%%%%%%%%
\begin{figure}
\begin{center}
\ifusefigs % then we include the figure

\begin{tikzpicture}

\begin{axis}[
xlabel={$\mu^2_\scE/Q^2$},
ylabel={$\mu^2_\scC/Q^2$},
ymin = -0.1,
ymax = 1.1,
xmin = -0.1,
xmax = 1.1,
legend cell align=left,
every axis legend/.append style = {
    at={(0.95,0.95)},
    anchor=north east}
]

\addplot [
scatter,only marks,point meta=explicit symbolic, 
scatter/classes={a={mark=*,black,mark size=2.5pt}
},
] table [meta=label] {
x y label 
0.0 0.0 a 
1.0 1.0  a 
0.0 1.0  a 
};

\node[]at({0.11,0.04}){$t = \infty$};
\node[]at(0.09,{0.95}){$t = 1$};
\node[]at({0.92,0.95}){$t = 0$};

\addplot [myorange, very thick,postaction={decorate, decoration={markings,
    mark=at position 0.5 with {\arrow[scale=1.0,>=Latex]{>};},}}] coordinates{(1,1) (0,1)};

\addplot [mygreen, very thick,postaction={decorate, decoration={markings,
    mark=at position 0.5 with {\arrow[scale=1.0,>=Latex]{>};},}}] coordinates{(0,1) (0,0)};

\end{axis}
\end{tikzpicture}

\else % We use the pdf figure.
\begin{center}
\includegraphics[width = 8.2 cm]{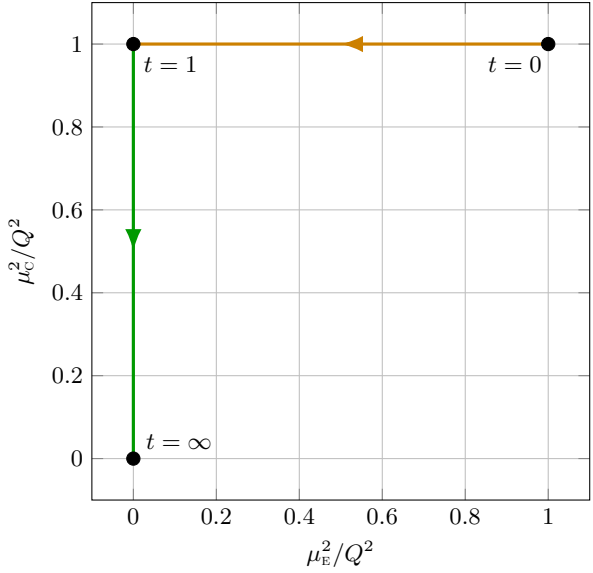}
\end{center}
\fi
\end{center}
\caption{
Evolution path with two segments.
}
\label{fig:evolution-path}
\end{figure}
%%%%%%%%%%%%%%%%%%% END FIGURE %%%%%%%%%%%%%%%%%%%%%%%%

Next, we need a path, $\vec\mu(t)$. We choose a path with two segments, illustrated in Fig.~\ref{fig:evolution-path}. In the first segment, for $0 < t < 1$, we choose $\mu_\scC^2 = Q^2$ and $\mu_\scE^2 = (1-t) Q^2$. On this segment of the path, the corner of the rectangle in Fig.~\ref{fig:unresolvedA2epsilon} is defined by $4z(1-z)$ decreasing from 1 to 0 and $\vartheta$ fixed at 1. In the second part, for $1 < t < t_\Lf = \infty$, we choose $\mu_\scC^2 = e^{-(t-1)} Q^2$ and $\mu_\scE^2 = 0$. On this segment of the path, the corner of the rectangle in Fig.~\ref{fig:unresolvedA2epsilon} is defined by $4z(1-z)$ fixed at 0 and $\vartheta$ decreasing from 1 to 0. Thus the path is
\begin{equation}
\begin{split}
\label{eq:2segmentpath}
\vec\mu(t)
    = {}&\left[
      \begin{array}{c}
      \mu_\scE(t)\\
      \mu_{\scC}(t)
      \end{array}
    \right]
    \\
    ={}& \theta(0<t<1)\,\sqrt{Q^2}\left[
      \begin{array}{c}
        \sqrt{1-t}\\
        1
      \end{array}
    \right]
    \\& + \theta(t>1)\,\sqrt{Q^2}\left[
      \begin{array}{c}
        0\\
        e^{(1-t)/2}
      \end{array}
    \right]
    \;.
\end{split}
\end{equation}
%

%%%%%%%%%%%%%%%%%%%% FIGURE %%%%%%%%%%%%%%%%%%%%%%%%%%%
\begin{figure}
\begin{center}
\ifusefigs % then we include the figure

\begin{tikzpicture}

\newcommand\al{2.0}
\newcommand\Avalue{0.99}
\newcommand\kappaA{0.0005}
\newcommand\ymaxinverse{(sqrt(\al)+sqrt(\al-1))^2}
\newcommand\xkappamin{0.5*(1 - sqrt(1 - 4*\al*\kappaA*\ymaxinverse))}
\newcommand\xkappamax{1 - \xkappamin}

\newcommand\coefA{\Avalue}
\newcommand\coefB{(-2*\al*(2*\al - 1)*\Avalue*\kappaA)}
\newcommand\coefC{(-\al^2*(\kappaA*(1 - \Avalue) - \Avalue*\kappaA^2))}
\newcommand\zomz{(-\coefB + sqrt(\coefB^2 - 4*\coefA*\coefC))/2/\coefA}
\newcommand\xcrossL{2*\zomz/(1 + sqrt(1 - 4*\zomz))}
\newcommand\xcrossR{1-\xcrossL}

\newcommand\anglekT{\al^2*\kappaA/((x*(1-x) + \al*\kappaA)^2 
+ \al^2*\kappaA*(1 - 4*x*(1-x)))};

\begin{axis}[
xlabel={$1-z$}, 
ylabel={$\vartheta$},
ymin = 0,
ymax = 1,
xmin = 0,
xmax = 1,
legend cell align=left,
every axis legend/.append style = {
    at={(0.95,0.95)},
    anchor=north east}
]

\addplot [name path=plotAleft,black,dashed,domain=0:\xcrossL,samples=30]
{\Avalue};
\addplot [name path=plotAcenter,blue,thick,domain=\xcrossL:\xcrossR,samples=30]
{\Avalue};
\addplot [name path=plotAright,black,dashed,domain=\xcrossR:1,samples=30]
{\Avalue};

\addplot [name path=plotkTleft,red,thick,samples=100,domain=\xkappamin:\xcrossL]
{\anglekT};
\addplot [name path=plotkTcenter,black,dashed,domain=\xcrossL:\xcrossR]
{\anglekT};
\addplot [name path=plotkTright,red,thick,domain=\xcrossR:\xkappamax]
{\anglekT};

\addplot [name path=plot0LL,black,domain=0:\xkappamin]{0};
\addplot [name path=plot0LC,black,domain=\xkappamin:\xcrossL]{0};
\addplot [name path=plot0center,black,domain=\xcrossL:\xcrossR]{0};
\addplot [name path=plot0CR,black,domain=\xcrossR:\xkappamax]{0};
\addplot [name path=plot0RR,black,domain=\xkappamax:1]{0};

\addplot [name path=plot1LL,blue,domain=0:\xkappamin]{1};
\addplot [name path=plot1LC,black,domain=\xkappamin:\xcrossL]{1};
\addplot [name path=plot1center,black,domain=\xcrossL:\xcrossR]{1};
\addplot [name path=plot1CR,black,domain=\xcrossR:\xkappamax]{1};
\addplot [name path=plot1RR,blue,domain=\xkappamax:1]{1};

\addplot[fill opacity=0.2, blue] fill between[of = plot0LL and plot1LL];
\addplot[fill opacity=0.2, blue] fill between[of = plot0LC and plotkTleft];
\addplot[fill opacity=0.2, blue] fill between[of = plot0center and plotAcenter];
\addplot[fill opacity=0.2, blue] fill between[of = plot0CR and plotkTright];
\addplot[fill opacity=0.2, blue] fill between[of = plot0RR and plot1RR];

\addplot[fill opacity=0.2, yellow] fill between[of = plotkTleft and plot1LC];
\addplot[fill opacity=0.2, yellow] fill between[of = plotAcenter and plot1center];
\addplot[fill opacity=0.2, yellow] fill between[of = plotkTright and plot1CR];

\addplot [red,ultra  thick] coordinates{(0.003,0.0) (0.003,1)};
\addplot [red,ultra  thick] coordinates{(0.0,0.003) (1.0,0.003)};

\end{axis}
\end{tikzpicture}

\else % We use the pdf figure.
\begin{center}
\includegraphics[width = 8.2 cm]{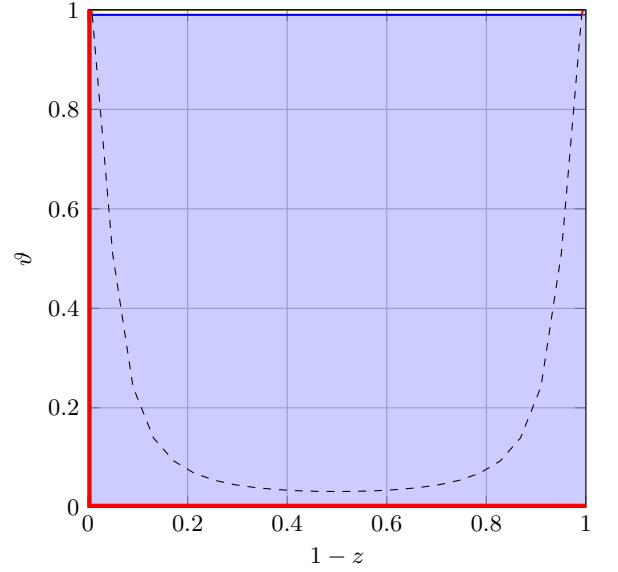}
\end{center}
\fi
\end{center}
\caption{
Unresolved region for $\cD^{[1,0]}_\mathrm{sc}(\vec\mu)$ at the end of the first stage of evolution. Here $a_l = 2$ and $m_\perp^2 = 0.0005\,Q^2$.
}
\label{fig:unresolvedstage1sc}
\end{figure}
%%%%%%%%%%%%%%%%%%% END FIGURE %%%%%%%%%%%%%%%%%%%%%%%%

%%%%%%%%%%%%%%%%%%%% FIGURE %%%%%%%%%%%%%%%%%%%%%%%%%%%
\begin{figure}
\begin{center}
\ifusefigs % then we include the figure

\begin{tikzpicture}

\newcommand\al{2.0}
\newcommand\kappaA{0.0005}
\newcommand\ymaxinverse{(sqrt(\al)+sqrt(\al-1))^2}
\newcommand\xxmin{0.5*(1 - sqrt(1 - 4*\al*\kappaA*\ymaxinverse))}
\newcommand\xxmax{1 - \xxmin}

\newcommand\anglekT{\al^2*\kappaA/((x*(1-x) + \al*\kappaA)^2 
+ \al^2*\kappaA*(1 - 4*x*(1-x)))};

\begin{axis}[
xlabel={$1-z$}, 
ylabel={$\vartheta$},
ymin = 0,
ymax = 1,
xmin = 0,
xmax = 1,
legend cell align=left,
every axis legend/.append style = {
    at={(0.95,0.95)},
    anchor=north east}
]

\addplot [name path=plot1,mygreen,thick,samples=100,domain=\xxmin:\xxmax]
{\anglekT};
\addplot [name path=plot2,black,domain=\xxmin:\xxmax]{0};
\addplot [name path=plot3,thin, black,domain=\xxmin:\xxmax]{1};

\addplot [name path=plot1L,blue,domain=0:\xxmin]{1};
\addplot [name path=plot2L,black,domain=0:\xxmin]{0};

\addplot [name path=plot1R,blue,domain=\xxmax:1]{1};
\addplot [name path=plot2R,black,domain=\xxmax:1]{0};
%\addlegendentry{$y = 0.3$}

\addplot[fill opacity=0.2, blue] fill between[of = plot1 and plot2];
\addplot[fill opacity=0.2, blue] fill between[of = plot1L and plot2L];
\addplot[fill opacity=0.2, blue] fill between[of = plot1R and plot2R];
\addplot[fill opacity=0.2, yellow] fill between[of = plot1 and plot3];

\addplot [red,ultra  thick] coordinates{(0.003,0.0) (0.003,1)};
%\addplot [red,ultra  thick] coordinates{(0.0,0.003) (1.0,0.003)};

\end{axis}
\end{tikzpicture}

\else % We use the pdf figure.
\begin{center}
\includegraphics[width = 8.2 cm]{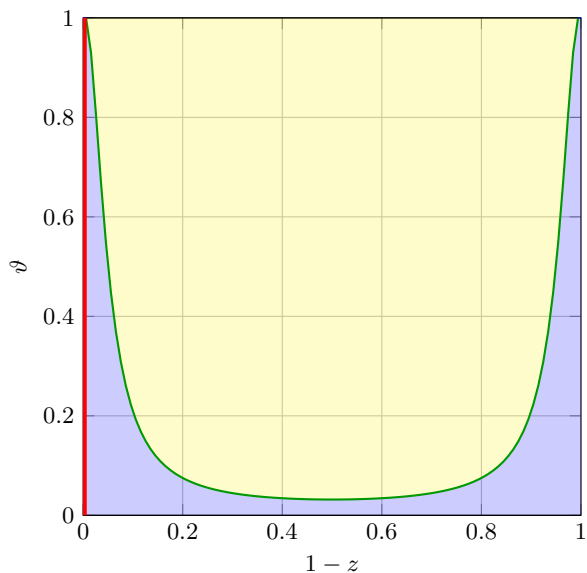}
\end{center}
\fi
\end{center}
\caption{
Unresolved region for $\cD^{[1,0]}_\mathrm{soft}(\vec\mu)$ at the end of the first stage of evolution. Here $a_l = 2$ and $m_\perp^2 = 0.0005\,Q^2$.
}
\label{fig:unresolvedstage1soft}
\end{figure}
%%%%%%%%%%%%%%%%%%% END FIGURE %%%%%%%%%%%%%%%%%%%%%%%%

The unresolved region for $\cD^{[1,0]}_\mathrm{sc}(\vec\mu)$ at the end of first segment of the path is shown in Fig.~\ref{fig:unresolvedstage1sc}. The same figure applies for any of our choices for C because $a_\scC(z, Q^2) \ge 1$ for $C = \perp$,  $\Lambda$, or $\angle$. There is no change in $\cD^{[1,0]}_\mathrm{sc}(\vec\mu)$ in this segment. Everything remains unresolved. The unresolved region for $\cD^{[1,0]}_\mathrm{soft}(\vec\mu)$ at the end of first segment of the path is shown in Fig.~\ref{fig:unresolvedstage1soft}. In this segment, $\cD^{[1,0]}_\mathrm{soft}(\vec\mu)$ changes substantially, so that at the end of this segment of the path, the unresolved region is only the region with $k_\LT^2 < m_\perp^2$. 

In the second segment of the path, the unresolved region for $\cD^{[1,0]}_\mathrm{soft}(\vec\mu)$ does not change at all. It remains as depicted in Fig.~\ref{fig:unresolvedstage1soft}. In this second segment, $\cD^{[1,0]}_\mathrm{sc}(\vec\mu)$ changes substantially, so that at the end of this segment of the path, the unresolved region is only the region with $k_\LT^2 < m_\perp^2$. This is the region that was already depicted in Fig.~\ref{fig:unresolvedstage1soft}, but now it applies to $\cD^{[1,0]}_\mathrm{sc}(\vec\mu)$.

%-------------------------------------------------
\section{Evolution with two scales}
\label{sec:EvolutionTwoScales}

The singular operator $\cD(\vec\mu)$ has a perturbative expansion (\ref{cDexpansion}). The shower generator $\cS_j(\vec\mu)$ is defined in Eq.~(\ref{eq:cSjdef}). The index $j \in \{\LE,\LC\}$ includes two scale choices. The shower generator has a perturbative expansion
\begin{equation}
\cS_j(\vec\mu) = \frac{\as}{2\pi}\,\cS_j^{(1)}(\vec\mu) + \cO(\as^2)
\;.
\end{equation}
From Eq.~(\ref{eq:cSjdef}), the first order contribution is
\begin{equation}
\label{eq:cSj1def}
\cS_j^{(1)}(\vec\mu) = 
\frac{\partial}{\partial \mu_j}\,
\cD^{(1)}(\vec\mu)
\;.
\end{equation}
In a first order shower, we truncate the expansion of $\cS_j(\vec\mu)$ at first order,
\begin{equation}
\cS_j(\vec\mu) = \frac{\as}{2\pi}\,\cS_j^{(1)}(\vec\mu)
\;.
\end{equation}

To obtain the shower evolution operator $\cU(t_2,t_1)$ following the chosen path $\vec\mu(t)$, we solve the differential equation Eq.~(\ref{eq:cUdiffeqn}). In general, this gives us the group multiplication property
\begin{equation}
\cU(t_2,t_1) = \cU(t_2,\tau)\,\cU(\tau,t_1)
\;.
\end{equation}
Our path has two segments, $0 < t < 1$ and $1 < t < \infty$. This gives
\begin{equation}
\label{eq:cUfactors}
\cU(\infty,0) = \cU(\infty,1)\,\cU(1,0)
\;.
\end{equation}

We have divided $\cD^{(1)}(\vec\mu)$ into a part $\cD^{(1)}_\mathrm{sc}(\vec\mu)$ with both soft and collinear singularities and a part $\cD^{(1)}_\mathrm{soft}(\vec\mu)$ with only soft singularities, as in Eq.~(\ref{eq:cDscandsoft}).  We recall that only $\mu_\scE$ changes on the first segment of the path and this change affects only $\cD^{(1)}_\mathrm{soft}(\vec\mu)$. We also recall that only $\mu_\scC$ changes on the second segment of the path and this change affects only $\cD^{(1)}_\mathrm{sc}(\vec\mu)$. Then
\begin{equation}
\label{eq:cU10}
\cU(1,0) = 
\mathbb{T}\exp\!\left\{-\int_{0}^{1}\! dt\,
\frac{d\mu_\scE(t)}{dt}\, \cS_\scE^\mathrm{soft}(\vec\mu(t))\right\}
\end{equation}
and
\begin{equation}
\label{eq:cUinfty1}
\cU(\infty,1) = 
\mathbb{T}\exp\!\left\{-\int_{1}^{\infty}\! dt\, 
\frac{d\mu_\scC(t)}{dt}\, \cS_\scC^\mathrm{sc}(\vec\mu(t))
\right\}
.
\end{equation}

In this formulation the parton shower, the result depends on what we choose for $\cD^{(1)}_\mathrm{soft}(\vec\mu)$ and $\cD^{(1)}_\mathrm{sc}(\vec\mu)$. We can choose $\cD^{(1)}_\mathrm{soft}(\vec\mu) = 0$. Then $\cD^{(1)}_\mathrm{sc}(\vec\mu)$ is all of $\cD^{(1)}(\vec\mu)$. When $\mu_\scC = \mu_\angle$ this gives us a simple angular ordered shower, as in Eq.~(\ref{eq:unresolvedangularordered}) and Fig.~\ref{fig:unresolvedAepsilon}. The only difference is conceptual. First, $\cD^{(1)}_\mathrm{sc}(\vec\mu)$ and its inverse $[\cD^{(1)}_\mathrm{sc}(\vec\mu)]^{-1}$ are well defined at the hard scale. This is important because $[\cD^{(1)}_\mathrm{sc}(\vec\mu)]^{-1}$ plays the role of removing infrared singularities from the hard scattering cross section calculated at next-to-leading order \cite{NSAllOrder}. Second, with the two scale formulation, we could have eliminated the $m_\perp^2$ cut. Then it would have been natural to choose a nonzero endpoint $\mu_{E,\Lf}^2 \sim 1 \GeV^2$ for the evolution in $\mu_\scE$ in the first segment of the path. This would leave us with no naked singularity in a natural way.

There are a number of nonzero choices we could make for $\cD^{(1)}_\mathrm{soft}(\vec\mu)$, letting $\cD^{(1)}_\mathrm{sc}(\vec\mu) = \cD^{(1)}(\vec\mu) - \cD^{(1)}_\mathrm{soft}(\vec\mu)$.  One possibility is to define $\cD^{(1)}_\mathrm{sc}(\vec\mu)$ so that, although it has a soft$\times$collinear double singularity, it has only a minimal wide-angle soft singularity. Whatever choice we make, the evolution $\cU(1,0)$, using $\cD^{(1)}_\mathrm{soft}(\vec\mu)$, comes first, followed by evolution $\cU(\infty,1)$, using $\cD^{(1)}_\mathrm{sc}(\vec\mu)$ with an ordering prescription such as angular ordering, $\Lambda$ ordering, or $k_\LT$ ordering.

This two scale formulation of a parton shower is reminiscent of soft-collinear effective theory (SCET). Suppose that we want to measure an observable that is nonzero when there are at least $N$ hard jets. We start with a hard scattering that produces $N$ hard jets. With a cut on $(N-1)$ jettiness \cite{Njettiness}, $\tau_{N-1} > \tau_\mathrm{min}$, we ensure that the hard partons constitute $N$ jets and not $N-1$ jets. With this as the hard state, the operator $\cS_\scE^\mathrm{soft}(\vec\mu(t))$ in $\cU(1,0)$ produces soft wide-angle radiation from the $N$ hard jets, analogously to the soft factor in SCET. In the second segment of the shower evolution, $\cU(\infty,1)$ can add more soft radiation. However, if $\cS_\scC^\mathrm{sc}(\vec\mu(t))$ (where $\LC = \angle, \Lambda, \perp$ or some other choice) is defined to have only minimal wide-angle soft singularities, it is the first segment, involving $\cS_\scE^\mathrm{soft}(\vec\mu(t))$ that will dominate the soft radiation between the jets. Then the second evolution segment acts as the collinear factor in a SCET analysis and fills in the collinear radiation for each jet.

%-------------------------------------------------
\section{Improved color with two scales}
\label{sec:BetterColor}

In this section, we describe how one might use the choices available when using the two scales, $\mu_\LC$ and $\mu_\LE$, to improve the treatment of color in the shower in a practical way.

First, we provide some background on color in parton showers. The most widely used parton shower event generators \cite{Herwig, Pythia, Sherpa} use the leading color (LC) approximation, which captures just the leading term in an expansion in powers of $1/N_\Lc^2$, where $N_\Lc = 3$ is the number of colors. Here one simply supplies a color factor $C_\LF = (N_\Lc^2-1)/(2 N_\Lc)$ or $C_\LA = N_\Lc$ for emission of a gluon from a quark or gluon line, respectively, or else a factor $T_\LR = 1/2$ for a gluon splitting to $q + \bar q$. To go beyond the LC approximation one needs to treat the color carried by quarks and gluons as fully quantum mechanical variables.

Throughout this paper, we have described color as fully quantum mechanical using a vector space for parton color with basis vectors $\sket{\{c,c'\}_m}$ \cite{NSI}. The basis vector $\sket{\{c,c'\}_m}$ represents a color density matrix $\ket{\{c\}_m}\bra{\{c'\}_m}$, where $\ket{\{c\}_m}$ is a basis vector for the space of quantum color states for $m$ partons. (\textsc{Deductor} uses the trace basis, but other choices are possible.) This description, with a somewhat different notation, is used in the recent papers \cite{AngelesMartinez:2018cfz, Forshaw:2019ver, Forshaw:2020wrq, DeAngelis:2020rvq, Holguin:2020joq, Hoche:2020pxj, Platzer:2020lbr} to study color in parton shower evolution, accounting approximately for both real emission graphs and virtual exchange graphs. Other papers have used the color density matrix, but for the description of just real emissions \cite{PlatzerSjodahl2012, PlatzerSjodahlThoren, Isaacson:2018zdi}. Ref.~\cite{HamiltonShowerSum} has worked to improve the treatment of color in parton showers without tying the description to the color density matrix.

One can express the evolution equations for a first order dipole shower so that it evolves with full color \cite{NSI}. However, some approximation is needed for a shower realized in computer code. The \textsc{Deductor} shower uses what we call the LC+ approximation\footnote{The LC+ approximation is defined using the trace basis for color. There is no equivalent approximation in the color flow basis.} for color \cite{NScolor}. This is an improvement over the LC approximation. The splitting operators with this approximation, $\cS^\mathrm{LC+}_j(\vec\mu)$, are, however, still approximate in color, leaving a difference
\begin{equation}
\label{eq:DeltacS}
\Delta \cS_j(\vec\mu) =
\cS_j(\vec\mu) -\cS^\mathrm{LC+}_j(\vec\mu)
\;.
\end{equation}
Simply using $\cS^\mathrm{LC+}_j$ would give us an uncontrolled approximation since we would not know the size of corrections from $\Delta \cS_j(\vec\mu)$. \textsc{Deductor} allows a systematically improvable approximation: the user can compute corrections proportional to powers $[\Delta \cS_j]^N$ of $\Delta \cS_j$ (with a single scale $\mu_\Ls$) \cite{NSNewColor, NSColoriPi, GapColor}. Any power $N$ is allowed. However, including powers of $\Delta \cS_j$ is computationally complicated and makes the program run more slowly. This leads to practical limits to the size of $N$. 

It would certainly be desirable to have particular choice of $\cS_\scE^\mathrm{soft}(\vec\mu(t))$ that results in making the inclusion of $\Delta \cS_j(\vec\mu)$ computationally simpler. With this in mind, we note that the LC+ approximation has an important property. At each splitting, the leading soft$\times$collinear singularity and the leading collinear singularity are treated exactly with respect to color \cite{NScolor}.  That is, $\Delta \cS_j(\vec\mu)$ has no collinear singularity. Thus we can set
\begin{equation}
\begin{split}
\label{eq:cSjLCplus}
\cS_j^\mathrm{sc}(\vec\mu(t)) ={}& \cS_j^\mathrm{LC+}(\vec\mu(t))
\;,
\\
\cS_j^\mathrm{soft}(\vec\mu(t)) ={}& \Delta\cS_j(\vec\mu(t))
\;.
\end{split}
\end{equation}
With this choice, $\cU(\infty,1)$ in Eq.~(\ref{eq:cUfactors}) is exact in color and the corrections to the LC+ approximation appear in the factor $\cU(1,0)$. This is significant for two reasons. First, the corrections to the LC+ approximation appear in one place, rather than appearing throughout the shower, interleaved with LC+ splittings, as in Refs.~\cite{NSNewColor, GapColor, NSColoriPi}. Second, the factor $\cU(1,0)$ operates on the hard scattering state with which the shower begins. This state is simple because it has few partons.

For $\cU(1,0)$, we can expand Eq.~(\ref{eq:cU10}) in powers of $\Delta\cS$,
\begin{equation}
\begin{split}
\label{eq:cU10expanded}
\cU(1,0) ={}& 
1
- \int_{0}^{1}\! dt\,
\frac{d\mu_\scE(t)}{dt}\, \Delta\cS_\scE(\vec\mu(t))
\\ & +
\int_{0}^{1}\! dt_2\,\frac{d\mu_\scE(t_2)}{dt_2} 
\int_{0}^{t_2}\! dt_1\,\frac{d\mu_\scE(t_1)}{dt_1}
\\&\quad \times
\Delta\cS_\scE(\vec\mu(t_2))\,
\Delta\cS_\scE(\vec\mu(t_1))
\\& + \cdots
\;,
\end{split}
\end{equation}
keeping terms up to order $[\Delta\cS_\scE]^N$, where $N$ is chosen by the user. A more elaborate treatment is possible, but, given the simplicity of the hard scattering state to which $\cU_\scC(1,0)$ is applied, this very simple treatment should suffice.

If we start with the simplest process in $e^+e^-$ annihilation, $e^+e^- \to q \bar q$, this is even simpler. Because the $q$ and $\bar q$ are each other's color connected partners, we have for the two parton $q\bar q$ state
\begin{equation}
\Delta\cS_\scE(\vec\mu(t))\sket{\{p, f, c, c'\}_2} = 0
\;.
\end{equation}
Thus $\cU(1,0) \sket{\{p, f, c, c'\}_2} = \sket{\{p, f, c, c'\}_2}$, so
\begin{equation}
\cU(\infty,0) \sket{\{p, f, c, c'\}_2} = 
\cU(\infty,1)\sket{\{p, f, c, c'\}_2}
\;.
\end{equation}
We can write this in more detail. We choose the evolution scale in $\cU(\infty,1)$ as $\mu_\scC = \mu_\angle$, $\mu_\Lambda$,  or $\mu_\perp$ according to our preference and use $\cS^\mathrm{LC+}(\mu_\scC) = \cS_\scC^\mathrm{LC+}(\mu_\scE,\mu_\scC)$ with $\mu_\scE = 0$. Then
\begin{equation}
\begin{split}
\cU(\infty,0)\sket{\{p, f, c, c'\}_2} ={}& 
\mathbb{T}\exp\!\left\{\int_{0}^{Q}\! d\mu_\scC\ 
\cS^\mathrm{LC+}(\mu_\scC)
\right\}
\\&\times\sket{\{p, f, c, c'\}_2}
\;.
\end{split}
\end{equation}

It is remarkable that the LC+ approximation for color gives the exact answer in this case. However, one should be careful about what ``exact'' means. A first order parton shower does not represent full QCD exactly. Two different choices for the choice of shower scale scheme will give two different parton shower algorithms. When we work within a framework that encompasses parton showers at any perturbative order \cite{NSAllOrder}, we see that the first of two algorithms can, in principle, be mapped into the second by adding order $\as^2$ and higher order terms to the splitting functions of the second. With both splitting functions truncated at order $\as$, the two algorithms give different results. The difference is a measure of the uncertainty inherent in using a first order shower. 

Thus it is indeed remarkable that the LC+ approximation for color is exact in this case, but the meaning of this statement is that differences from the LC+ approximation in the one scale treatment can be absorbed into terms in the shower splitting functions that are higher order in $\as$ in the two scale treatment. 

We emphasize that $e^+e^-$ annihilation with $e^+e^- \to q \bar q$ as the hard process is a special case. A hard scattering process with $m$ final state partons with $m > 2$ will lead to $\cU(1,0)\sket{\{p, f, c, c'\}_m}$ being nontrivial. Then one will need to use Eq.~(\ref{eq:cU10expanded}) for $\cU(1,0)$. 

%-------------------------------------------------
\section{More complex contour}
\label{sec:morecomplex}

One might argue that the two segment contour is too extreme since we put all the wide angle soft contributions just after the hard interaction. This might provide a good approximation if we consider a measurement that examines just the jets created by the initial hard partons, so that we wish to have the shower generate soft gluons that can see only the initial hard jets.

%%%%%%%%%%%%%%%%%%%% FIGURE %%%%%%%%%%%%%%%%%%%%%%%%%%%
\begin{figure}
\begin{center}
\ifusefigs % then we include the figure

\begin{tikzpicture}

\begin{axis}[
xlabel={$\mu^2_\scE/Q^2$},
ylabel={$\mu^2_\scC/Q^2$},
ymin = -0.1,
ymax = 1.1,
xmin = -0.1,
xmax = 1.1,
legend cell align=left,
every axis legend/.append style = {
    at={(0.95,0.95)},
    anchor=north east}
]

\addplot [
scatter,only marks,point meta=explicit symbolic, 
scatter/classes={a={mark=*,black,mark size=2.3pt}
},
] table [meta=label] {
x y label 
1.0 1.0 a 
0.6 1.0  a 
0.6 0.6  a 
0.0 0.6  a
0.0 0.0  a
};

\node[]at({0.11,0.04}){$t = \infty$};
\node[]at({0.09,0.64}){$t = 3$};
\node[]at({0.68,0.64}){$t = 2$};
\node[]at({0.50,0.95}){$t = 1$};
\node[]at({0.92,0.95}){$t = 0$};

\addplot [myorange, very thick,postaction={decorate, decoration={markings,
    mark=at position 0.6 with {\arrow[scale=1.0,>=Latex]{>};},}}] coordinates{(1,1) (0.6,1)};

\addplot [mygreen, very thick, postaction={decorate, decoration={markings,
    mark=at position 0.6 with {\arrow[scale=1.0,>=Latex]{>};},}}] coordinates{(0.6,1) (0.6,0.6)};

\addplot [myorange, very thick,postaction={decorate, decoration={markings,
    mark=at position 0.55 with {\arrow[scale=1.0,>=Latex]{>};},}}] coordinates{(0.6,0.6) (0,0.6)};

\addplot [mygreen, very thick,postaction={decorate, decoration={markings,
    mark=at position 0.55 with {\arrow[scale=1.0,>=Latex]{>};},}}] coordinates{(0,0.6) (0,0)};

\end{axis}
\end{tikzpicture}

\else % We use the pdf figure.
\begin{center}
\includegraphics[width = 8.2 cm]{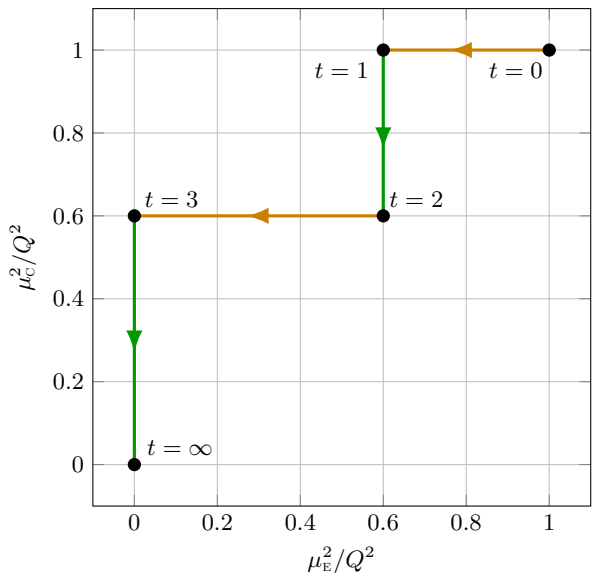}
\end{center}
\fi
\end{center}
\caption{
Evolution path with four segments.
}
\label{fig:4seg-evol-path}
\end{figure}
%%%%%%%%%%%%%%%%%%% END FIGURE %%%%%%%%%%%%%%%%%%%%%%%%

But what happens if our observable is sensitive to the structure of extra jets in addition to the initial hard jets. With the two-segment path, these jets are not corrected by any wide angle soft emissions beyond those generated within the LC+ approximation. We can adapt the evolution path for such an observable by using a four segment contour as illustrated in Fig.~\ref{fig:4seg-evol-path}. This path can be parameterized similarly to Eq.~(\ref{eq:2segmentpath}).

On the first segment of the contour, the evolution operator is     
\begin{equation}
\cU(1,0) = 
\mathbb{T}\exp\!\left\{-\int_{0}^{1}\! dt\,
\frac{d\mu_\scE(t)}{dt}\, \Delta\cS_\scE(\vec\mu(t))\right\}\;.
\end{equation}
Only the wide angle soft operator $\Delta\cS_\scE(\vec\mu)$ contributes. This comes right after the hard stage and tries to add partons with rather large energy and large emission angle. Small angle radiation is suppressed in $\Delta\cS_\scE(\vec\mu)$ and small energy emissions are not allowed because $\mu_\scE^2$ is never small on this path segment. 

On the second segment of the contour, the evolution operator is
\begin{equation}
\cU(2,1) = 
\mathbb{T}\exp\!\left\{-\int_{1}^{2}\! dt\, 
\frac{d\mu_\scC(t)}{dt}\, \cS_\scC^\mathrm{LC+}(\vec\mu(t))
\right\}\;.
\end{equation}
We have evolution in $\mu_\scC^2$ in the LC+ approximation for color, with a condition on the energy of the emitted parton, $4z(1-z)Q^2 > 0.6\, Q^2$ (in this example). Since $\mu_\scC^2$ is never small on this path segment, the radiation produced is neither very soft nor very collinear. We expect jets from this segment that are resolvable from each other at a fairly large scale. 

The evolution operator for the third part of the shower evolution is  
\begin{equation}
\cU(3,2) = 
\mathbb{T}\exp\!\left\{-\int_{2}^{3}\! dt\, 
  \frac{d\mu_\scE(t)}{dt}\,
    \cS_\scE(\vec\mu(t)) 
  \right\}\;.
\end{equation}
This is different than the evolution on the first segment. Here the whole splitting operator $\cS_\scE(\vec\mu) = \cS^{\rm LC+}_\scE(\vec\mu) + \Delta\cS_\scE(\vec\mu)$ contributes to the soft evolution. It is still only wide angle soft effect. For emissions created by $\cS_\scC^\mathrm{LC+}$, the emission angle is bounded from below because $\mu^2_\scC = 0.6$ (in this example). For emissions created by $\Delta\cS_\scE(\vec\mu)$, we do not have a direct lower bound on the emission angle, but the small angle emissions are suppressed by the splitting function. We expect that this part of the evolution could be treated perturbatively as in Eq.~(\ref{eq:cU10expanded}).

The evolution operator for the fourth part of the shower evolution is
\begin{equation}
\cU(\infty,3) = 
\mathbb{T}\exp\!\left\{-\int_{3}^{\infty}\! dt\, 
\frac{d\mu_\scC(t)}{dt}\, \cS_\scC^\mathrm{LC+}(\vec\mu(t))
\right\}\;.
\end{equation}
This gives soft-collinear evolution using the LC+ approximation, just as in the two segment case. We expect the emissions from the smallest values of $\mu_\scC^2$ on this segment to be unresolved by the observable considered.

%-------------------------------------------------
\section{Comparisons for $e^+e^-$ annihilation at 10 TeV}
\label{sec:10TeV}

In this section, we study $e^+e^-$ annihilation at $\sqrt{Q^2} = 10 \TeV$, with the aims of demonstrating the practical application of the methods described in this paper, exploring the differences among the choices $\mu_\scC^2 = \mu_\perp^2$,
$\mu_\Lambda^2$, and $\mu_\angle^2$, and testing the dependence on the treatment color.

The hard scattering process is $e^+e^- \to q \bar q$, with more partons being provided by the parton shower. There are no data at such a large $Q^2$, but with a large $Q^2$, there is more room for shower evolution between the hard scale and the roughly 1 GeV scale at which we stop the shower. We use the \textsc{Deductor} parton shower to examine two jet production as a function of the resolution parameter $y_\mathrm{cut}$ using the Cambridge jet algorithm \cite{CambridgeJets}. 

The fraction of events with exactly two jets is $(1/\sigma_\mathrm{tot})\,\sigma(2\ \mathrm{jets},y_\mathrm{cut})$. For each event, there is a value $y_{23}$ of the resolution parameter at which the event changes a two jet event to a three jet event. The distribution of $\log(y_{23})$ is
\begin{equation}
\frac{y_{23}}{\sigma_\mathrm{tot}}\frac{d\sigma}{dy_{23}}=
\left[\frac{y_\mathrm{cut}}{\sigma_\mathrm{tot}}\,
\frac{d\sigma(2\ \mathrm{jets},y_\mathrm{cut})}
{d y_\mathrm{cut}}\right]_{y_\mathrm{cut} = y_{23}}
\;.
\end{equation}
We will study the behavior of this distribution.

We use a version\footnote{This version, \textsc{Deductor} v.~3.4.99, is available at \href{http://www.desy.de/~znagy/deductor/}{http://www.desy.de/$\sim$znagy/deductor/} and \href{http://pages.uoregon.edu/soper/deductor}{http://pages.uoregon.edu/soper/deductor}.} of \textsc{Deductor} that is designed to include $k_\LT$ ordering, $\Lambda$ ordering, and angular ordering so that only the ordering variable changes among the three choices.

%%%%%%%%%%%%%%%%%%%% FIGURE %%%%%%%%%%%%%%%%%%%%%%%%%%%
\begin{figure}
\begin{center}
\ifusefigs % then we include the figure

\begin{tikzpicture}
\begin{semilogxaxis}[title = {y23 distribution},
   xlabel={$y_{23}$}, 
   ylabel={$(y_{23}/\sigma_\mathrm{tot})\,d\sigma/d y_{23}$},
%  xmin=5*10^(-7), 
%  xmax=1.0,
%  ymin=, ymax=,
  legend cell align=left,
  every axis legend/.append style = {
  at={(0.22,0.05)},
  anchor=south west}
]

\errorband[blue,semithick]{fill=blue!30!white, opacity=0.5}
{data/y23kT.dat}{1}{3}{4};
\addlegendentry{\textsc{Deductor}-$k_\LT$}

\errorband[red,semithick]{fill=red!30!white, opacity=0.5}
{data/y23Lambda.dat}{1}{3}{4};
\addlegendentry{\textsc{Deductor}-$\Lambda$}

\errorband[myorange,semithick]{fill=myorange!30!white, opacity=0.5}
{data/y23angular.dat}{1}{3}{4};
\addlegendentry{\textsc{Deductor}-$\vartheta$}

%
%\errorband[blue,semithick,dashed]{fill=blue!30!white, opacity=0.5}
%{data/y23kTDES.dat}{1}{3}{4};
%
%\errorband[red,semithick,dashed]{fill=red!30!white, opacity=0.5}
%{data/y23LambdaDES.dat}{1}{3}{4};
%
%\errorband[myorange,semithick,dashed]{fill=myorange!30!white, opacity=0.5}
%{data/y23angularDES.dat}{1}{3}{4};
%

\addplot [black,thick,dashed,domain=0.000001:0.22]
{-0.043537 - 0.0252632*ln(x) + 0.00230913*ln(x)^2 - 0.000304404*ln(x)^3 - 0.000173146*ln(x)^4 - 0.0000143419*ln(x)^5 - 3.51321*10^(-7)*ln(x)^6};

\addlegendentry{Analytical NLL}

\end{semilogxaxis}
\end{tikzpicture}

\else % We use the pdf figure.
\begin{center}
\includegraphics[width = 8.2 cm]{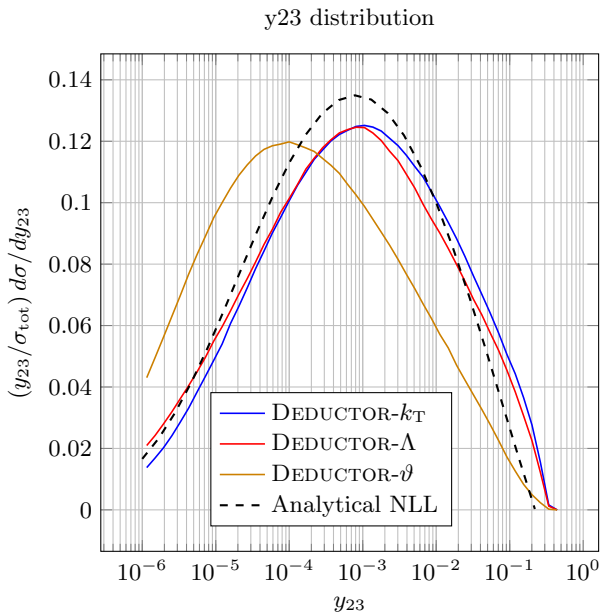}
\end{center}
\fi
\end{center}
\caption{
The $y_{23}$ distribution with the Cambridge algorithm.
}
\label{fig:twojet}
\end{figure}
%%%%%%%%%%%%%%%%%%% END FIGURE %%%%%%%%%%%%%%%%%%%%%%%%

We use the two segment scheme, Eq.~(\ref{eq:2segmentpath}) and Fig.~\ref{fig:evolution-path}, with three choices for the primary ordering scale, $\mu_\scC = \mu_\perp$ for $k_\LT$ ordering, $\mu_\scC = \mu_\Lambda$ for $\Lambda$ ordering, and $\mu_\scC = \mu_\angle$ for angular ordering. In each case, the primary evolution uses the LC+ approximation for color, so that the soft splitting operator is the difference, $\Delta S_j(\vec\mu)$, between splitting with full color and splitting with the LC+ approximation for color, Eq.~(\ref{eq:cSjLCplus}). Since we start with just a $q \bar q$ state and the LC+ approximation is exact for such a state, there is no evolution on the first segment of the path. For each choice of ordering scale, we let the $m_\perp^2$ cut end the shower. We choose $m_\perp^2 = 1 \GeV^2$. We do not provide a hadronization stage for the shower.

With the LC+ approximation in \textsc{Deductor}, the shower can generate contributions with values greater than zero of a parameter called the color suppression index, $I$ \cite{NScolor}. These contributions are suppressed by a factor of at least $1/N_\Lc^I$. The user can choose a value $I_\mathrm{max}$ such that values of $I$ greater than $I_\mathrm{max}$ are not generated \cite{NSNewColor}. We choose $I_\mathrm{max} = 4$.

The nominal renormalization scale according to the formulation given above for $\cS_\scC^\mathrm{LC+}(\vec\mu)$ is $\mu_\scR = \mu_\scC$ or, more generally, some function of the scales $\vec\mu$, Eq.~(\ref{eq:muR}). However, \textsc{Deductor} attempts to incorporate some contributions from higher order splitting functions by evaluating $\as$ in the splitting functions at $\mu^2 = k_\LT^2/z = (1-z) 2\, \hat p_l\cdot \hat p_{\mpone}$.

In Fig.~\ref{fig:twojet}, we show the results for $(y_{23}/\sigma_\mathrm{tot})\,d\sigma/d y_{23}$ as a function of $y_{23}$ for $k_\LT$ ordering, $\Lambda$ ordering, and angular ($\vartheta$) ordering in the second segment of the two segment path in Fig.~\ref{fig:evolution-path}. We also show the next-to-leading-log (NLL) analytic expectation \cite{JetsNLL,pinkbook} for this quantity. We see that the distribution for $\Lambda$ ordering lies between the distributions for $k_T$ ordering and for angular ordering. This was to be expected because, according to Eqs.~(\ref{eq:kTsq}) and (\ref{eq:AfromLambda}), $k_\LT^2 < \Lambda^2 < \vartheta Q^2$ for any splitting. The results for $k_\LT$ ordering and $\Lambda$ ordering are close to each other and are quite close to the NLL analytic expectation. The angular ordering result is substantially different from the $k_\LT$ ordering and $\Lambda$ ordering results and the NLL analytic expectation. We do not have a satisfying explanation for this behavior, but we note that an analysis in Appendix \ref{sec:I2angle} along the lines of Ref.~\cite{NSThrustSum} indicates that for the thrust, $T$, distribution, the angular ordered version of the algorithm fails to sum large logarithms of $1-T$ at the NLL or even LL level.

%%%%%%%%%%%%%%%%%%%% FIGURE %%%%%%%%%%%%%%%%%%%%%%%%%%%
\begin{figure}
\begin{center}
\ifusefigs % then we include the figure

\begin{tikzpicture}
\begin{semilogxaxis}[title = {y23 distribution},
   xlabel={$y_{23}$}, 
   ylabel={$(y_{23}/\sigma_\mathrm{tot})\,d\sigma/d y_{23}$},
%  xmin=, xmax=,
%  ymin=, ymax=
  legend cell align=left,
  every axis legend/.append style = {
  at={(0.22,0.05)},
  anchor=south west}
]
\errorband[red,semithick]{fill=red!30!white, opacity=0.5}
{data/y23C0.dat}{1}{3}{4};
\addlegendentry{LC+}

\errorband[blue,dashed,thick]{fill=blue!30!white, opacity=0.5}
{data/y23C2.dat}{1}{3}{4};
\addlegendentry{LC+ \& $[\Delta S_\Lc]^2$}

%\errorband[black,dashed,thick]{fill=blue!30!white, opacity=0.5}
%{data/y23Lambda.dat}{1}{3}{4};
%\addlegendentry{New version}

\end{semilogxaxis}
\end{tikzpicture}

\else % We use the pdf figure.
\begin{center}
\includegraphics[width = 8.2 cm]{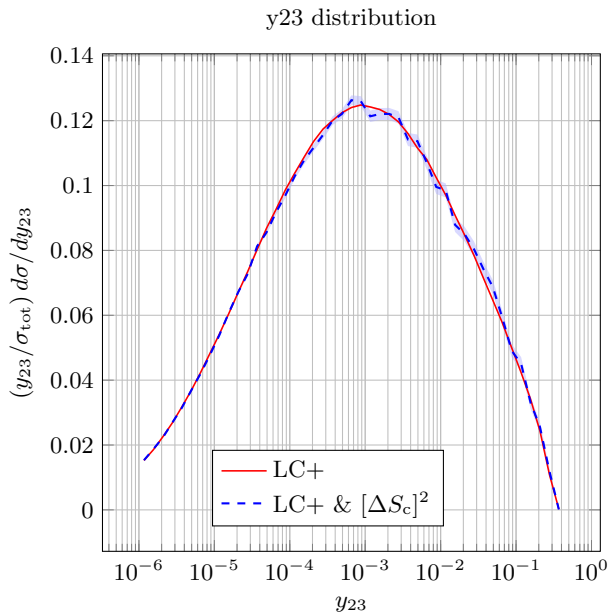}
\end{center}
\fi
\end{center}
\caption{
The $y_{23}$ distribution with the Cambridge algorithm using a calculation with just one scale $\mu_\scS$. Results with just the LC+ approximation and with up to two units of the color correction operator $\Delta S_\Lc$ are compared.
}
\label{fig:twojetcolor}
\end{figure}
%%%%%%%%%%%%%%%%%%% END FIGURE %%%%%%%%%%%%%%%%%%%%%%%%

As discussed in Sec.~\ref{sec:BetterColor}, because of the nature of the LC+ approximation and the simple nature of the $q\bar q$ hard state, the results in Fig.~\ref{fig:twojet} are exact in color. That is, whatever is lacking in the color treatment would be corrected up to order $\as^3$ if we had $\as^2$ corrections to the shower splitting functions. It seems a reasonable conjecture that the color dependence of these $\as^2$ corrections are numerically unimportant. To address this question, we can use the previous version (v.~(3.0.3)) of \textsc{Deductor}, in which there is a single shower scale $\mu_\scS$. The default shower uses the LC+ approximation, but the user can add powers of $\Delta S_\Lc$ perturbatively, interleaved with the LC+ evolution \cite{NSNewColor}. Our conjecture implies that the effect of adding $\Delta S_\Lc$ powers is not numerically important in the present case of $e^+e^-$ annihilation with a $q\bar q$ hard state.

To test this conjecture, we choose $\Lambda$ ordering in \textsc{Deductor}-(3.0.3) and compare the result with the LC+ approximation with the result with up to two powers of $\Delta S_\Lc$ added. The result is shown in Fig.~\ref{fig:twojetcolor}. We see that adding $\Delta S_\Lc$ powers makes the program run more slowly and thus increases the statistical errors. However, within the statistical errors (indicated by the band in Fig.~\ref{fig:twojetcolor}), adding $\Delta S_\Lc$ powers makes no difference to the result.

%-------------------------------------------------
\section{Summary and outlook}
\label{sec:conclusions}

In a parton shower, the state of many partons evolves as partons split with increasing ``shower time'' $t$. For a first order shower, one parton can split into two partons as dictated by three splitting variables such as $(k_\LT,z,\phi)$. We have taken the view that for any $t$ there is a resolvable region and an unresolvable region in the space of parton splitting variables. As $t$ increases, more splittings become resolvable, so that there is a probability for a newly resolvable splitting to occur. The surface that divides the two regions can be parameterized by variables $\vec\mu$, with functions $\vec\mu(t)$ specifying the progression of boundary surfaces.

One could have any number of parameters $\mu_n$ to describe a surface in the space of parton splitting variables. Parton shower algorithms typically use one. In this paper, we use two parameters. We choose the first to be a parameter $\mu_\scE$ that provides a cut on the energy of an emitted parton. We choose the second parameter to be $\mu_\scC$, which could be any of $\mu_\perp$, $\mu_\Lambda$, or $\mu_\angle$. In single variable evolution, this would correspond, respectively, to $k_\LT$, $\Lambda$, or angular ordering, as described in Sec.~\ref{sec:UnresolvedOneScale}.

The probability for parton splitting in a first order shower is determined (using Eq.~(\ref{eq:Dhatldef})) by splitting operators $\bm D_l$, where $l$ is the index of the parton that splits and where $\bm D_l$ is an operator on the color space of the partons and a function of the parton momenta and flavors. These operators can be decomposed into two terms
\begin{equation}
\bm D_l = \bm D_l^\mathrm{sc} + \bm D_l^\mathrm{soft}
\;,
\end{equation}
where $\bm D_l^\mathrm{soft}$ is singular for soft emissions but is not singular for collinear emissions (or for soft$\times$collinear emissions). We have adopted an especially useful way to do this by defining 
\begin{equation}
\bm D_l^\mathrm{sc} = \bm D_l^\mathrm{LC+}
\;.
\end{equation}
In $\bm D_l^\mathrm{LC+}$, we approximate $\bm D_l$ using the LC+ approximation \cite{NScolor} for color. Then the second term is
\begin{equation}
\label{eq:Slsoftdef}
\bm D_l^\mathrm{soft} = \bm D_l - \bm D_l^\mathrm{LC+}
\;.
\end{equation}
As we see in Appendix \ref{sec:deductor}, the LC+ approximation is exact in the limit of collinear emissions \cite{NScolor}. That is, $\bm D_l - \bm D_l^\mathrm{LC+}$ has no collinear singularity. 

This decomposition is important because we can define different treatments of the unresolvable region for the two contributions to $\bm D_l$. For splittings derived from $\bm D_l^\mathrm{LC+}$, there are three cuts used to determine when a splitting is unresolved. First, we impose a fixed infrared cutoff by defining a splitting to be unresolved whenever $k_\LT^2 < m_\perp^2$, where $m_\perp^2$ is of order $1\GeV^2$. Second, there is a cut that depends on the energy scale $\mu_\scE^2$: a splitting is unresolved whenever $4z(1-z) < \mu_\scE^2/Q^2$. Third, there is a cut that depends on $\mu_\scC^2$. If $\LC = {\perp}$, a splitting is unresolved whenever $k_\LT^2 < \mu_\perp^2$. If $\LC = \Lambda$, a splitting is unresolved whenever $\Lambda^2 < \mu_\Lambda^2$, where $\Lambda^2$ is the default ordering variable in \textsc{Deductor} and is proportional to the virtuality in the splitting. If $\LC = \angle$, a splitting is unresolved whenever $\vartheta\, Q^2 < \mu_\angle^2$ where $\vartheta = [1-\cos (\theta)]/2$ and $\theta$ is the angle between the daughter parton momenta in the rest frame of $Q$. 

We treat splittings derived from $\bm D_l^\mathrm{soft}$ differently. The soft singularity is controlled by the cut $4z(1-z) < \mu_\scE^2/Q^2$. Since for these splittings there is no collinear singularity, we can omit the cut based on $\mu_\scC^2$.

The final ingredient in the formulation presented in this paper is the choice of a path $(\mu_\scE(t), \mu_\scC(t))$. The path we choose is has two segments, as shown in Fig.~\ref{fig:evolution-path}. In the first segment, with $0 < t < 1$, $\mu_\scC^2$ is fixed at $Q^2$ and $\mu_\scE$ decreases from $Q^2$ to 0. In the second segment, with $1 < t < \infty$, $\mu_\scE^2$ is fixed at $0$ and $\mu_\scC^2$ decreases from $Q^2$ to 0. In the first path segment, there is no unresolved region available for $\bm D_l^\mathrm{LC+}$ because of the cut  imposed by $\mu_\scC^2$. However, this cut does not apply for $\bm D_l^\mathrm{soft}$, so there is a contribution from $\bm D_l^\mathrm{soft}$. In the second part, $\bm D_l^\mathrm{LC+}$ contributes, but the unresolved region does not change for $\bm D_l^\mathrm{soft}$, so $\bm D_l^\mathrm{soft}$ does not contribute. This gives Eq.~(\ref{eq:cUfactors}) for the complete evolution:
\begin{equation}
\label{eq:cUfactorsencore}
\cU(\infty,0) = \cU(\infty,1)\,\cU(1,0)
\;.
\end{equation}
The second factor here, $\cU(\infty,1)$, is a complete shower using the LC+ approximation for color and either $k_\LT$, $\Lambda$, or angular ordering. The first factor provides an evolution in parton energy using the soft operator $\bm D_l^\mathrm{soft}$.

In the case of angular ordering, the formulation presented here provides a way to understand an angular ordered shower in which the only cutoff on soft emissions is provided by the fixed cutoff $k_\LT^2 > m_\perp^2$. If we were to set $m_\perp^2$ to zero, we would have a naked singularity in the resolved region. With a fixed value of $m_\perp^2$, we do not find infinities in the results, but we can find large logarithms, $\log(Q^2/m_\perp^2)$, that are not summed by a renormalization group equation. In the two scale treatment, the large logarithms are absorbed into $\cU(1,0)$.

The LC+ shower provided by $\cU(\infty,1)$ is corrected by the operator $\cU(1,0)$ that is built from $\bm D_l^\mathrm{soft}$, Eq.~(\ref{eq:Slsoftdef}). The splitting operator $\bm D_l^\mathrm{soft}$ has a complicated color structure, making numerical calculations based on this operator difficult. However, this operator tends to be numerically small because it starts with a factor $1/N_\Lc^2 \sim 1/10$ and because it lacks a collinear singularity. Thus one can attack the numerical evaluation by expanding $\cU(1,0)$ in powers of $\bm D_l^\mathrm{soft}$. We have done this in Ref.~\cite{NSNewColor}, with splittings according to $\bm D_l^\mathrm{soft}$ interleaved with LC+ evolution. The numerical evidence suggests that an expansion in powers of $\bm D_l^\mathrm{soft}$ is adequate. With the shower formulation presented in this paper, the needed calculations are simpler because $\cU(1,0)$ is applied to the initial hard scattering state, denoted by $\sket{\rho_\scH}$, which has few partons. The needed calculations are also simpler because the splittings from $\bm D_l^\mathrm{soft}$ do not need to be interleaved with LC+ evolution, which we found to be complicated and computationally expensive.

In the case of $e^+e^-$ annihilation with a color singlet $q\bar q$ state $\sket{\rho_\scH}$ to start the shower, the calculations are, in fact, trivial. Because the space of $q\bar q \Lg$ color states is just one dimensional, $\bm D_l^\mathrm{soft}$ applied to $\sket{\rho_\scH}$ vanishes. Thus $\cU(1,0)\sket{\rho_\scH} = \sket{\rho_\scH}$ and no numerical calculation is needed.

Application of the formulation of this paper to hadron-hadron collisions is left to future work. Here, we note that for the Drell-Yan process at the Born level, the initial state with a color singlet $q\bar q$ is like a $q\bar q$ final state in $e^+e^-$ annihilation, so that  $\cU(1,0)\sket{\rho_\scH} = \sket{\rho_\scH}$ However, for jet production in hadron-hadron collisions, $\cU(1,0)\sket{\rho_\scH} \ne \sket{\rho_\scH}$. Then a perturbative expansion of $\cU(1,0)$ will be needed. However, this expansion should be much simpler than when powers of $\bm D_l^\mathrm{soft}$ are interleaved with the LC+ shower in the style of Ref.~\cite{NSNewColor}.

Finally, we offer the speculation that using multiple scales may prove useful in developing a parton shower algorithm with splitting functions defined at order $\as^2$ instead of just $\as$. At order $\as^2$, one can have two real emissions, one real emission together with a virtual exchange, or two virtual exchanges. For the case of two real emissions, both can be soft, one can be soft and one collinear with an existing parton, two can be collinear to two existing partons, or two can be collinear with one existing parton. The resulting singular surfaces are much more complicated than they are in a first order shower. It may well be useful to employ different scale parameters to describe an unresolved region that includes all of the singularities.

%-------------------------------------------------
\acknowledgments{ 
This work was supported in part by the United States Department of Energy under grant DE-SC0011640. The work benefited from access to the University of Oregon high performance computer cluster, Talapas.  
}

%-------------------------------------------------
\appendix
%-------------------------------------------------

%-------------------------------------------------
\section{About the \textsc{Deductor} shower}
\label{sec:deductor}

In this appendix, we specify details of the \textsc{Deductor} shower kinematics  \cite{NSI, NSThreshold} and splitting functions \cite{NSI, NSII, NScolor, NSspin} used in the main text. We adopt a notation that is different from that in Refs.~\cite{NSI, NSII, NScolor, NSspin, NSThreshold} and emphasizes some of the features that are important in this paper. We concentrate on the singular operators ${\cal D}^{[1,0]}_l(\mur^2,\vec\mu)$ and ${\cal D}^{[0,1]}_l(\mur^2,\vec\mu)$ from which the splitting functions used in the shower are derived \cite{NSAllOrder} since these operators carry more information than the shower splitting functions.

%-------------------------------------------------
\subsection{The form of $\cD^{[1,0]}_l$}
\label{sec:cD10}

To define the singular operator for a final state splitting, we begin with the kinematic variables. Before the splitting, there are incoming partons labelled $\La,\Lb$ and final state partons $1,2,\dots,m$. For electron-positron annihilation, the incoming partons do not participate in the shower since they carry no color charge. The final state partons have momenta $\{p\}_m = \{p_1,\dots,p_m\}$ and flavors $\{f\}_m = \{f_1,\dots,f_m\}$. The total momentum of the final state partons is $Q$. Then also $Q = p_\La + p_\Lb$. 

Now, for a final state splitting, a parton labelled $l \in \{1,\dots,m\}$ splits. The size of $p_l$ is conveniently described using the auxiliary variable $a_l$, Eq.~(\ref{eq:aldef}). It is also useful to define an auxiliary lightlike vector $n_l$ in the plane of $p_l$ and $Q$, Eq.~(\ref{eq:nldef}). Parton $l$ splits into a new parton with label $l$ and momentum $\hat p_l$ and a new parton with label $\mpone$ and momentum $\hat p_\mpone$. We use a scaled virtuality variable $y$, Eq.~(\ref{eq:ydef}), and a momentum fraction $z$, Eq.~(\ref{eq:zdef}), to specify the splitting. We also define an azimuthal angle $\phi$ of the splitting using the part, $k_\perp$, of $\hat p_l$ that is orthogonal to $p_l$ and $n_l$. The three splitting variables $y$, $z$, and $\phi$ determine $\hat p_l$ and $\hat p_\mpone$ using
\begin{equation}
\begin{split}
\label{eq:daughtermomenta}
\hat p_l ={}& 
z\,h_+(y)\, p_l
+ (1 - z) h_-(y)\, n_l
+k_\perp
\;,
\\
\hat p_{m+1} ={}& 
(1-z)\, h_+(y)\, p_l
+ z  h_-(y)\, n_l
- k_\perp
\;,
\end{split}
\end{equation}
where
\begin{equation}
\begin{split}
\label{eq:lambdahpmdefs}
h_\pm(y) ={}& \frac{1}{2}\,[1 + y \pm \lambda(y)]
\;,
\\
\lambda(y) ={}& 
\sqrt{(1 + y)^2 - 4 a_l y}
\;.
\end{split}
\end{equation}
The magnitude of the transverse momentum $k_\perp$ is given by Eq.~(\ref{eq:kTsq}),
\begin{equation}
- \frac{k_\perp^2}{2p_l\cdot Q} =
z (1-z)y
\;.
\end{equation}
For $i \notin \{l,\mpone\}$, the momenta $\hat p_i$ are related to the momenta $p_i$ before the splitting by a Lorentz transformation,  $\hat p_i^\mu = \Lambda^\mu_\nu p_i^\nu$ \cite{NSI}. This Lorentz transformation is a boost in the plane of $p_l$ and $Q$ and allows $\sum_{i=1}^{m+1} \hat p_i = Q$.

For a final state splitting, we need the singular operator $\cD^{[1,0]}_l(\mur,\vec\mu)$ that appears in Eqs.~(\ref{cDexpansion}) and (\ref{eq:cD10cD01}). Here $\mur$ is the renormalization scale and $\vec\mu$ is the shower scale, which may have more than one component, as in Eq.~(\ref{eq:cDfullexpansion}). The operator $\cD^{[1,0]}_l(\mur,\vec\mu)$ has both soft and collinear singularities. We do not now divide it into two parts that get different treatments, as in Eq.~(\ref{eq:cDscandsoft}).

We can now state what $\cD^{[1,0]}_l(\mur,\vec\mu)$ contains. We apply $\cD^{[1,0]}_l(\mur,\vec\mu)$ to an $m$-parton state and write the result in the form
\begin{equation}
\begin{split}
\label{eq:Dhatldef}
{\cal D}^{[1,0]}_l&(\mur,\vec\mu)
\sket{\{p,f,c,c'\}_{m}}
\\
={}& 
\int\!d\{\hat p,\hat f\}_\mpone\
\sket{\{\hat p,\hat f\}_{\mpone}} 
\\&\times
\frac{\as(\mur^2)}{2\pi}
\hat {\bm D}_l(\{\hat p,\hat f\}_\mpone,\{p,f\}_{m})
\sket{\{c,c'\}_{m}}
\;.
\end{split}
\end{equation}
Here $\hat {\bm D}_l$ is a function of the momenta and flavors before and after the splitting and is an operator that maps the color space with $m$ final state partons, into the color space with $\mpone$ final state partons.

The operator $\hat {\bm D}_l$ has the form derived from Eq.~(5.7) of Ref.~\cite{NScolor} and Eq.~(8.20) of Ref.~\cite{NSI} with dimensional regulation added,
\begin{align}
\label{eq:D10l}
\hat {\bm D}_l(&\{\hat p,\hat f\}_\mpone,\{p,f\}_{m};\epsilon)
\\\notag={}&
\left(\frac{\mur^2}{2p_l\cdot Q}\right)^\epsilon
\frac{(4\pi)^\epsilon}{\Gamma(1-\epsilon)}\,
\int_0^1\!dz\
[z(1-z)]^{-\epsilon}
\\\notag&\times
\int_0^1\!\frac{dy}{y}\
y^{-\epsilon}
\left[\lambda(y)\right]^{1-2\epsilon}\,
\theta(\lambda^2(y) > 0)
\sum_{\hat a \in \mathbb{S}(a)}
\\\notag&\times
\int\!\frac{d^{1-2\epsilon}\phi}{S(2-2\epsilon)}\
\delta\big(\{\hat p, \hat f\}_\mpone - R_l(y,z,\phi, 
 \hat a;\{p,f\}_{m})\big)
\\\notag&\times
\Theta\big((y,z) \in U(\vec\mu)\big)\,
\\\notag&\times
\sum_{k}\frac{1}{2}\bigg[
\theta(k = l)\,
\frac{1}{N(\hat a, a)}\, \hat P^{\hat a a}(z,y,a_l,\epsilon)
\\\notag&\qquad -
\theta(k \ne l)\,
\delta_{\hat a a}\,
\frac{2}{1-z}\, 
W_0\!\left(\xi_{l k}, a_l, z, y, \phi - \phi_k \right)
\bigg]
\\\notag&\times
\{
t^\dagger_l(f_l \to \hat f_l\!+\!\hat f_\mpone)\otimes
t_k(f_k \to \hat f_k \!+\! \hat f_\mpone)
\\\notag&\quad
+
t^\dagger_k(f_k \to \hat f_k \!+\! \hat f_\mpone)\otimes
t_l(f_l \to \hat f_l \!+\! \hat f_\mpone)
\}
\;.
\end{align}
There are dimensionally regulated integrations over splitting variables $y$, $z$, and $\phi$. The variable $\phi$ is a unit vector in the $2 - 2 \epsilon$ dimensional transverse momentum space and represents the azimuthal angle of $\hat p_l$ around the direction of $p_l$. The integration over $\phi$ is an integration over a unit sphere that is a $1 - 2\epsilon$ dimensional surface. The function $S(2 - 2\epsilon)$ is the surface area of this sphere, so that
\begin{equation}
\int\!\frac{d^{1-2\epsilon}\phi}{S(2-2\epsilon)}\ 1 = 1
\;.
\end{equation}
There is also a sum over the flavor $\hat a$ of parton $l$ after the splitting, which we use as a splitting variable that specifies the flavor content of the splitting. The set of allowed values of $\hat a$, $\mathbb{S}(a)$, depends on the flavor $a \equiv f_l$ of the parton that splits. For all $a$, $a \in \mathbb{S}(a)$. This corresponds to a splitting $a \to a + \Lg$, where \textsc{Deductor} labels the daughter gluon as $\mpone$. For $a = \Lg$, also $q \in \mathbb{S}(a)$ for any quark flavor $q$. This corresponds to a splitting $\Lg \to q + \bar q$, where \textsc{Deductor} labels the daughter quark as $l$.

After the integrations, there is a delta function that sets $\{\hat p, \hat f\}_\mpone$ to the momenta and flavors obtained from a splitting with variables $(y,z,\phi, \hat a)$ applied to partons with momenta and flavors $\{p,f\}_{m}$ according to \textsc{Deductor} conventions.

The idea of the singular operator ${\cal D}^{[1,0]}$ is that it integrates over splittings that are arbitrarily close to the soft and collinear limits, but with an ultraviolet cutoff that depends on scale parameters $\vec\mu$. The region of $(y,z)$ allowed by the cutoff is called the unresolved region and is denoted by $U(\vec\mu)$. We therefore insert a theta function that specifies that $(y,z)$ lies in the unresolved region.

\textsc{Deductor} is a dipole shower. In the following factor, there is a sum over dipole partner partons $k$. In the first term, the partner parton is the same as the emitting parton, $k = l$. This term contains a color factor $N(\hat a, a)$ defined by 
\begin{equation}
\begin{split}
\label{eq:NaahatFS}
N(q, \Lg) ={}& T_\LR \;,
\\
N(q, q) ={}& C_\LF \;,
\\
N(\Lg, \Lg) ={}& C_\LA \;,
\end{split}
\end{equation}
where $q$ is any quark or antiquark flavor.  Then there is a splitting function $\hat P^{\hat a a}(z,y,a_l,\epsilon)$. In the case of a $\Lg \to q \bar q$ splitting, where $q$ is a quark flavor, the label $l$ after the splitting is assigned to the quark. Thus we have $a = \Lg$ and $\hat a = q$. Then $\hat P^{q \Lg}$ is related to the function $\overline w_{ll}(\{\hat p,\hat f\}_\mpone)$ that appears in Eq.~(A.1) of Ref.~\cite{NSII} by
\begin{equation}
\label{eq:PaadefI}
\frac{4\pi\as(\mur^2)}{y\, p_l\cdot Q}\,
\frac{\hat P^{q \Lg}(z,y,a_l,\epsilon)}{N(q, \Lg)} = 
\overline w_{ll}(\{\hat p,\hat f\}_\mpone)
\;.
\end{equation}
In all other splittings, one of the partons after the splitting is a gluon. The label $\mpone$ is assigned to the gluon. Then parton $l$ can be a quark, antiquark, or gluon and $\hat a = a$. In this case, $\hat P^{a a}$ is related to the functions $\overline w_{\La\La}(\{\hat p,\hat f\}_\mpone)$ and $\overline w_{\La\La}^{\rm eikonal}(\{\hat p,\hat f\}_\mpone)$ that appear in Eqs.~(2.23) and (2.58) of Ref.~\cite{NSII} by
\begin{equation}
\begin{split}
\label{eq:PaadefII}
\frac{4\pi\as(\mur^2)}{y\, p_l\cdot Q}\, &
\frac{\hat P^{a a}(z,y,a_l,\epsilon)}{N(a, a)}
\\
={}& 
\overline w_{ll}(\{\hat p,\hat f\}_\mpone)
- \overline w_{ll}^{\rm eikonal}(\{\hat p,\hat f\}_\mpone)
\\&
+ \frac{4\pi\as(\mur^2)}{y p_l\cdot Q}\, 
\left[\frac{2}{1 - z + a_l y} - 2
\right]
\;.
\end{split}
\end{equation}
In Eq.~(\ref{eq:PaadefI}) and Eq.~(\ref{eq:PaadefII}), we calculate $\overline w_{ll}$ in $4 - 2\epsilon$ dimensions by counting the number of spin states of a gluon as $2 - 2\epsilon$ instead of just 2.

The functions $\hat P^{a \hat a}(z,y,a_l,\epsilon)$ are somewhat complicated. 
It is helpful to express these functions using the variables 
\begin{equation}
\begin{split}
\label{eq:xlthetal}
x(y) ={}& \frac{1+y-\lambda(y)}{1+y+\lambda(y)}\;\;,
\\
\vartheta(y) ={}& \frac{x(y)}{[z+(1-z)x(y)]\,[1-z+z x(y)]}
\;.
\end{split}
\end{equation}
The variable $x$ vanishes for $y \to 0$: $x(y)\sim a_l y + \cO(y^2)$. The variable $\vartheta$ is the angle variable for the splitting defined in Eq.~(\ref{eq:vartheta}). Then we find, 
\begin{align}
\label{eq:hatP}
\hat P^{q \Lg}(&z,y,a_l,\epsilon) = 
T_\LR \bigg[1 - \frac{2 z(1-z)}{1-\epsilon}\bigg]
\;,
\\
\hat P^{q q}(&z,y, a_l,\epsilon)
\notag  
\\={}& C_\LF
\bigg[
\frac{2}{1 - z + a_l y}-2
+ (1-\epsilon)\,(1-z) h_+(y)
\notag
\\&\qquad
+2z(1-z)\,
\frac{[h_+(y) - 1 + x(y)][1-x(y)]}{(1-z+z x(y))^2}
\bigg]
\;,
\notag
\\
\hat P^{\Lg \Lg}(&z,y,a_l,\epsilon)
\notag
\\={}& C_\LA
\bigg[
\frac{2}{1 - z + a_l y}-2
\notag
\\&\qquad
+ z(1-z)\left(1-\frac{2\vartheta(y)[1-\vartheta(y)]}{1-\epsilon}\right)
\bigg]
\notag
\;.
\end{align}
In $\hat P^{q \Lg}$, $q$ can be any flavor of quark, while in $\hat P^{q q}$, $q$ can be any flavor of quark or antiquark. For $\epsilon = 0$, these functions are given in Eqs.~(A.1) and (2.23) of Ref.~\cite{NSII} or Appendix B of Ref.~\cite{NSThreshold}.

The functions $\hat P^{\hat a a}(z,y,a_l,\epsilon)$ are simple at $\epsilon = 0$, $y = 0$:
\begin{equation}
\begin{split}
\hat P^{q \Lg}(z,0,a_l,0) ={}& 
T_\LR \left[1 - 2 z(1-z)\right]
\;,
\\
\hat P^{q q}(z,0,a_l,0)  ={}& C_\LF
\frac{1+z^2}{1-z}
\;,
\\
\hat P^{\Lg \Lg}(z,0,a_l,0)  ={}& C_\LA
\left[\frac{2z}{1-z} + z(1-z)\right]
\;.
\end{split}
\end{equation}
The first two of these are the standard DGLAP parton evolution kernels. In $\hat P^{\Lg \Lg}$, both parton $l$ and parton $m+1$ after the splitting are gluons. Parton $l$ carries momentum fraction $z$, while parton $m+1$ carries momentum fraction $1-z$. The \textsc{Deductor} algorithm breaks the symmetry between these two gluons. The total probability to produce a gluon with momentum fraction $z$ is given by the standard DGLAP parton evolution kernel,
\begin{equation}
\begin{split}
\hat P^{\Lg \Lg}(z,0,a_l,0)& + \hat P^{\Lg \Lg}(1-z,0,a_l,0)  
\\= {}&
2C_\LA
\left[\frac{z}{1-z} + \frac{1-z}{z} + z(1-z)\right]
\;.
\end{split}
\end{equation}

Next in Eq.~(\ref{eq:D10l}) is a term proportional to a function $W_0\!\left(\xi_{l k}, a_l, z, y, \phi - \phi_k \right)$. This term comes from interference between emission of a gluon from parton $l$ and emission from dipole partner parton $k$ with $k \ne l$. We write the momentum of parton $k$ before the splitting as
\begin{equation}
\begin{split}
\label{eq:partnermomentumbefore}
p_k ={}& 
F_{lk}\Big[(1-\xi_{lk})\,p_l + \xi_{lk}\, n_l
\\&
+ \sqrt{\xi_{lk} (1-\xi_{lk}) Q^2/a_l^2} \,u_{\perp}\Big]
\;.
\end{split}
\end{equation}
The variable $\xi_{l k}$ is $(1 - \cos\theta_{l, k})/2$ where $\theta_{l, k}$ is the angle between $p_k$ and $p_l$ as measured in the rest frame of $Q$. The vector $u_\perp$ is a transverse unit vector, $p_l\cdot u_\perp = n_l\cdot u_\perp = 0$ and $u_\perp^2 = - 1$. The azimuthal angle of $u_\perp$ is $\phi_k$ and the azimuthal angle $\phi$ of $k_\perp$ is defined by
\begin{equation}
k_\perp \cdot u_\perp
= - \sqrt{k_T^2}\, \cos(\phi-\phi_k)
\;.
\end{equation}
In order to conserve momentum in the splitting, \textsc{Deductor} makes a small Lorentz transformation on all of the final state momenta except for $\hat p_l$ and $\hat p_{m+1}$ \cite{NSI}. This Lorentz transformation changes $p_k$ to
\begin{equation}
\begin{split}
\label{eq:partnermomentumafter}
\hat p_k ={}& 
F_{lk}\Big[e^{\omega(y)} (1-\xi_{lk})\,p_l 
+ e^{-\omega(y)}\xi_{lk}\, n_l
\\&
+ \sqrt{\xi_{lk} (1-\xi_{lk}) Q^2/a_l^2} \,u_{\perp}\Big]
\;.
\end{split}
\end{equation}
The boost angle $\omega$ is given by 
\begin{equation}
e^{-\omega(y)} = 1 - \sqrt{x(y)\, y/a_l}
\;.
\end{equation}
Thus $\omega(y) \to 0$ when $y \to 0$: $\omega(y) \sim y + \cO(y^2)$.

The function $W_0$ is defined by
\begin{equation}
\begin{split}
\label{eq:Wlk0-fin-definition}
\frac{2}{1-z}\,
W_0\!&\left(\xi_{lk}, a_l, y, z, \phi - \phi_k\right) 
\\={}&
\frac{p_l\!\cdot\! Q\,y}{4\pi\as(\mur^2)}
A'_{l k}(\{\hat p\}_{m+1})\, 
\overline w_{lk}^\mathrm{dipole}(\{\hat p\}_{m+1})
\\&\qquad
- \left(\frac{2}{1-z+a_l y} - 2 \right)
\;.
\end{split}
\end{equation}
The function $\overline w_{l k}^\mathrm{dipole}$ is the familiar dipole radiation function that appears in Ref.~\cite{NScolor}, Eq.~(5.3),
\begin{equation}
\begin{split}
\label{eq:dipole}
\overline w_{l k}^\mathrm{dipole}(\{\hat p\}_{m+1}) ={}& 
4\pi\as(\mur^2)\,\frac{2 \hat p_l\cdot \hat p_k}
{\hat p_\mpone\cdot \hat p_l\, \hat p_\mpone\cdot \hat p_k}
\;.
\end{split}
\end{equation}
This represents the interference between emission of a gluon with momentum $\hat p_\mpone$ from parton $l$ and emission of this gluon from parton $k$.

To use $\overline w_{l k}^\mathrm{dipole}$ in a partitioned dipole shower, we multiply by $1 = A'_{l k} + A'_{kl}$, where $A'_{l k}$ is the dipole partitioning function from Ref.~\cite{NSspin}, Eq.~(7.12),
\begin{equation}
\begin{split}
\label{eq:partitioning}
A'_{l k}(\{\hat p\}_{m+1}) ={}& 
\frac{\hat p_\mpone\!\cdot\!\hat p_k\ \hat p_l\! \cdot\! Q}
{\hat p_\mpone\!\cdot\! \hat p_k\, \hat p_l\! \cdot\! Q
+ \hat p_\mpone\!\cdot\! \hat p_l\, \hat p_k\! \cdot\!  Q}
\;,
\\
A'_{kl}(\{\hat p\}_{m+1}) ={}& 
\frac{\hat p_\mpone\!\cdot\! \hat p_l\ \hat p_k \!\cdot\! Q}
{\hat p_\mpone\!\cdot\! \hat p_k\, \hat p_l \! \cdot\! Q
+ \hat p_\mpone\!\cdot \!\hat p_l\, \hat p_k\! \cdot\!  Q}
\;.
\end{split}
\end{equation}
Then $A'_{l k} \overline w_{l k}^\mathrm{dipole}$ is associated with emission from parton $l$ and $A'_{kl} \overline w_{l k}^\mathrm{dipole}$ is associated with emission from parton $k$. Crucially, $A'_{l k} = 0$ when $\hat p_{m+1}$ is collinear with $\hat p_k$, $\hat p_{m+1} \cdot \hat p_k = 0$. Thus the pole $1/\hat p_{m+1} \cdot \hat p_k$ in $\overline w_{l k}^\mathrm{dipole}$ is cancelled.  When $\hat p_{m+1}$ is collinear with $\hat p_l$, we have $A'_{kl} = 0$ so $A'_{lk} = 1$.

What is $\overline w_{l k}^\mathrm{dipole}$ when $\hat p_{m+1}$ becomes collinear with $\hat p_l$? This is the limit $\vartheta \to 0$ with fixed $z$, or, equivalently, $y \to 0$ with fixed $z$. In this limit, we have $A'_{lk} \to 1$ and
\begin{equation}
\frac{\hat p_l\cdot \hat p_k}{\hat p_\mpone\cdot \hat p_k}
\to 
\frac{\hat p_l\cdot n_l}{\hat p_\mpone\cdot n_l}
=\frac{z}{1-z}
\;.
\end{equation}
Using $2\hat p_\mpone\cdot \hat p_l = 2 p_l\cdot Q\, y$, we find
\begin{equation}
\begin{split}
\frac{p_l\!\cdot\! Q\,y}{4\pi\as(\mur^2)}\,&
A'_{l k}(\{\hat p\}_{m+1})\,
\overline w_{l k}^\mathrm{dipole}(\{\hat p\}_{m+1}) 
\\ \sim{}&
\frac{2z}{1-z}
= \frac{2}{1-z}-2
\\ \sim{}&
\frac{2}{1-z + a_l y}-2
\;.
\end{split}
\end{equation}
In Eq.~(\ref{eq:Wlk0-fin-definition}), we have subtracted the value of the first line of the right-hand side in this collinear limit.\footnote{Using a denominator $(1 - z + a_l y)$ instead of just $(1-z)$ does not change the behavior of the subtraction in the collinear limit, but avoids adding singular behavior that is not present in $A'_{l k} \overline w_{l k}^\mathrm{dipole}$ in the integration region $(1-z) \ll a_l y \ll 1$. This region corresponds to the emitted soft gluon moving opposite to the mother parton direction.}  Thus in the collinear limit, $W_0 \to 0$. 

The function $y A'_{lk} \overline w_{l k}^\mathrm{dipole}$ is singular in limit of soft emissions, $(1-z) \to 0$ with fixed $\vartheta$. In this limit, $\hat p_{m+1} \sim (1-z)\,\hat p_{m+1}^{(0)}$ with $\hat p_{m+1}^{(0)}$ fixed in the soft limit and with $\hat p_{l} \to p_l$ and $\hat p_{k} \to p_k$ in the soft limit. One then obtains a result of the form
\begin{equation}
y A'_{lk} \overline w_{l k}^\mathrm{dipole}
= \frac{f(\vartheta)}{1-z} + \cO((1-z)^0)
\;.
\end{equation}
The subtraction in Eq.~(\ref{eq:Wlk0-fin-definition}) eliminates the leading $\vartheta \to 0$ behavior, leaving
\begin{equation}
\frac{2}{1-z}\,W_0
= \frac{f(\vartheta)-f(0)}{1-z} + \cO((1-z)^0)
\;.
\end{equation}
Thus $W_0$ has a finite limit as $(1-z) \to 0$ at fixed $\vartheta$.

A convenient method to evaluate $W_0(\xi_{lk}, a_l, y, z, \phi - \phi_k)$ is to write the vectors involved as functions of $y,z,\phi$ and evaluate the vector dot products in Eqs.~(\ref{eq:dipole}) and (\ref{eq:partitioning}).

Finally in Eq.~(\ref{eq:D10l}) there is a factor with color operators. The operator  $t^\dagger_l(f_l \to \hat f_l\!+\!\hat f_\mpone)$, acting on the ket color state $\ket{\{c\}_m}$, gives the new color state $\ket{\{\hat c\}_\mpone}$ that one gets after emitting the new parton $\mpone$ from parton $l$ with flavor $f_l = a$, giving a new parton $l$ with flavor $\hat f_l = \hat a$. This operator is described in some detail in Ref.~\cite{NSI}. Similarly,   $t_k(f_k \to \hat f_k\!+\!\hat f_\mpone)$, acting on the bra color state $\bra{\{c'\}_m}$, gives the new color state $\bra{\{\hat c'\}_\mpone}$ that one gets after emitting the new parton $\mpone$ from parton $k$ with flavor $f_k$.

In the case that parton $\mpone$ is a gluon, the color operators obey the identity
\begin{equation}
\sum_{k = 1}^m t_k(f_k \to f_k\!+\!\Lg)
= 0
\;.
\end{equation}
This identity arises from the fact that the parton color state is an overall color singlet, so that attaching a color generator matrix $T_k^c$ to all of the parton lines $k$ in the state, including $k = l$, gives zero. We have used this identity to add the same term, proportional to $[2/(1-z+a_l y) - 2]$, to both the $k = l$ term and the $k \ne l$ terms in Eq.~(\ref{eq:D10l}). We have added this term in both places in order to move the soft$\times$collinear singularity from the $k \ne l$ terms to the $k = l$ term. After this change, the $k \ne l$ terms, proportional to $W_0$, have a soft singularity but not a collinear singularity.

%-------------------------------------------------
\subsection{The form of $\cD^{[0,1]}_l$}
\label{sec:1cD01l}

As in Eq.~(\ref{eq:cD10cD01}), the singular operator $\cD^{[1]}_l$ associated with parton $l$ consists of two parts, $\cD^{[1,0]}_l$ that specifies real splittings of parton $l$ and $\cD^{[0,1]}_l$, in which a virtual parton is exchanged. We have described $\cD^{[1,0]}_l$. We now would like to define the real part of ${\cal D}^{[0,1]}_l(\mur,\vec\mu)$. 

This operator comes from virtual graphs, in which we integrate over a momentum $q$ that flows around a loop. The operator ${\cal D}^{[0,1]}_l$ captures the infrared singularities when $q \to 0$ or $q$ becomes collinear with $p_l$ \cite{NSAllOrder}. Since ${\cal D}^{[0,1]}_l$ simply captures the singularities, it is defined to leave parton momenta and flavors unchanged: ${\cal D}^{[0,1]}_l \sket{\{p,f,c,c'\}_{m}}$ is defined to be a linear combination of states $\sket{\{p,f,\hat c, \hat c'\}_{m}}$ with the same momenta and flavors. The operator ${\cal D}^{[0,1]}_l$ does, however, change colors. It contains two kinds of terms. First, there are terms with the color structure of self-energy insertions on one of the parton legs. These terms are proportional to the unit operator on the color space. Second, there are terms with the color structure of gluon exchanges between two parton legs, $l$ and $k$.  The gluon line attaches to line $l$ with a color generator matrix $T_l^c$ in the  $\bm 8$, $\bm 3$ or $\bar {\bm 3}$ representation according to the flavor of parton $l$. The gluon line attaches to line $k$ with the appropriate generator matrix $T_k^c$. Then we sum over the gluon color index $c$. The result can be denoted by $\bm T_k \cdot \bm T_l$. Thus the gluon exchange terms are proportional to either $[\bm T_k \cdot \bm T_l \otimes 1]$ for a virtual graph on the ket amplitude or $[1 \otimes \bm T_k \cdot \bm T_l]$ for a virtual graph on the bra amplitude. The virtual graphs have $1/\epsilon^2$ and $1/\epsilon$ poles. By using color identities, we can arrange that the terms proportional to the unit operator on the color space have $1/\epsilon^2$ and $1/\epsilon$ poles, while the terms with $[\bm T_k \cdot \bm T_l \otimes 1]$ and $[1 \otimes \bm T_k \cdot \bm T_l]$ color operators have only $1/\epsilon$ poles that arise from the exchange of a soft gluon.

Since ${\cal D}^{[0,1]}_l$ leaves parton momenta and flavors unchanged but can change the $m$-parton color state, it has the form\footnote{There are imaginary contributions to the virtual graphs in ${\cal D}^{[0,1]}_l$, although the imaginary contributions from final state virtual exchanges with a final state emitting parton cancel \cite{NScolor}.}
\begin{equation}
\begin{split}
\label{eq:Gammaldef}
\mathrm{Re}\,&{\cal D}^{[0,1]}_l(\mur,\vec\mu)
\sket{\{p,f,c,c'\}_{m}}
\\& =  \sket{\{p,f\}_{m}}
\frac{\as(\mur^2)}{2\pi}\,
\bm \Gamma_{l}(\{p, f\}_m, \epsilon) \sket{\{c,c'\}_{m}}
.
\end{split}
\end{equation}
Now, we need to define $\bm \Gamma_{l}$. We will do this by relating ${\cal D}^{[0,1]}_l(\mur,\vec\mu)$ to the inclusive splitting probability produced by ${\cal D}^{[1,0]}_l(\mur,\vec\mu)$.

The probability associated with a basis state $\sket{\{\hat p,\hat f, \hat c, \hat c'\}_\mpone}$ is
\begin{equation}
\begin{split}
\label{eq:probabilityencore}
\sbrax{1}\sket{\{\hat p,\hat f, \hat c, \hat c'\}_\mpone} 
= \sbrax{1_\mathrm{pf}}\sket{\{\hat p,\hat f\}_\mpone}
\sbrax{1_\mathrm{color}}\sket{\{\hat c, \hat c'\}_\mpone} 
\end{split}
\end{equation}
with
\begin{equation}
\begin{split}
\label{eq:probabilitypartsencore}
\sbrax{1_\mathrm{pf}}\sket{\{\hat p,\hat f\}_\mpone} 
={}& 1
\;,\\
\sbrax{1_\mathrm{color}}\sket{\{\hat c, \hat c'\}_\mpone} 
={}& \brax{\{\hat c'\}_\mpone}\ket{\{\hat c\}_\mpone}
\;.
\end{split}
\end{equation}
Thus the probability corresponding to ${\cal D}^{[1,0]}_l$ applied to the state $\sket{\{p,f,c,c'\}_{m}}$ is
\begin{align}
\label{eq:1D10lstart}
\sbra{1}
{\cal D}^{[1,0]}_l&(\mur,\vec\mu)
\sket{\{p,f,c,c'\}_{m}}
\\ \notag
={}& 
\int\!d\{\hat p,\hat f\}_\mpone\
\frac{\as(\mur^2)}{2\pi}
\\ \notag &\times
\sbra{1_\textrm{color}}
\hat {\bm D}_{l}(z;\{\hat p,\hat f\}_\mpone,\{p,f\}_{m};\epsilon)
\sket{\{c,c'\}_{m}}
\;.
\end{align}
We write this using another operator $\hat {\bm P}_l$ as
\begin{equation}
\begin{split}
\label{eq:1D10lA}
\sbra{1}
{\cal D}^{[1,0]}_l&(\mur,\vec\mu)
\sket{\{p,f,c,c'\}_{m}}
\\
={}& 
\frac{\as(\mur^2)}{2\pi}
\sbra{1_\textrm{color}}
\hat {\bm P}_l(\{p,f\}_{m};\epsilon)
\sket{\{c,c'\}_{m}}
\;,
\end{split}
\end{equation}
where $\isbra{1_\textrm{color}}$ times the operator $\hat {\bm P}_l$ is
\begin{align}
\label{eq:1Plstart}
\sbra{1_\mathrm{color}}&
\hat {\bm P}_l(\{p,f\}_{m};\epsilon)
\sket{\{c,c'\}_{m}}
\\\notag ={}&
\left(\frac{\mur^2}{2 p_l\cdot Q}\right)^{\!\epsilon}
\frac{(4\pi)^\epsilon}{\Gamma(1-\epsilon)}\,
\int_0^1\!dz\ [z(1-z)]^{-\epsilon}
\\\notag&\times
\int_0^1\!\frac{dy}{y}\
y^{-\epsilon}\,
[\lambda(y)]^{1-2\epsilon}
\theta(\lambda^2(y) > 0)
\sum_{\hat a \in \mathbb{S}(a)}
\\\notag&\times
\int\!\frac{d^{1-2\epsilon}\phi}{S(2-2\epsilon)}\
\Theta\big((y,z) \in U(\vec\mu)\big)\,
\\\notag&\times
\sum_{k}\frac{1}{2}\bigg[
\theta(k = l)\,
\frac{1}{N(\hat a, a)}\, \hat P^{\hat a a}(z,y,\epsilon)
\\\notag&\qquad -
\theta(k\ne l)\,
\delta_{\hat a a}\,
\frac{2}{1-z}\, 
W_0\!\left(\xi_{l k}, a_l, z, y, \phi - \phi_k \right)
\bigg]
\\\notag&\times
\bra{\{c'\}_{m}}
t_k(f_k \to \hat f_k \!+\! \hat f_\mpone)
t^\dagger_l(f_l \to \hat f_l\!+\!\hat f_\mpone)
%\ket{\{c\}_{m}}
\\\notag&\quad +
%\bra{\{c'\}_{m}}
t_l(f_l \to \hat f_l \!+\! \hat f_\mpone)
t^\dagger_k(f_k \to \hat f_k\!+\!\hat f_\mpone)
\ket{\{c\}_{m}}
\;.
\end{align}
Here we have used the momentum conserving delta function in $\hat {\bm D}_l$ to eliminate the integration over $\{\hat p,\hat f\}_\mpone$. In the color factor, we have used the instruction in Eq.~(\ref{eq:probabilitypartsencore}) to take the trace of
\begin{align}
\sum_{\hat c,\hat c'}&\rho(\{\hat c,\hat c'\}_\mpone)
\ket{\{\hat c\}_\mpone} 
\bra{\{\hat c'\}_\mpone}
\\\notag
={}& t^\dagger_l(f_l \to \hat f_l\!+\!\hat f_\mpone)\ket{\{c\}_{m}}
\bra{\{c'\}_{m}}
t_k(f_k \to \hat f_k \!+\! \hat f_\mpone)
\end{align}
and the analogous color density matrix with $l \leftrightarrow k$.

We can simplify the color here. In the case that $k = l$,
\begin{equation}
\begin{split}
t_l(f_l \to \hat f_l \!+\! \hat f_\mpone)
t^\dagger_l(f_l \to \hat f_l\!+\!\hat f_\mpone)
= N(\hat a, a)
\;,
\end{split}
\end{equation}
where $N(\hat a, a)$ is the Casimir eigenvalue (\ref{eq:NaahatFS}) appropriate to the flavor content of the splitting. When $k \ne l$, the emitted parton $\mpone$ is always a gluon. Thus for $k \ne l$,
\begin{equation}
\begin{split}
t_k(f_k \to \hat f_k \!+\! &\hat f_\mpone)
t^\dagger_l(f_l \to \hat f_l\!+\!\Lg)
\\
={}& t_l(f_l \to \hat f_l \!+\! \Lg)
t^\dagger_k(f_k \to \hat f_k\!+\!\Lg)
\\
={}& \bm T_k \cdot \bm T_l
\;.
\end{split}
\end{equation}

These simplifications give us
\begin{equation}
\begin{split}
\label{eq:1PlA}
\sbra{1_\mathrm{color}}&
\hat {\bm P}_l(\{p,f\}_{m};\epsilon)
\sket{\{c,c'\}_{m}}
\\={}&
\left(\frac{\mur^2}{2 p_l\cdot Q}\right)^{\!\epsilon}
\frac{(4\pi)^\epsilon}{\Gamma(1-\epsilon)}\,
\int_0^1\!dz\ [z(1-z)]^{-\epsilon}
\\&\times
\int_0^1\!\frac{dy}{y}\
y^{-\epsilon}\,
[\lambda(y)]^{1-2\epsilon}
\theta(\lambda^2(y) > 0)
\\&\times
\int\!\frac{d^{1-2\epsilon}\phi}{S(2-2\epsilon)}\
\Theta\big((y,z) \in U(\vec\mu)\big)\,
\\&\times
\bigg[
\sum_{\hat a \in \mathbb{S}(a)}
\hat P^{\hat a a}(z,y,\epsilon)
\brax{\{c'\}_{m}}\ket{\{c\}_{m}}
\\&\qquad -
\sum_{k\ne l}
\frac{2}{1-z}\, 
W_0\!\left(\xi_{l k}, a_l, z, y, \phi - \phi_k \right)
\\&\qquad\quad \times
\bra{\{c'\}_{m}}\bm T_k \cdot \bm T_l\ket{\{c\}_{m}}
\bigg]
\;.
\end{split}
\end{equation}

This specifies $\sbra{1_\mathrm{color}} \hat {\bm P}_l(\{p,f\}_{m};\epsilon)$ but not the operator $\hat {\bm P}_l(\{p,f\}_{m};\epsilon)$. We need to specify the color content of $\hat {\bm P}_l(\{p,f\}_{m};\epsilon)$. We make a choice that matches the color structure of the virtual exchange operator $\cD^{[0,1]}_l$. We note that
\begin{equation}
\begin{split}
\bra{\{c'\}_{m}}\bm T_k &\cdot \bm T_l\ket{\{c\}_{m}}
\\
 ={}& \mathrm{Tr}\left[
 \bm T_k \cdot \bm T_l\ket{\{c\}_{m}}\bra{\{c'\}_{m}}
 \right]
\\
={}& \mathrm{Tr}\left[
 \ket{\{c\}_{m}}\bra{\{c'\}_{m}}\bm T_k \cdot \bm T_l
 \right]
.
\end{split}
\end{equation}
Thus we can define the color content of $\hat {\bm P}_l(\{p,f\}_{m};\epsilon)$ by
\begin{equation}
\begin{split}
\label{eq:1PlB}
\hat {\bm P}_l(&\{p,f\}_{m};\epsilon)
\\={}&
\left(\frac{\mur^2}{2 p_l\cdot Q}\right)^{\!\epsilon}
\frac{(4\pi)^\epsilon}{\Gamma(1-\epsilon)}\,
\int_0^1\!dz\ [z(1-z)]^{-\epsilon}
\\&\times
\int_0^1\!\frac{dy}{y}\
y^{-\epsilon}\,
[\lambda(y)]^{1-2\epsilon}
\theta(\lambda^2(y) > 0)
\\&\times
\int\!\frac{d^{1-2\epsilon}\phi}{S(2-2\epsilon)}\
\\&\times
\Theta\big((y,z) \in U(\vec\mu)\big)\,
\\&\times
\bigg[
\sum_{\hat a \in \mathbb{S}(a)}
\hat P^{\hat a a}(z,y,\epsilon)
\\&\qquad -
\sum_{k\ne l}
\frac{2}{1-z}\, 
W_0\!\left(\xi_{l k}, a_l, z, y, \phi - \phi_k \right)
\\&\qquad\quad\times
\frac{1}{2}\left\{
[\bm T_k \cdot \bm T_l \otimes 1]
+[1 \otimes \bm T_k \cdot \bm T_l]
\right\}
\bigg]
.
\end{split}
\end{equation}

This enables us to define the operator $\bm \Gamma_{l}$ that appears in Eq.~(\ref{eq:Gammaldef}) for $\mathrm{Re}\,{\cal D}^{[0,1]}_l$. Because of the familiar real-virtual cancellations, poles in $\bm \Gamma_{l}(\{p, f\}_m,\epsilon)$ match the poles in $- \hat{\bm P}_l(\{p, f\}_m, \epsilon)$:\footnote{For details, see Ref.~\cite{jetcalc1992}, for example.}
\begin{equation}
\Big[\bm \Gamma_{l}(\{p, f\}_m,\epsilon)\Big]_\mathrm{poles}
= - \Big[\hat{\bm P}_l(\{p, f\}_m, \epsilon)\Big]_\mathrm{poles}
\;.
\end{equation}
This leaves the finite part of $\bm \Gamma_{l}(\{p, f\}_m,\epsilon)$ undefined. It is not evident how to impose an ultraviolet cutoff on the unresolved region for virtual graphs that matches the cutoff that we used for real emission graphs. In Ref.~\cite{NSThreshold} we proposed a method for this. Here, we propose a simpler method that gives the same result. We define
\begin{equation}
\label{eq:GammafromP}
\bm \Gamma_{l}(\{p, f\}_m,\epsilon)
= - \hat{\bm P}_l(\{p, f\}_m, \epsilon)
\;.
\end{equation}

Eq.~(\ref{eq:GammafromP}) gives us 
\begin{equation}
\sbra{1}
{\cal D}^{[1]}_l
= \sbra{1}
\left[{\cal D}^{[1,0]}_l + 
{\cal D}^{[0,1]}_l 
\right]
= 0
\;.
\end{equation}
This is significant because the shower splitting operators $S_j$ for a first order shower are defined by Eq.~(\ref{eq:cSjdef}),
\begin{equation}
\cS_j(\mur,\vec\mu) = 
\frac{\partial}{\partial \mu_{_\scS,j}}\,
\cD^{[1]}(\mur,\vec\mu)
\;.
\end{equation}
This gives us $\isbra{1}\cS_j(\mur,\vec\mu) = 0$. Then the shower evolution operator $\cU(t_2,t_1)$, Eq.~(\ref{eq:cUexponential}), is probability preserving:\footnote{The situation is more subtle when there are one or two hadrons in the initial state because then the shower evolution involves the evolution of the parton distribution functions \cite{NSThreshold, NSThresholdII}.}
\begin{equation}
\sbra{1}\cU\big(t_2,t_1\big) = \sbra{1}
\;.
\end{equation}
%

%-------------------------------------------------
\subsection{The form of $\cD^{[1]}_{l,\mathrm{soft}}$}
\label{sec:cD1lsoft}

In the main text, we have used a decomposition, Eq.~(\ref{eq:scsoft}), of $\bm D_l(\{\hat p,\hat f\}_\mpone;\{p,f\}_m)$ into a part with both soft and collinear singularities and a part with only soft singularities:
\begin{equation}
\begin{split}
\label{eq:scsoftencore}
\bm D_l
={}& 
\bm D_l^\mathrm{sc} 
+ \bm D_l^\mathrm{soft}
\;.
\end{split}
\end{equation}
In Eq.~(\ref{eq:cSjLCplus}), this decomposition was achieved using the LC+ approximation for color:
\begin{equation}
\begin{split}
\bm D_l^\mathrm{sc} ={}& \bm D_l^\mathrm{LC+}
\;,
\\
\bm D_l^\mathrm{soft}={}& \bm D_l - \bm D_l^\mathrm{LC+}
\;.
\end{split}
\end{equation}

The LC+ approximation \cite{NScolor} is simple. To define $\hat {\bm D}_l^\mathrm{LC+}$, we start with $\hat {\bm D}_l$ in Eq.~(\ref{eq:D10l}) and drop some contributions. We keep all of the contributions for $k = l$. In the contributions for $k\ne l$ (for which parton $\mpone$ is a gluon), we expand $\bra{\{c'\}_m} t_k(f_k \to f_k + \Lg)$ and $t^\dagger_k(f_k \to f_k + \Lg) \ket{\{c\}_m}$ in color basis vectors and retain all contributions in which parton $\mpone$ is color connected to parton $l$, dropping all other contributions.

The corresponding expression for $\hat {\bm P}_l^\mathrm{LC+}$  is obtained from $\hat {\bm P}_l$ in  Eq.~(\ref{eq:1PlB}) by retaining the terms proportional to $\hat P^{\hat a a}(z,y,\epsilon)$ times the unit color matrix. Then for each $k \ne l$ term that was retained in $\hat {\bm D}_l^\mathrm{LC+}$, the color matrix $\bm T_k \cdot \bm T_l$ is replaced by $C_\LA/2$ or $C_\LF$ times the unit color matrix \cite{NScolor}.

The result of this is that $\hat {\bm D}_l^\mathrm{soft}$ and $\hat {\bm P}_l^\mathrm{soft}$ are given by expressions analogous to the $\hat {\bm D}_l$ and $\hat {\bm P}_l$ that contain only terms proportional to $W_0$ times color operators. Recall that $W_0$ has soft singularities but no collinear or soft$\times$collinear singularities. We conclude that $\hat {\bm D}_l^\mathrm{soft}$ and $\hat {\bm P}_l^\mathrm{soft}$ have only soft singularities.

In the formulation of a shower with two scales as presented in the main text, we take $\vec\mu = (\mu_\scE, \mu_\scC)$, where the collinear sensitive scale $\mu_\scC$ is one of $\mu_\angle$, $\mu_\Lambda$, or $\mu_\perp$. Then $\mu_\scE$ controls the soft singularity according to Eq.~(\ref{eq:aEdef}). Then for $\bm D_l^\mathrm{LC+}$ and $\bm P_l^\mathrm{LC+}$, we use the unresolved region $U(\vec\mu) = U(\mu_\scE, \mu_\scC)$ as defined by Eq.~(\ref{eq:twoscaleangle}). However, for $\bm D_l^\mathrm{soft}$ and $\bm P_l^\mathrm{soft}$, there is no collinear singularity so we can use the unresolved region $U(\mu_\scE, 0)$  as defined by Eq.~(\ref{eq:unresolvedsoft}).

%-------------------------------------------------
\section{Thrust logarithms for angular ordering}
\label{sec:I2angle}

%%%%%%%%%%%%%%%%%%%% FIGURE %%%%%%%%%%%%%%%%%%%%%%%%%%%
\begin{figure}
\begin{center}
\ifusefigs % then we include the figure

\begin{tikzpicture}
\begin{axis}[title = {$\Lambda$ and $k_\LT$ ordering},
   xlabel=$\log(\nu)$, 
   ylabel={{$\langle I^{[2]}_2(\nu)\rangle$}},
   xmin=0, xmax=16,
%  ymin=, ymax=
  legend cell align=left,
  every axis legend/.append style = {
  at={(0.1,0.1)},
  anchor=south west}
]

\errorband[red,semithick]{fill=red!30!white, opacity=0.5}
{data/I2Lambda-20.dat}{0}{1}{2}
\addlegendentry{$\langle I^{[2]}_2(\nu)\rangle$\ ($\Lambda$)}
%\errorband[blue,semithick]{fill=blue!30!white, opacity=0.5}
%{data/I2Lambda-20.dat}{0}{3}{4}
%\addlegendentry{$d \langle I^{[2]}_2(\nu)\rangle/d\log\nu$\ ($\Lambda$)} 

\errorband[blue,  semithick]{fill=blue!30!white, opacity=0.5}
{data/I2kT-20.dat}{0}{1}{2}
\addlegendentry{$\langle I^{[2]}_2(\nu)\rangle$\ ($k_\LT$)}
%\errorband[blue, dashed, semithick]{fill=blue!30!white, opacity=0.5}
%{data/I2kT-20.dat}{0}{3}{4}
%\addlegendentry{$d \langle I^{[2]}_2(\nu)\rangle/d\log\nu$ ($k_\LT$)}    

\end{axis}
\end{tikzpicture}

\else % We use the pdf figure.
\begin{center}
\includegraphics[width = 8.2 cm]{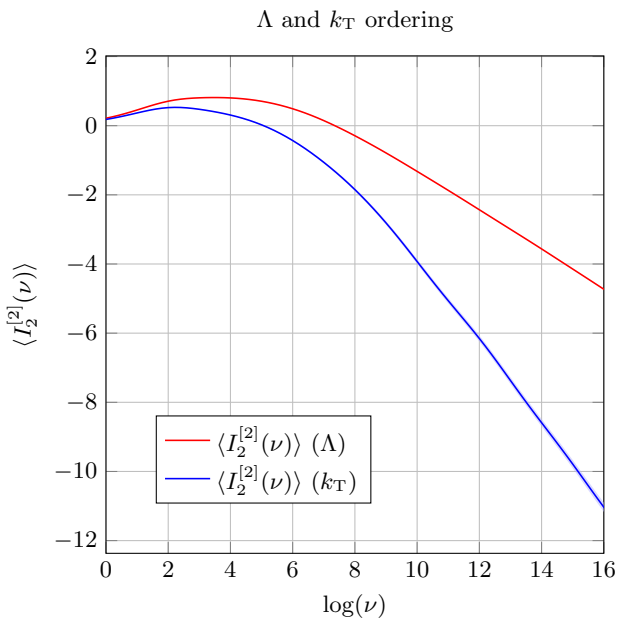}
\end{center}
\fi
\end{center}
\caption{
$\langle I^{[2]}_2(\nu)\rangle$, as in Ref.~\cite{NSThrustSum}, versus the Laplace parameter $\nu$ for the thrust distribution for $\Lambda$ ordering and $k_\LT$ ordering.
}
\label{fig:I22Lambda}
\end{figure}
%%%%%%%%%%%%%%%%%%% END FIGURE %%%%%%%%%%%%%%%%%%%%%%%%

In Fig.~\ref{fig:twojet}, we illustrated the application of the methods of this paper to the two jet cross section with the Cambridge algorithm in $e^+e^-$ annihilation. This is quite simple since the contribution from the first component of the two component path is just the unit operator when one starts with just a $q \bar q$ state. A surprising (at least to us) outcome was that with angular ordering for the second component of the path, the results were quite different than with $\Lambda$ ordering or $k_\LT$ ordering for the second component.

Although the question of why this is lies outside of the main topic of this paper, we investigate in this appendix whether leaving everything the same in the \textsc{Deductor} code used for this paper and simply changing from $\Lambda$ or $k_\LT$ ordering to angular ordering might change the accuracy with which the shower sums large logarithms. 

For this purpose, we consider the thrust distribution, which we had previously investigated \cite{NSThrustSum} (although not with angular ordering).  The thrust, $T$, distribution is strongly peaked at small $1-T$. It contains a factor $1/(1-T)$ and large logarithms of $(1-T)$. To investigate these logarithms, one takes the Laplace transform $\tilde g(\nu)$ of the $(1-T)$ distribution, with Laplace transform variable $\nu$. For large $\nu$, this function contains contributions proportional to $a_s^n \log^k(\nu)$ with $k \le 2 n$. In QCD, $\tilde g(\nu)$ exponentiates in the sense that $\log[\tilde g(\nu)]$ contains contributions proportional to $a_s^n \log^k(\nu)$ with $k \le n+1$. The terms with $k=n+1$ are the leading-log (LL) terms and the terms with $k=n$ are the next-to-leading-log (NLL) terms. These terms are calculated analytically in Ref.~\cite{thrustsum}. Ref.~\cite{NSThrustSum} provides both analytical and numerical methods for investigating whether a parton shower reproduces those terms. In this appendix, we use one of the numerical methods. We calculate certain quantities $\langle I^{[J]}_n(\nu)\rangle$ that are based on operating $J$ times with the shower splitting operator and calculating its contribution at order $\as^n$  to $\log[\tilde g(\nu)]$ minus what $\log[\tilde g(\nu)]$ should be according to the analytic result. 

We calculate $\langle I^{[2]}_2(\nu)\rangle$ for the \textsc{Deductor} splitting functions with exact SU(3) color. In Fig.~\ref{fig:I22Lambda}, we show the results with $\Lambda$ ordering and $k_\LT$ ordering.\footnote{The result in this figure is close that of Figs.~1 and 6 of Ref.~\cite{NSThrustSum}. There are small differences because the revised code in this paper treats the running coupling $\as$ slightly differently from the code in Ref.~\cite{NSThrustSum}.} This is an order $\as^2$ contribution, so the NLL term in the analytical result is proportional to $\log^2(\nu)$. If the parton shower is giving a result correct to NLL, then $\langle I^{[2]}_2(\nu)\rangle$ should {\em not} contain a $\log^2(\nu)$ contribution for large $\nu$. Thus, for NLL accuracy, the curves representing $\langle I^{[2]}_2(\nu)\rangle$ should be a linear functions of $\log(\nu)$, as indeed they are. 

\begin{figure}
\begin{center}
\ifusefigs % then we include the figure

\begin{tikzpicture}
\begin{axis}[title = {Angular ordering},
   xlabel=$\log(\nu)$, 
   ylabel={{$\langle I^{[2]}_2(\nu)\rangle$}},
   xmin=0, xmax=16,
%  ymin=, ymax=
  legend cell align=left,
  every axis legend/.append style = {
  at={(0.1,0.9)},
  anchor=north west}
]

\errorband[red,semithick]{fill=red!30!white, opacity=0.5}
{data/I2angle-20.dat}{0}{1}{2}
\addlegendentry{$\langle I^{[2]}_2(\nu)\rangle$}
\errorband[blue,semithick]{fill=blue!30!white, opacity=0.5}
{data/I2angle-20.dat}{0}{3}{4}
\addlegendentry{$d \langle I^{[2]}_2(\nu)\rangle/d\log\nu$}    
    
\errorband[red,dashed,semithick]{fill=red!30!white, opacity=0.5}
{data/I2Lambda-20.dat}{0}{1}{2}
\addlegendentry{$\langle I^{[2]}_2(\nu)\rangle$, $\Lambda$ ordered}    

%\errorband[blue, dashed, semithick]{fill=blue!30!white, opacity=0.5}
%{data/I2kT-20.dat}{0}{1}{2}
%\addlegendentry{$\langle I^{[2]}_2(\nu)\rangle$\ ($k_\LT$)}

\end{axis}
\end{tikzpicture}

\else % We use the pdf figure.
\begin{center}
\includegraphics[width = 8.2 cm]{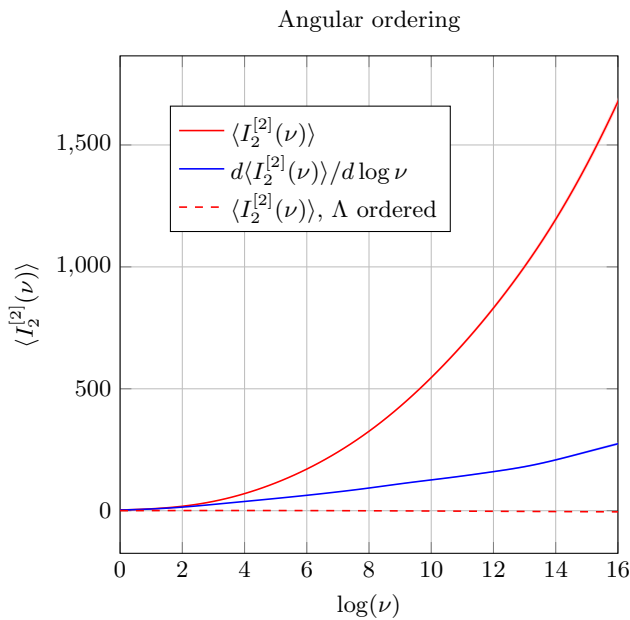}
\end{center}
\fi
\end{center}
\caption{
$\langle I^{[2]}_2(\nu)\rangle$, as in Ref.~\cite{NSThrustSum}, versus the Laplace parameter $\nu$ for the thrust distribution for angular ordering. The $\Lambda$-ordered result for $\langle I^{[2]}_2(\nu)\rangle$ is also shown as a dashed line.
}
\label{fig:I22angle}
\end{figure}

Now we try the same calculation with angular ordering. We display the result in Fig.~\ref{fig:I22angle}. We see, first, that $\langle I^{[2]}_2(\nu)\rangle$ is much larger in magnitude than the same quantity with $\Lambda$ ordering, which is shown as a dashed line. This suggests a failure of cancellation of large contributions. For NLL accuracy, $\langle I^{[2]}_2(\nu)\rangle$ should be a linear function of $\log(\nu)$ for large $\nu$ but it is not. The blue curve shows $d\langle I^{[2]}_2(\nu)\rangle /d \log(\nu)$. For LL accuracy, this curve should be a linear function of $\log(\nu)$ for large $\nu$. The numerical evidence is perhaps not definitive, but this evidence suggests a failure of the angular ordered shower to achieve LL accuracy. We emphasize that the code for Figs.~\ref{fig:I22Lambda} and \ref{fig:I22angle} is the same except for changing the ordering variable.

%%%%%%%%%%%%%%%%%%%% FIGURE %%%%%%%%%%%%%%%%%%%%%%%%%%%
%%%%%%%%%%%%%%%%%%% END FIGURE %%%%%%%%%%%%%%%%%%%%%%%%

%-------------------------------------------------
%-------------------------------------------------

%%%%%%%%%%%%%%%%%%%%%%%%%%%%%%%%%%%%%%%%%%%%%%%%%%%%%%%%%%%%%%%%%%%%%%%%%%%%%%%%
%%%%%%%%%%%%%%%%%%%%%%%%%%%%%%%%%%%%%%%%%%%%%%%%%%%%%%%%%%%%%%%%%%%%%%%%%%%%%%%%%%%%

%-------------------------------------------------------------------

\end{document}